\documentclass[aps,amssymb,amsmath,notitlepage,superscriptaddress,pra,twocolumn,nofootinbib]{revtex4-1}

\usepackage{amsfonts}
\usepackage{graphicx,graphics,epsfig,times,bm,bbm,amssymb,amsmath,amsfonts,mathrsfs}
\usepackage[normalem]{ulem}
\usepackage{wrapfig}
\usepackage{setspace}
\usepackage{subfigure}
\usepackage{dsfont}
\usepackage{braket}
\usepackage[pdftex]{color}
\usepackage[pdfstartview=FitH]{hyperref}
\usepackage{upgreek }
\usepackage[x11names]{xcolor}
\usepackage{tikz}
\usepackage{natbib}
\usepackage{chngcntr}

\hypersetup{
    colorlinks=true,       
    linkcolor=cyan,          
    citecolor=magenta,        
    filecolor=magenta,      
    urlcolor=cyan,           
    runcolor=cyan
}

\newcommand{\bes} {\begin{subequations}}
\newcommand{\ees} {\end{subequations}}
\newcommand{\bea} {\begin{eqnarray}}
\newcommand{\eea} {\end{eqnarray}}
\newcommand{\beq} {\begin{equation}}
\newcommand{\eeq} {\end{equation}}

\newcommand{\mc}{\mathcal}

\def\>{\rangle}
\def\<{\langle}
\def\Tr{\mathrm{Tr}}

\newcommand{\Var}{\mathrm{Var}} 
\newcommand{\abs}[1]{\lvert #1 \rvert}
\newcommand{\ketbra}[2]{|{#1}\>\<#2|}

\newcommand{\eps}{\varepsilon}
\newcommand{\ident}{\mathds{1}}
\newcommand{\ignore}[1]{}

\begin{document}
\title{Consistency Tests of Classical and Quantum Models for a Quantum Annealer}
\author{Tameem Albash}
\affiliation{Department of Physics and Astronomy, University of Southern California, Los Angeles, California 90089, USA}
\affiliation{Center for Quantum Information Science \& Technology, University of Southern California, Los Angeles, California 90089, USA}
\affiliation{Information Sciences Institute, University of Southern California, Marina del Rey, CA 90292}
\author{Walter Vinci}
\affiliation{Center for Quantum Information Science \& Technology, University of Southern California, Los Angeles, California 90089, USA}
\affiliation{London Centre for Nanotechnology, University College London, WC1H 0AH London, UK}
\affiliation{Department of Computer Science, University College London, WC1E 6BT London, UK}
\author{Anurag Mishra}
\affiliation{Department of Physics and Astronomy, University of Southern California, Los Angeles, California 90089, USA}
\affiliation{Center for Quantum Information Science \& Technology, University of Southern California, Los Angeles, California 90089, USA}
\author{Paul A. Warburton}
\affiliation{London Centre for Nanotechnology, University College London, WC1H 0AH London, UK}
\affiliation{Department of Electronic \& Electrical Engineering, University College London, WC1E 7JE London, UK}
\author{Daniel A. Lidar}
\affiliation{Department of Physics and Astronomy, University of Southern California, Los Angeles, California 90089, USA}
\affiliation{Center for Quantum Information Science \& Technology, University of Southern California, Los Angeles, California 90089, USA}
\affiliation{Department of Electrical Engineering, University of Southern California, Los Angeles, California 90089, USA}
\affiliation{Department of Chemistry, University of Southern California, Los Angeles, California 90089, USA}

\begin{abstract}
Recently the question of whether the D-Wave processors exhibit large-scale quantum behavior or can be described by a classical model has attracted significant interest. In this work we address this question by studying a $503$ qubit D-Wave Two device in the ``black box" model, i.e., by studying its input-output behavior. Our work generalizes an approach introduced in Boixo \textit{et al.} [\href{http://dx.doi.org/10.1038/ncomms3067}{Nat. Commun. \textbf{4}, 2067 (2013)}], and uses groups of up to $20$ qubits to realize a transverse Ising model evolution with a ground state degeneracy whose distribution acts as a sensitive probe that distinguishes classical and quantum models for the D-Wave device.  Our findings rule out all classical models proposed to date for the device and provide evidence that an open system quantum dynamical description of the device that starts from a quantized energy level structure is well justified, even in the presence of relevant thermal excitations and a small value of the ratio of the single-qubit decoherence time to the annealing time.
\end{abstract}
\maketitle
%
\section{Introduction}
%
How can one determine whether a given ``black box" is quantum or classical \cite{Reichardt:2013db}? A case in point are the devices built by D-Wave \cite{Johnson:2010ys,Berkley:2010zr,Harris:2010kx}. These devices are commercial computers that the user can only access via an input-output interface. Reports \cite{DWave-16q,q-sig,q108} that the D-Wave devices implement quantum annealing (QA) with hundreds of qubits have attracted much attention recently, and have also generated considerable debate \cite{Aaronson-blog,Smolin,comment-SS,SSSV,SSSV-comment}. At stake is the question of whether the experimental evidence suffices to rule out classical models, and whether a quantum model can be found that is in full agreement with the evidence. 

It is our goal in this work to distinguish several classical models (simulated annealing, spin dynamics \cite{Smolin}, and hybrid spin-dynamics Monte Carlo \cite{SSSV}) and a quantum adiabatic master model \cite{ABLZ:12-SI} of the D-Wave device, and to decide which of the models survives a comparison with the experimental input-output data of a ``quantum signature" test. This test is not entanglement-based and does not provide a Bell's-inequality-like \cite{Bell-book} no-go result for classical models. Instead, our approach is premised on the standard notion of what defines a ``good theory:'' it should have strong predictive power. That is, if the theory has free parameters then these can be fit once, and future predictions cannot require that the free parameters be adjusted anew. It is in this sense that we show that we can rule out the classical models, while at the same time we find that the adiabatic quantum master equation passes the ``good theory" test.

The D-Wave devices operate at a non-zero temperature that can be comparable to the energy gap from the ground state, so one might expect that thermal excitations act to drive the system out of its ground state, potentially causing the annealing process to be dominated by thermal fluctuations rather than by quantum tunneling.  Furthermore, the coupling to the environment should cause decoherence, potentially resulting in the loss of any quantum speedup. This issue was recently studied in Refs.~\cite{q-sig,q108}, where data from a $108$-qubit D-Wave One (DW1) device was compared to numerical simulations implementing classical simulated annealing (SA), simulated quantum annealing (SQA) using quantum Monte Carlo, and a quantum adiabatic master equation (ME) derived in Ref.~\cite{ABLZ:12-SI}. These studies demonstrated that SA correlates poorly with the experimental data, while the ME (in Ref.~\cite{q-sig}) and SQA (in Ref.~\cite{q108}) are in good agreement with the same data. 
Specifically, the eight-qubit ``quantum signature'' Hamiltonian introduced in Ref.~\cite{q-sig} has a $17$-fold degenerate ground state that splits into a single ``isolated" state and a $16$-fold degenerate ``cluster,'' with the population in the former suppressed relative to the latter according to the ME but enhanced according to SA; the experiment agreed with the ME prediction \cite{q-sig}. 
Subsequently, Ref.~\cite{q108} rejected SA on much larger problem sizes by showing that the ground state population (``success probability") distribution it predicts for random Ising instances on up to $108$ spin variables is unimodal, while the experimental data and SQA both give rise to a bimodal distribution. This was interpreted as positive evidence for the hypothesis that the device implements QA.  

However, interesting objections to the latter interpretation were raised in Refs.~\cite{Smolin,SSSV}, where it was argued that there are other classical models that also agree with the experimental data of Refs.~\cite{q-sig,q108}. First, Smolin and Smith \cite{Smolin} pointed out that a classical spin-dynamics (SD) model of O(2) 
rotors could be tuned to mimic the suppression of the isolated ground state found in Ref.~\cite{q-sig} and the bimodal success probability histograms for random Ising instances found in Ref.~\cite{q108}. Shortly thereafter this classical model was rejected in Ref.~\cite{comment-SS} by demonstrating that the classical SD model correlates poorly with the success probabilities measured for random Ising instances, while SQA correlates very well. In response, a new hybrid model where the spin dynamics are governed by Monte Carlo updates that correlates at least as well with the DW1 success probabilities for random Ising instances as SQA was very recently proposed by Shin, Smolin, Smith, and Vazirani (SSSV) \cite{SSSV}. In this model the qubits are replaced by O(2) rotors with classical Monte Carlo updates along the annealing schedule of the D-Wave device. 
This can also be interpreted as a model of qubits 
without any entanglement, updated at each time step to the classical thermal equilibrium state determined by the instantaneous Hamiltonian.  
Moreover, the hybrid model correlates almost perfectly with SQA, suggesting that the SSSV model is a classical analog of a mean-field approximation to SQA, and that this approximation is very accurate for the set of problems solved by the DW1 in Ref.~\cite{q108}.\footnote{We note that phases of quantum models often have an accurate mean-field description, and that SQA is a classical  simulation method obtained by mapping a quantum spin model to a classical one after the addition of an extra spatial dimension of extent $\upbeta$ (the inverse temperature). Moreover, SQA scales polynomially in problem size, which is the reason that Ref.~\cite{q108} was able to use SQA to predict the experimental outcomes of $108$ qubit problem instances.}

At this point it is important to note that recent work already established that eight-qubit entangled ground states are formed during the course of the annealing evolution in experiments using a D-Wave Two device \cite{DWave-entanglement}. This demonstration of entanglement was done outside of the ``black box" paradigm we are considering here\footnote{The experiment had access to the internal workings of the D-Wave device, in particular the ability to perform qubit tunneling spectroscopy and thus obtain the instantaneous energy spectrum.} and is, of course, a crucial demonstration of non-classicality. However, it does not necessarily imply that non-classical effects play a role in deciding the \emph{final outcome} of a computation performed by the D-Wave devices. It is the latter that we are concerned with in this work, and it is the fundamental reason we are interested in the ``black-box" paradigm. 

Using a physically motivated noise model, we show that none of the three classical models introduced to date matches new data we obtained from the D-Wave Two (DW2) device using a generalized ``quantum signature'' Hamiltonian on up to $20$ qubits.  At the same time the ME matches the new data well. Thus, our results confirm the earlier rejection of the SA and SD models---this is of independent interest since the ``quantum signature" provided in Ref.~\cite{q-sig} for the DW1 had remained in question in light of the SD-based critique of Ref.~\cite{Smolin}---and also serve to reject the new SSSV model \cite{SSSV} for system sizes of up to $20$ qubits. Of course, this still leaves open the possibility that a classical model can be found that will match the experimental data while satisfying the ``good theory" criteria.  Since, as mentioned above, our quantum signature-based test does not provide a no-go result for classical models, the distinction we demonstrate between a natural quantum model and fine-tuned classical models is perhaps the best that can be hoped for within our approach.

The ``quantum signature'' Hamiltonian we consider here 
is defined in Sec.~\ref{sec:background} and is a direct generalization of the Hamiltonian introduced in Ref.~\cite{q-sig}. We introduce a controllable overall energy scale, or an effective (inverse temperature) ``noise control knob''. Decreasing the energy scale amounts to increasing thermal excitations, enabling us to drive the D-Wave processor between qualitatively distinct regimes. At the largest energy scale available, the annealing process appears to be dominated by coherent quantum effects, and thermal fluctuations are negligible.  As the energy scale is decreased, thermal excitations become more relevant, and for a sufficiently small energy scale, the system behaves more like a classical annealer based on incoherent Ising spins.  
Nevertheless, at all energy scales the system is very well described by the ME. This suggests that an open system quantum dynamical description of the D-Wave device is well justified, even in the presence of relevant thermal excitations and a small single-qubit decoherence time to annealing time ratio, at least for the class of Hamiltonians studied here.

The structure of this paper is as follows. We provide theoretical background on the quantum signature Hamiltonian---our workhorse in this study---in Sec.~\ref{sec:background}.  We describe a noise model for the D-Wave device in Sec.~\ref{sec:cross-talk}, which includes both stochastic and systematic components, the latter being dominated by spurious qubit cross-talk. We analyze the effect of tuning the thermal noise via the magnitude of the final Hamiltonian in Sec.~\ref{sec:tuning}. Our first set of main results is presented in Sec.~\ref{sec:sim}, where we demonstrate a clear difference between the behavior of the classical model and that of the quantum model in the absence of cross-talk.  We then include the cross-talk and establish in Sec.~\ref{sec:calib-noise} the input-output characteristics of the D-Wave device that allow us to critically assess the classical models, and confirm the agreement with an open quantum system description via the adiabatic quantum ME.  We achieve a close match between the ME and the experimental data, while rejecting the SSSV model, the strongest of the classical models. We demonstrate that the ME predicts that an entangled ground state is formed during the course of the annealing evolution in Sec.~\ref{sec:ent}.  We provide a discussion and conclusions in Sec.~\ref{sec:conc}. The appendices provide further technical details and experimental and numerical results.

\section{Theoretical background}
\label{sec:background}
Quantum and classical annealing are powerful techniques for solving hard optimization problems, whether they are implemented as numerical algorithms or on analog (physical) devices.
The general \emph{simulation} strategy is to implement an ``escape'' rule from local minima of an energy or penalty function to reach the global minimum, representing a solution of the optimization problem \cite{kirkpatrick_optimization_1983,finnila_quantum_1994,kadowaki_quantum_1998}. The \emph{physical} strategy is to use a natural system or build a device whose physical ground state represents the sought-after solution \cite{brooke_quantum_1999,brooke_tunable_2001,2002quant.ph.11152K,Dwave}. In both cases, by progressively reducing the escape probability, the system is allowed to explore its configuration space and eventually ``freeze'' in the global minimum with some probability. 

\subsection{Quantum annealing Hamiltonian}
The QA Hamiltonian is given by
\beq
H(t) = A(t) H_X + B(t)  H_{\mathrm{I}} \ ,
\label{eq:adiabatic}
\eeq
where $H_X = - \sum_i \sigma_i^x$ (with $\sigma_i^x$ being the Pauli matrix acting on qubit $i$) is the transverse field, $H_{\mathrm{I}}$ is the classical Ising Hamiltonian,
\beq 
H_{\mathrm{I}} = -\sum_{i\in \mc{V}} h_i \sigma_i^z - \sum_{(i,j)\in \mc{E}} J_{ij} \sigma_i^z \sigma_j^z \ ,
\label{eq:problem}
\eeq
and the time-dependent functions $A(t)$ and $B(t)$ control the annealing schedule. Typically $A(t_f) = B(0) = 0$, where $t_f$ is the total annealing time, and $A(t)$ [$B(t)$] decreases (increases) monotonically. The local fields $\{h_i\}$ and couplings $\{J_{ij}\}$ are fixed. The qubits occupy the vertices $\mc{V}$ of a graph $G=\{\mc{V},\mc{E}\}$ with edge set $\mc{E}$.

A \emph{spin configuration} is one of the $2^N$ elements of a set of $\pm 1$ eigenvalues of all the Pauli matrices $\{\sigma_i^z\}_{i=1}^N$, which we denote without risk of confusion by $\vec{\sigma}^z = (\sigma_1^z,\dots,\sigma_N^z)$. The goal is to find the minimal energy spin configuration of $H_{\mathrm{I}}$, i.e., $\textrm{argmin}_{\vec{\sigma}^z}H_{\mathrm{I}}$. In QA, the non-commuting field $H_X$ \cite{finnila_quantum_1994,kadowaki_quantum_1998,morita:125210,RevModPhys.80.1061} allows quantum tunneling out of local minima. This ``escape probability'' is reduced by turning off this non-commuting field adiabatically, i.e., the time-scale of the variation of the $A(t)$ and $B(t)$ functions must be slow compared to the inverse of the minimal energy gap of $H(t)$. In a physical device implementation of QA there is always a finite temperature effect, and hence one should consider both tunneling and thermal barrier crossing \cite{Lee:1999lq,childs_robustness_2001,PhysRevLett.95.250503,TAQC}. 
 
\begin{figure}[t] 
\includegraphics[width=0.95\columnwidth]{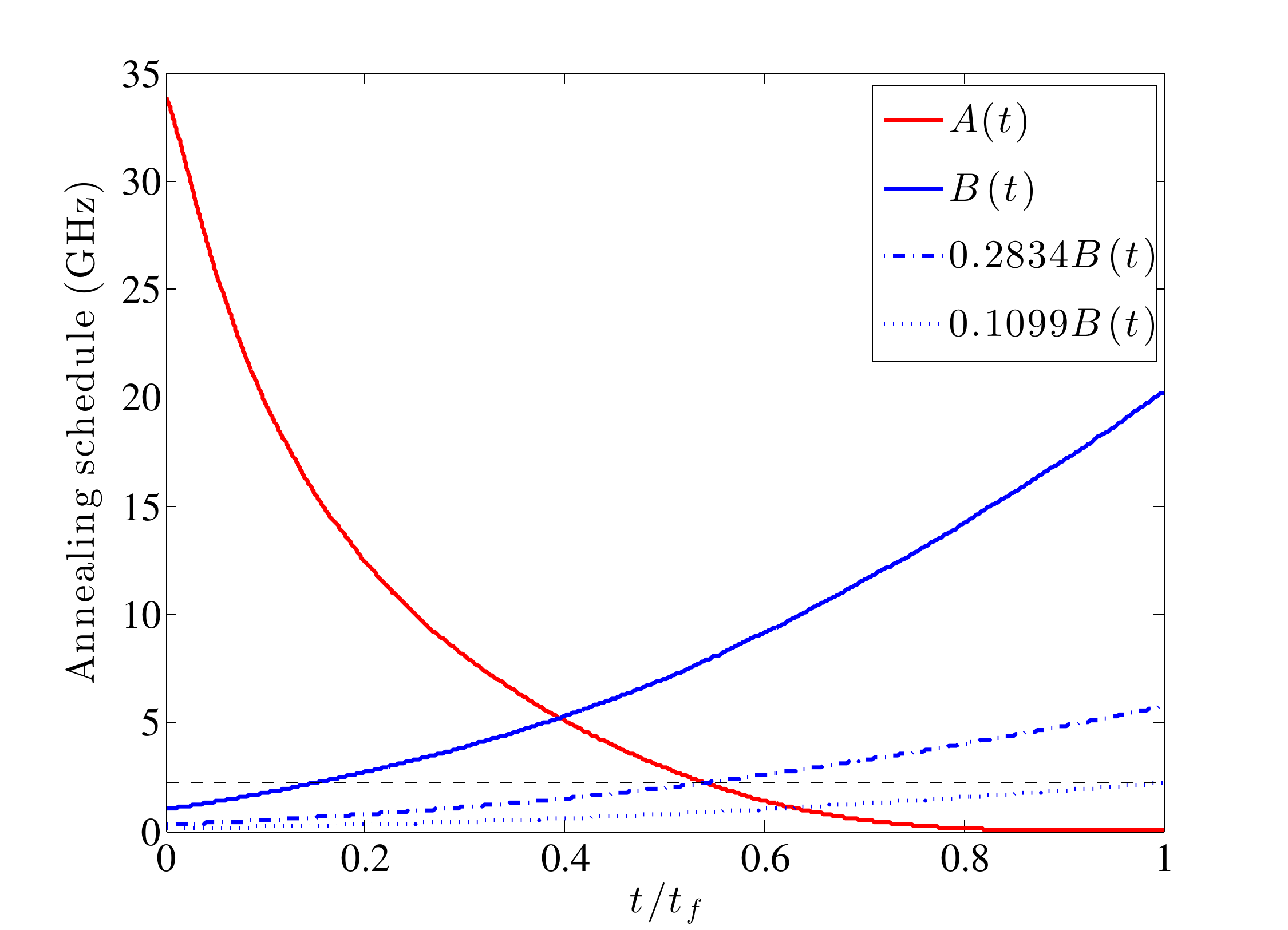}
\caption{ DW2 annealing schedules $A(t)$ and $B(t)$ along with the operating temperature of $T=17$mK (black dashed horizontal line). The large $A(0)/(k_B T)$ value ensures that the initial state is the ground state of the transverse field Hamiltonian. The large $B(t_f)/(k_B T)$ value ensures that thermal excitations are suppressed and the final state reached is stable. Also shown are the attenuated $\alpha B(t)$ curves for (a) the value of $\alpha$ at which the intersection between $A(t)$ and $\alpha B(t)$ coincides with the operating temperature (blue dot-dashed curve), and (b) the largest $\alpha$ such that $\alpha B(t)$ remains below the temperature line for the entire evolution (blue dotted curve).}
\label{fig:AnnealingSchedule}
\end{figure}

Such physical QA devices, operating at $\sim 20\, \mathrm{mK}$ using superconducting flux technology, have been built by D-Wave~\cite{Johnson:2010ys,Berkley:2010zr,Harris:2010kx}. The qubits occupy the vertices of the ``Chimera'' graph (shown in the Appendix \ref{app:device}). Excluding the coupling to the thermal bath, the Hamiltonian driving the device is well-described by Eq.~\eqref{eq:adiabatic}, with 
the functions $A(t)$ and $B(t)$ depicted in Fig.~\ref{fig:AnnealingSchedule}.

\begin{figure}[t] 
\centering
 \includegraphics[height=1.75in]{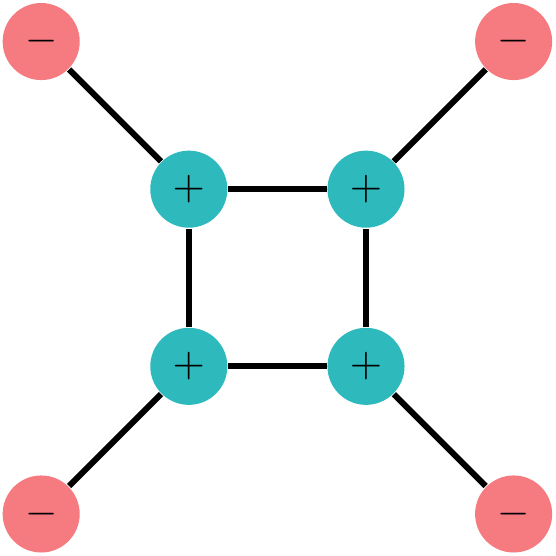} 
\caption{ The eight-spin Ising Hamiltonian. The inner ``core" spins (green circles) have local fields $h_i = +1$ [using the convention in Eq.~\eqref{eq:problem}] while the outer spins (red circles) have $h_i = -1$.  All  couplings are ferromagnetic: $J_{ij} = 1$ (black lines).}
\label{fig:8spinIsing}
\end{figure}

\subsection{The quantum signature Hamiltonian}
\begin{figure*}[t] 
\centering
 \includegraphics[height=1.5in]{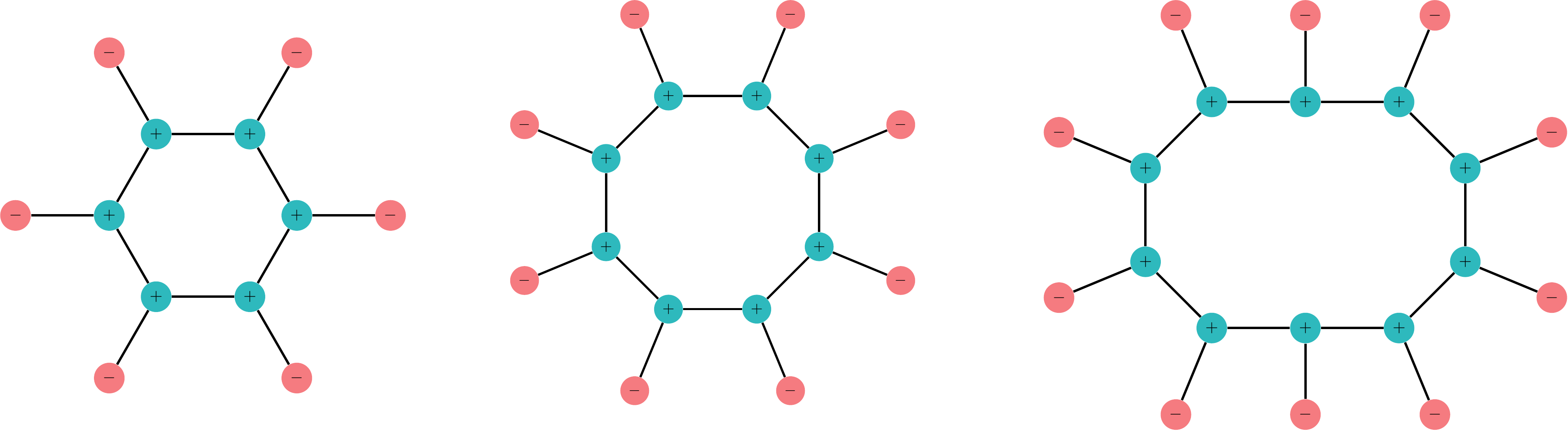} 
\caption{ Schematic representation of the $12$, $16$, and $20$ spin Hamiltonians used in our tests.  Extensions to larger $N$ follow the same pattern, with $N/2$ qubits in the inner ring and $N/2$ qubits in the outer ring.  Notation conventions are as in Fig.~\ref{fig:8spinIsing}.}
\label{fig:12-16-20spinIsing}
\end{figure*}
Reference~\cite{q-sig} introduced an eight-qubit ``quantum signature Hamiltonian," schematically depicted in Fig.~\ref{fig:8spinIsing}, designed to distinguish between SA and QA.  The eight-spin problem comprises four spins connected in a ring, which we refer to as \emph{core} spins, and four additional spins connected to each core spin, which we refer to as \emph{outer} spins.  One special property of this Hamiltonian is that it has a $17$-fold degenerate ground state. Of these, $16$ states form a subspace of spin configurations connected via single flips of the outer spins, hence we refer to them as the \emph{clustered} (C) ground states, or just the ``cluster-states," or ``cluster." There is one additional state, which we call the \emph{isolated} (I) ground state, connected to the cluster-states via four core spin flips:
\bes
\label{eq:C-I}
\begin{align}
{\rm C}&:\{\ket{0000\,0000},\ket{0001\,0000},\dots,
\ket{1111\,0000}\} \ ,  \\
{ \rm I}&: \{\ket{\underbrace{1111}_{\text{outer}}\, \underbrace{1111}_{\text{core}}}\}  \ ,
\end{align}
\ees
where $\ket{0}$ and $\ket{1}$ are, respectively, the $+1$ and $-1$ eigenstates of $\sigma^z$. 
This structure of the ground state manifold is easily verified by inspection of the Hamiltonian of Fig.~\ref{fig:8spinIsing}. 

The clustered ground states arise from the frustration of the outer spins, due to the competing effects of the ferromagnetic coupling and local fields. This frustration arises only when the core spins have eigenvalue $+1$, which is why there is only a single additional (isolated) ground state where all spins have eigenvalue $-1$ (for a more detailed discussion of the energy landscape of this eight-spin Hamiltonian, see Ref.~\cite{q-sig}.)
Here we also consider quantum signature Hamiltonians with larger numbers of spins $N$, as depicted in Fig.~\ref{fig:12-16-20spinIsing}. These Hamiltonians share the same qualitative features, but the degeneracy of the ground state grows exponentially with $N$:
\begin{eqnarray}
\begin{split}
{\rm Degeneracy \, of \, C}&:&   N_\textrm{C}=2^{N/2} \ ;  \\
 {\rm Degeneracy \, of \, I}&:& N_\textrm{I}=1  \ .
\end{split}
\label{eq:degen}
\end{eqnarray}

At the end of any evolution, be it quantum or classical, at $t=t_f$, there is a certain probability of finding each ground state.  Let us denote the observed population of the isolated state at $t=t_f$ by $P_\textrm{I}$, and the average observed population in the cluster at $t=t_f$ by 
\beq
P_\textrm{C} = \frac{1}{N_\textrm{C}}\sum_{c=1}^{N_\textrm{C}} P_c\ ,
\label{eq:P_C}
\eeq 
where $N_\textrm{C} = 16$ for the eight-spin case, and where $P_c$ is the population of cluster-state number $c$. As shown in Ref.~\cite{q-sig} for the eight-spin case, SA and QA can be distinguished because they give opposite predictions for the population ratio $P_\textrm{I}/P_\textrm{C}$.  For SA, the isolated state population is enhanced relative to any cluster state's population, i.e., $P_\textrm{I}/P_\textrm{C} \geq 1$, whereas for QA, the isolated state population is suppressed relative to any cluster state's population, i.e $P_\textrm{I}/P_\textrm{C} \approx 0$. This conclusion also holds for the $N>8$ cases, as we show in detail in Appendix \ref{app:generalized}. 

These two starkly different predictions for QA and SA allowed Ref.~\cite{q-sig} to rule out SA as an explanation of  the experimental results obtained from the DW1 using the eight-spin Hamiltonian. In addition, Ref.~\cite{q-sig} demonstrated that the ME \cite{ABLZ:12-SI} correctly predicts the suppression of the isolated state, including the dependence on the annealing time $t_f$, thus providing evidence that the DW1 results correlate well with the predictions of open system quantum evolution. However, as we discuss in detail and demonstrate with data from the DW2, the suppression of the isolated state can change as a function of the thermal noise, and suppression can turn into enhancement at sufficiently high noise levels. Yet, this does \emph{not} imply that the system admits a classical description.

One limitation of the analysis so far is that the quantity $P_\mathrm{C}$ as defined in Eq.~\eqref{eq:P_C} is an average over the cluster state populations, so it does not account for variations in individual cluster state populations.  In the absence of any non-idealities in the quantum Hamiltonian of Eq.~\eqref{eq:adiabatic}, SA and QA (under a closed system evolution) predict that all the cluster states end up with equal populations.  Therefore, the cluster state populations alone cannot be used to distinguish between SA and QA, yet, as we will show, not all classical models preserve the cluster state symmetry, making it a useful feature to take into account.  However, when we make $|J|$ and $|h|$ unequal or introduce additional spurious couplings between qubits, the symmetry between the cluster states is broken in all models, and we must be careful to model such noise sources accurately.  

%
\section{D-Wave Control Noise Sources}
\label{sec:cross-talk}
%
So far we have not considered the effect of control noise on the local fields and couplings in $H_{\textrm {I}}$, which are important effects on the D-Wave processor.  For each annealing run of a given  problem Hamiltonian, the values of $\{h_i,\,J_{ij}\}$ are set with a Gaussian distribution centered on the intended value, and with standard error of about $5\%$~\cite{Trevor}. In our experiments we used averaging techniques described in Appendix \ref{app:experiment} to minimize the effects of this noise. However, it is important to test its effects on the quantum and classical models as well. 

Besides this random noise source on the $\{h_i,\,J_{ij}\}$, there is a spurious cross-talk between qubits. The model accounting for this cross-talk is
\bes
\label{eq:cross-talk}
\begin{align}
h_i &\mapsto h_{i} - \chi \sum_{k\neq i} J_{ik} h_k \\
J_{ij} &\mapsto J_{ij} + \chi \sum_{k\neq i,j} J_{ik} J_{jk} \ ,
\end{align}
\ees
where $\chi$ is the qubit background susceptibility multiplied by the mutual inductance \cite{Trevor}.

To understand the correction to $J_{ij}$ consider first the case where $i$ and $j$ are nearest-neighbor qubits (i.e., are perpendicular, intersecting superconducting loops in the same unit cell of the Chimera hardware graph), and $k$ is an index of qubits that are next nearest neighbors of $i$ (i.e., are parallel, non-intersecting loops in the same unit cell of the Chimera hardware graph) \cite{harris_flux_qubit_2010}. Then both $J_{ij}$ and $J_{jk}$ are existing ``legitimate" couplings in the unit cell, but $\chi J_{jk}$ is a spurious next nearest neighbor coupling. Next consider the case where qubits $i$ and $j$ are next-nearest neighbors (i.e., parallel loops), so that nominally $J_{ij}=0$.  Now the correction is due to qubits $k$ that are nearest neighbors of (i.e., perpendicular to) both $i$ and $j$. The sign is determined by the following rule of thumb: a ferromagnetic chain or ring is strengthened by the background $\chi$, i.e., the intermediate qubit is mediating an effective ferromagnetic interaction between next-nearest neighbors. 

While in reality $\chi$ is time and distance dependent, for simplicity we model it as constant and separately fit $\chi$ for the ME and SSSV to the DW2 cluster-state populations.  The sum in Eq.~\eqref{eq:cross-talk} extends over the unit cell of the Chimera graph, i.e., in the sum over $k$ we include all the couplings between physically parallel qubits. We set $J_{ik}$ and $J_{jk}$ equal to the unperturbed $J_{ij}$. To account in addition for the effect of Gaussian noise on $h$ and $J$, we apply the cross-talk perturbation terms after the Gaussian perturbation. 
\ignore{

}
\section{Introducing an energy scale}
\label{sec:tuning}
\begin{figure}[b] 
   \centering
     \includegraphics[width=0.98\columnwidth]{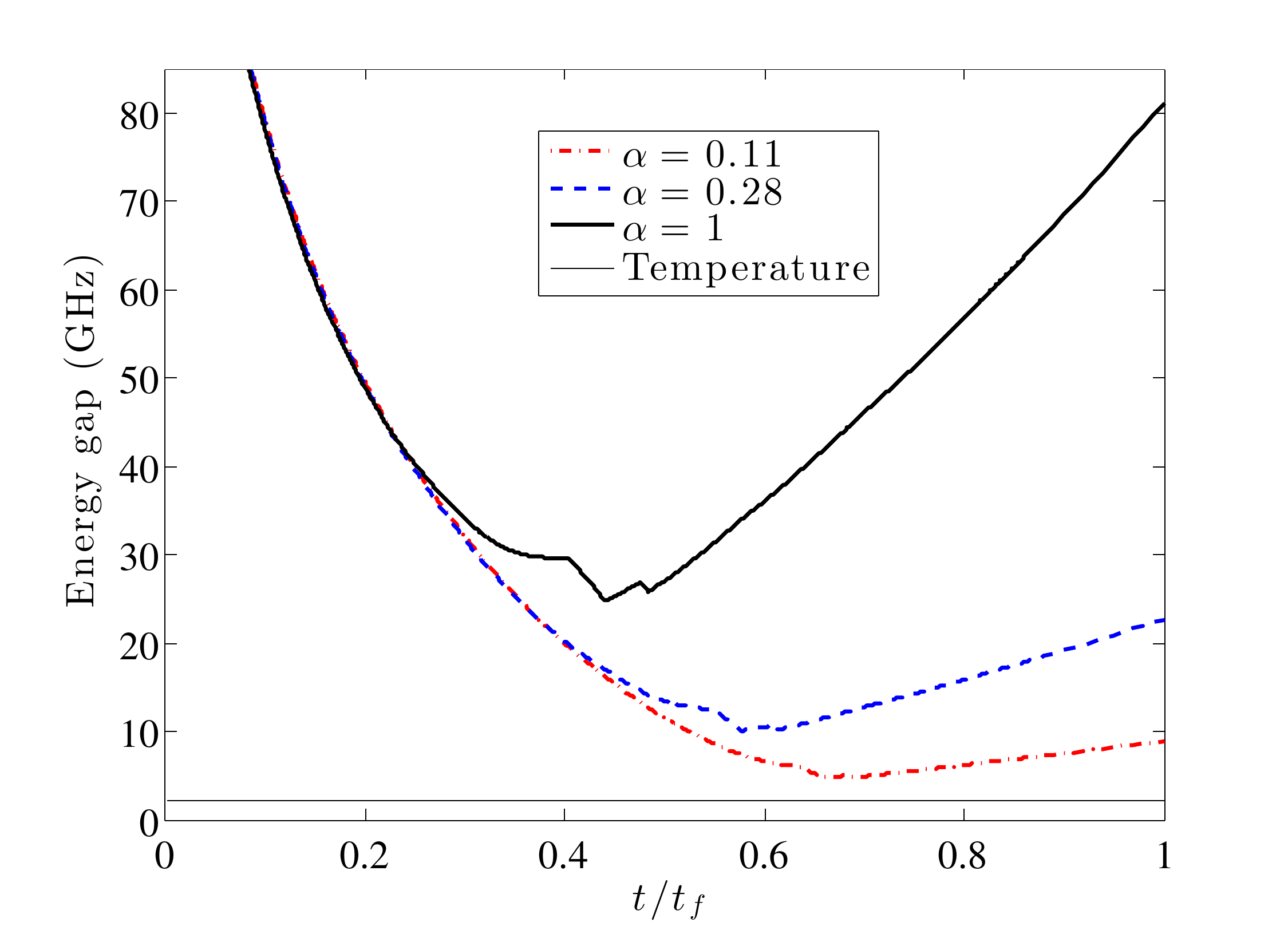}
   \caption{ Numerically calculated evolution of the gap between the instantaneous ground state and the $17$-th excited state (which becomes the first excited state at $t=t_f$), for the eight-spin Hamiltonian in Eq.~\eqref{eq:adiabatic}, following the annealing schedule of the DW2 device (Fig.~\ref{fig:AnnealingSchedule}). The gap value is shown for some interesting  values of $\alpha$ (see Fig.~\ref{fig:AnnealingSchedule}). The kinks are due to energy level crossings, as explained in Appendix \ref{app:kink}. A reduction in $\alpha$ results in a reduction of the size of the minimal gap and delays its appearance.}
   \label{fig:GapVsTime}
\end{figure}
The QA and SA protocols represent two opposite extremes: in the former quantum fluctuations dominate while in  the latter thermal fluctuations dominate.  Can we interpolate between these regimes on a physical annealer? Since we are unable to directly change the temperature on the DW2 device\footnote{On-chip variability of the SQUID critical currents leads to uncertainty in both the qubit biases $h$ and the qubit coupling strengths $J$. These uncertainties are calibrated out each time the chip is thermally cycled. The conditions required for optimum calibration are however temperature-dependent. We therefore conduct our experiments at a fixed operating temperature of $17$ mK.}, our strategy to answer this question is to indirectly modify the relative strength of thermal effects during the annealing process. As we now discuss, this can be done by modifying the overall energy scale. 

A straightforward way to tune the thermal noise indirectly is to change the overall energy scale of the problem Hamiltonian $H_\mathrm{I}$ by rescaling the local fields and couplings by an overall dimensionless factor denoted by $\alpha$:
\begin{equation}
(J_{ij},h_i) \, \mapsto \, \alpha \,(J_{ij},h_i)\,
\end{equation}
In the notation above, $\alpha = 1$ corresponds to implementing the largest allowed value of the physical couplings on the device (assuming $|h_\mathrm{max}|=|J_\mathrm{max}|=1$ in dimensionless units).  
The scale of the transverse field $H_X$ is not changed.
Due to the form of the cross-talk corrections this scales the cross-talk corrections by $\alpha^2$.  

For SA, reducing $\alpha$ is tantamount to increasing the temperature.  Since this does not change the energy spectrum, the earlier arguments for SA remain in effect, and we expect to have $P_\mathrm{I} \geq P_\mathrm{C}$ for all $\alpha > 0$ values. This is confirmed in our numerical simulations as shown in Appendix \ref{app:sims}.

For QA, decreasing $\alpha$ from the value $1$ has two main effects, as can be clearly seen in Fig.~\ref{fig:GapVsTime}.   First, the minimal gap between the instantaneous ground state and the $17$th excited state is reduced (the lowest $17$ states become degenerate at the end of the evolution as explained previously). Since thermal excitations are suppressed by a factor of $e^{- \upbeta \Delta}$ (with $\upbeta = 1/ k_B T$ the inverse temperature and $\Delta$ the energy gap), a reduction in the gap will increase the thermal excitation rate \cite{ABLZ:12-SI}. One might expect that, by sufficiently reducing $\alpha$, it is possible to make the gap small enough that the competition between non-adiabaticity and thermalization becomes important. However, simulations we have performed for the closed system case with $\alpha \in [0.01,1]$ show that $P_\textrm{I}/P_\textrm{C}$ is essentially $0$ over the entire range.  Therefore we do not expect non-adiabatic transitions to play a role over the entire range of $\alpha$'s we studied.
Second, reducing $\alpha$ delays the appearance of the minimal gap. This effectively prolongs the time over which thermal excitations can occur, thus also increasing the overall loss of the ground state population. Hence we see that by changing $\alpha$ we expect to move from a regime where thermal fluctuations are negligible ($\alpha \simeq 1$), to a regime where they are actually dominant ($\alpha \lesssim 0.1$), when the minimal gap is comparable to or even smaller than the physical temperature of the device.  

\begin{figure}[t] 
\centering
\includegraphics[width=0.98\columnwidth]{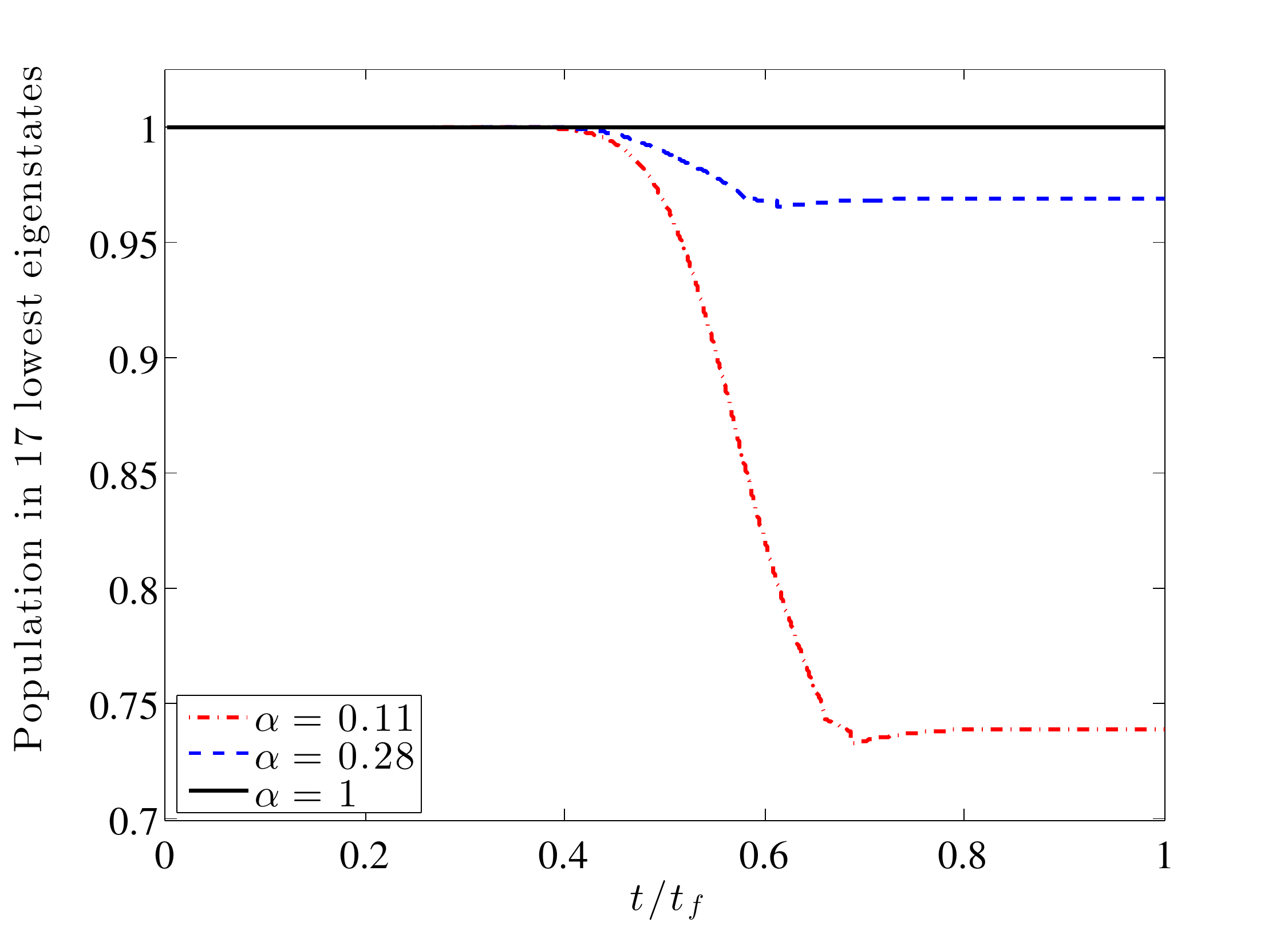} \label{fig:17stateVsTime}
\caption{ ME simulation for the time-dependence of the probability of being in the lowest $17$ energy eigenstates, for different values of $\alpha$.  Simulation parameters are $t_f = 20 \, \mu s$ (the minimal annealing time of the DW2) and $\kappa = 1.27 \times 10^{-4}$, where $\kappa$ is an effective, dimensionless system-bath coupling strength defined in Appendix \ref{app:ME}. The chosen value of $\kappa$ allows us to reliably probe the small $\alpha$ regime.}
\label{fig:MEIntuition}
\end{figure}

In agreement with these considerations, the effect that the position and size of the minimal gap have on the probability of being in a given energy eigenstate is shown in Fig.~\ref{fig:MEIntuition}. This figure shows the total population of the $17$ lowest energy eigenstates (i.e., the subspace that eventually becomes the ground state manifold), computed using the ME. As is clear from Fig.~\ref{fig:MEIntuition}, as $\alpha$ decreases, the increasingly delayed and smaller minimum gap causes this subspace to lose more population due to the increased rate and duration of thermal excitations. This behavior is interrupted by kinks caused by level crossing, whose position is a function of $\alpha$, with the kinks occurring later for smaller $\alpha$ (see Appendix \ref{app:kink}).
%

%
\section{Numerical simulations without cross-talk}
\label{sec:sim}

We have performed extensive numerical simulations using SA (described in 
Appendix \ref{app:SA}), SD (Appendix \ref{app:SD}), the ME (Appendix \ref{app:ME}), and the SSSV model~\cite{SSSV}, which we describe below.  
Since experimental evidence for rejection of SA and the SD models has already been presented in Refs.~\cite{q-sig,q108}, while the SSSV model  
presents a particularly interesting challenge since it nicely reproduces the success probability correlations that were used in Ref.~\cite{q108} to reject both SA and SD, we focus on the SSSV model here, and present our discussion of SA and SD in Appendix \ref{app:MoreNoiselessData}.

The starting point of the SSSV model is a classical Hamiltonian inspired by the original QA model \eqref{eq:adiabatic}:
\begin{align}
\label{eqt:SA-SD}
H(t) &= -A(t) \sum_i \sin \theta_i  \\
&+ B(t) \left( - \sum_i h_i \cos \theta_i + \sum_{i,j} J_{ij} \cos \theta_i \cos \theta_j \right)\, , \notag
\end{align}
i.e., each qubit $i$ is replaced by a classical O(2) spin $\vec{M}_i=(\sin\theta_i,0,\cos\theta_i)$ and the annealing schedules $A(t)$ and $B(t)$ are the same as those of the D-Wave device. However, the time evolution is now governed by a Metropolis algorithm.  In particular, at each discrete time step a certain number of Monte Carlo update steps are performed, as follows. Starting from the initial condition $\theta_i = \pi/2$, one variable at a time a random angle $\theta_i \in [0, \pi]$ is drawn with uniform probability.  If this new angle does not increase the energy [as given by the Hamiltonian in Eq.~\eqref{eqt:SA-SD}] it is accepted. If the new angle increases the energy it is accepted only if $p < \exp\left(- \upbeta \Delta E \right)$, where $p\in [0,1]$ is drawn with uniform probability and $\Delta E > 0$ is the change in energy.  Each complete time evolution following the entire annealing schedule constitutes one run. For the $n$th run out of a total of $N_r$ runs we obtain a set of angles $\{ \theta_j^{(n)} \}$, which is interpreted in terms of a state in the computational basis according to the sign of $\cos (\theta_j^{(n)})$, i.e., if $0\leq \theta_j^{(n)} \leq \pi/2$ then it is the $\ket{0}$ state, whereas if $\pi/2 < \theta_j^{(n)} \leq \pi$ then it is the $\ket{1}$ state.

\begin{figure*}[th]
        \subfigure[\ ME, no noise]{ \includegraphics[width=0.32\textwidth]{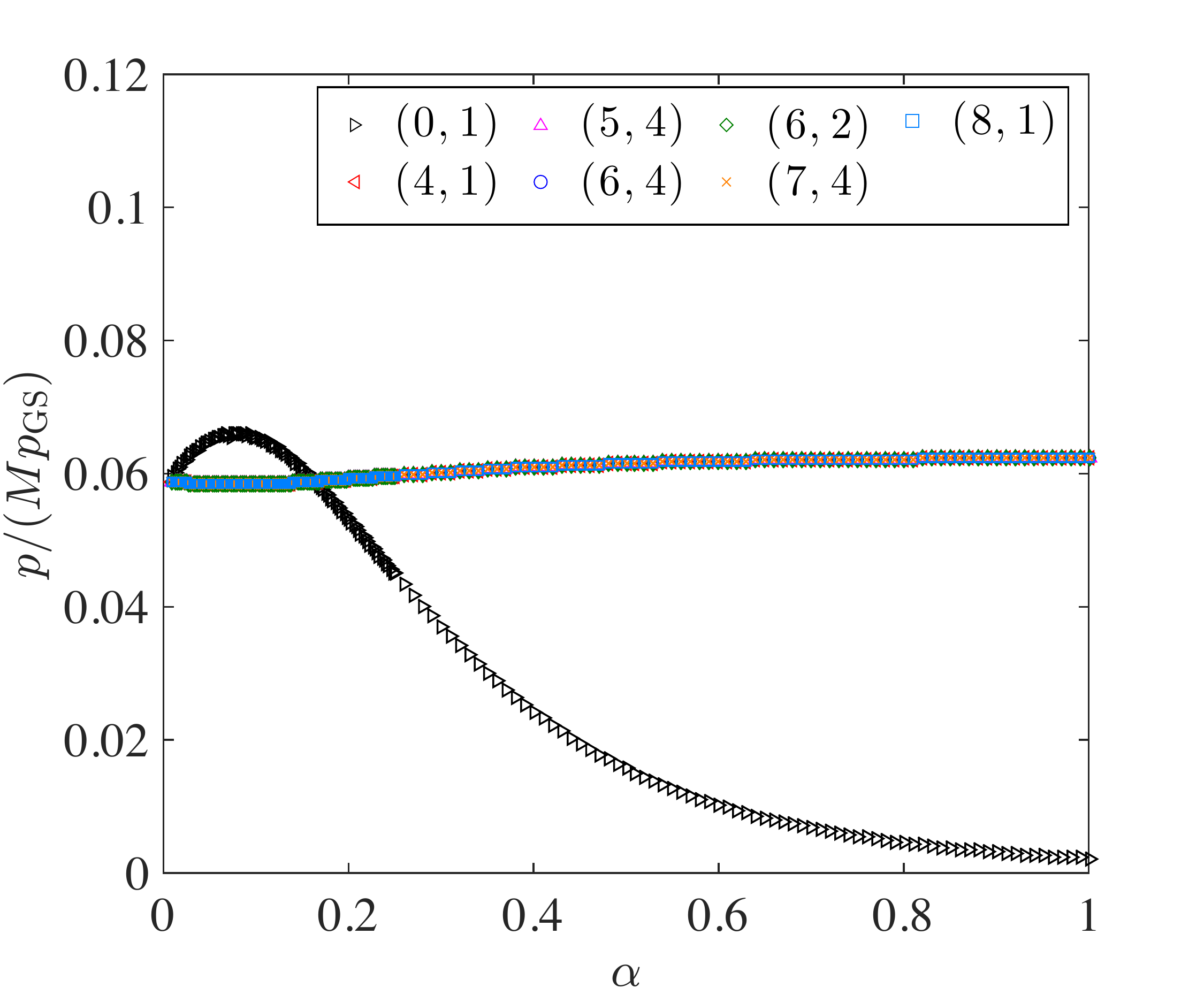} \label{fig:ME-Uncalibrated}} 
   \subfigure[\ SSSV, no noise]{ \includegraphics[width=0.32\textwidth]{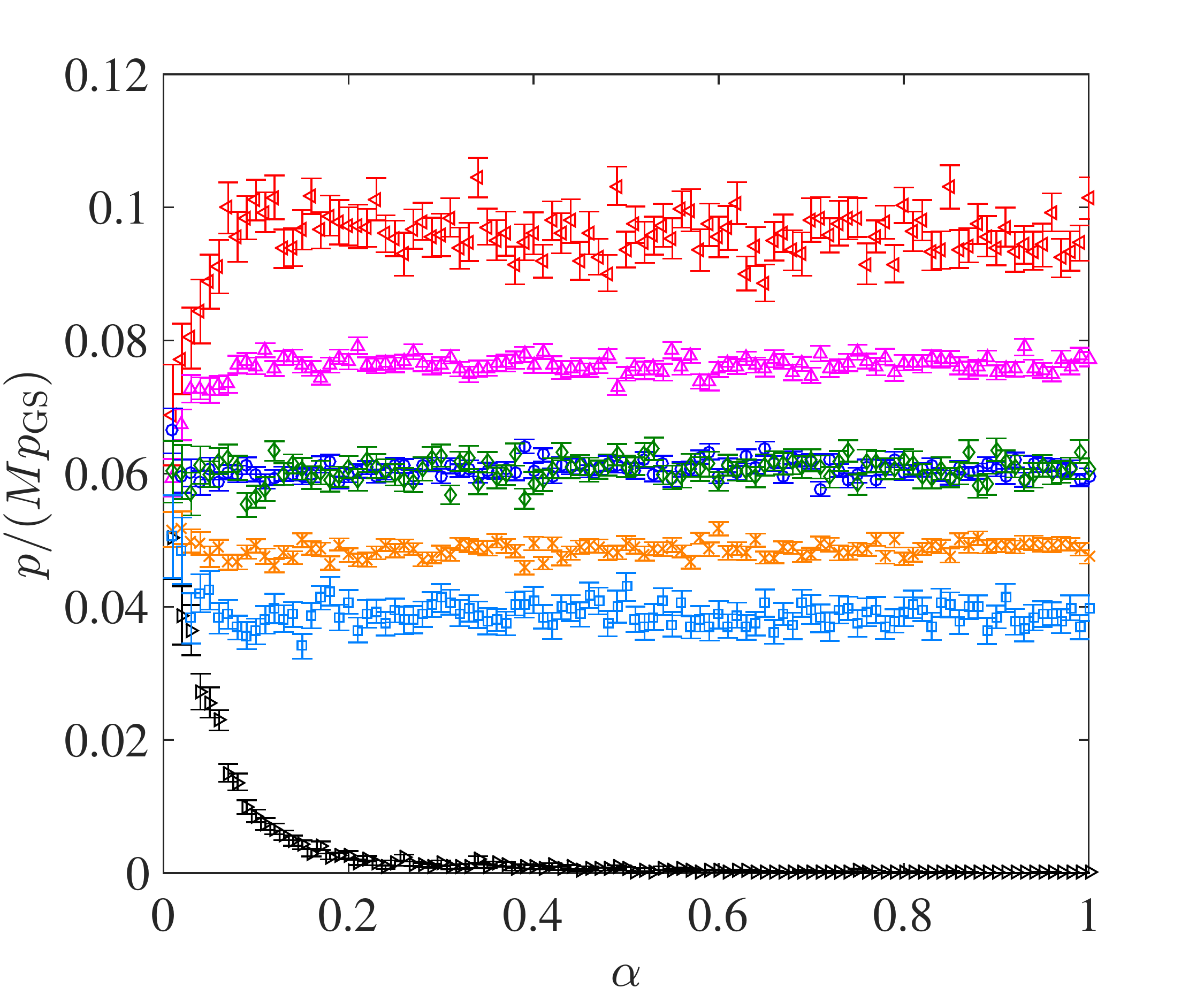} \label{fig:SSSV-nonoise}} 
      \subfigure[\ SSSV, noisy]{ \includegraphics[width=0.32\textwidth]{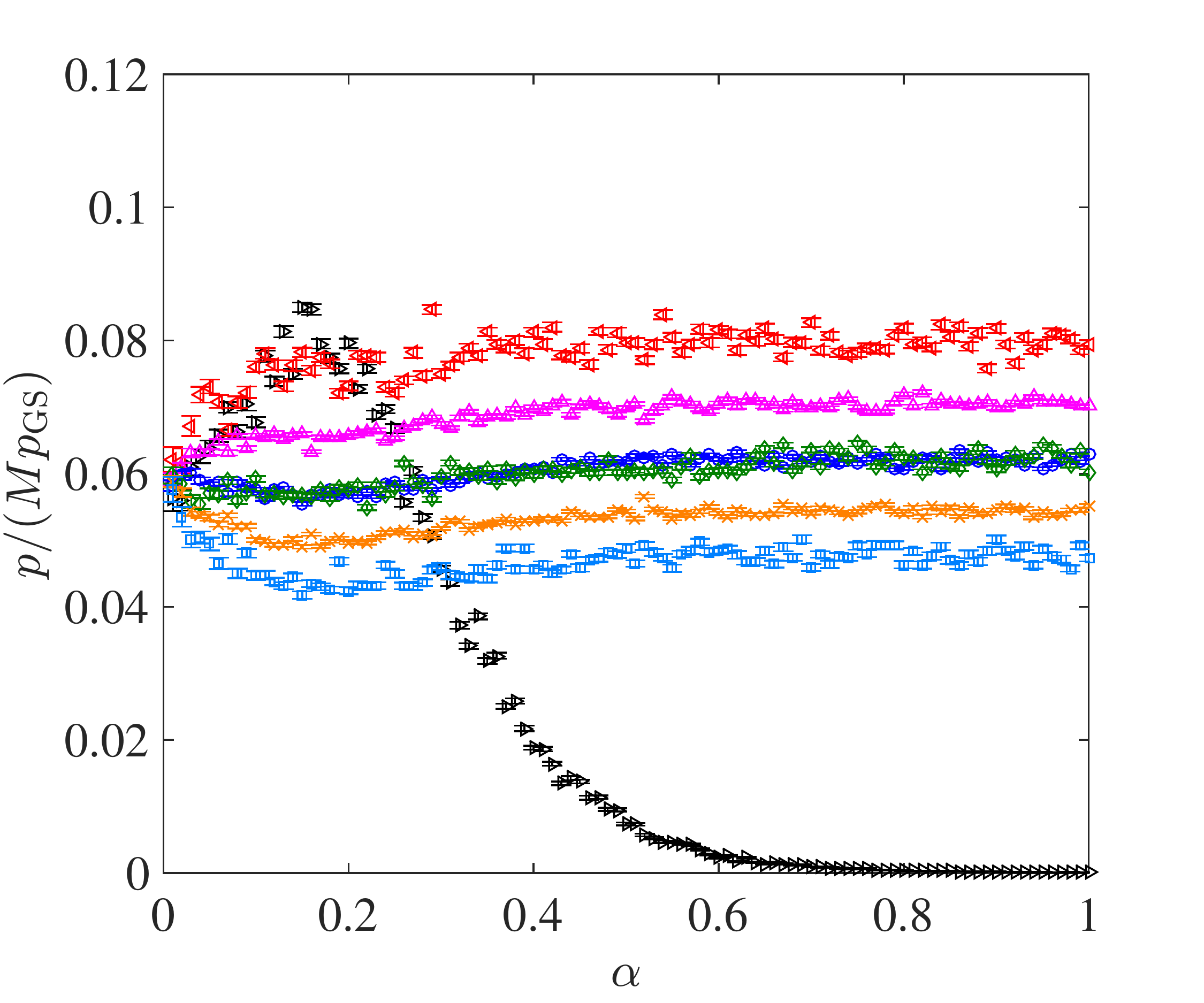} \label{fig:SSSV-Uncalibrated}} 
\caption{ Distribution of the ground states for $N=8$ for (a) ME with no noise on $\{h_i,\,J_{ij}\}$, (b) SSSV with no noise on $\{h_i,\,J_{ij}\}$, and (c) SSSV with $\{h_i,\,J_{ij}\}$ noise using $\sigma = 0.085$. The cluster states are labeled by their Hamming distance $H$ from the isolated state, and by their multiplicity $M$ for a given value of $H$.  The vertical axis is the final probability $p$ of a given $(H,M)$ set, divided by its multiplicity and the total ground state probability.  The data symbols ($\circ$, etc.) are the mean values of the bootstrapped \cite{Efron:1994qp}  distributions, and the error bars are two standard deviations below and above the mean representing the 95\% confidence interval.  Note that the SSSV model prefers the $\ket{1111\,0000}$ cluster state, whereas the ME gives a uniform distribution over all cluster states.  SSSV parameters are $T=10.56$mK and $1 \times 10^5$ Monte Carlo step updates per spin (``sweeps"). The same parameters are used in all subsequent SSSV figures. These results do not include the cross-talk correction.}
   \label{fig:NoCalibration}
\end{figure*}  

 We note that our ME simulations have only one adjustable parameter, $\kappa$, an effective, dimensionless system-bath coupling strength (defined in Appendix \ref{app:ME}). The SSSV model has two: the temperature and the number of Monte Carlo update steps. In addition, as we discuss in detail below, it requires the addition of stochastic noise to the local fields and couplings in order to match the ME and the experimental results, which introduces a third free parameter in the form of the noise standard deviation. When we discuss the effect of cross-talk in the next section, both the SSSV model and the ME will require the susceptibility $\chi$ as an additional free parameter.

We now present our first set of numerical findings, where we do not include the cross-talk correction discussed in Sec.~\ref{sec:cross-talk}, but focus instead on the role of the stochastic noise on the local fields and couplings. 
The results in this section will help to clarify the roles played by these various sources of imperfection in the experiment.

\subsection{Ratio of the populations of the isolated state to the cluster states}
Figure~\ref{fig:NoCalibration} shows the distribution of cluster states and isolated state for the entire range of $\alpha$ values.  First, we observe that in the absence of noise on $\{h_i,\,J_{ij}\}$ the behavior of the isolated state is strikingly different between SSSV and the ME.  Whereas SSSV shows a monotonic increase with decreasing $\alpha$,  the ME result for the isolated state is non-monotonic in $\alpha$; see Fig.~\ref{fig:ME-Uncalibrated}. Initially, as $\alpha$ is decreased from its largest value of $1$, the ratio of isolated to cluster state population increases and eventually becomes larger than $1$; i.e., the population of the isolated state becomes enhanced rather than suppressed.  For sufficiently small $\alpha$, the ME isolated state population turns around and decreases towards $1$.  

Thus, the SSSV model captures the suppression of the isolated state at high $\alpha$ but does not capture the ground state population inversion at low $\alpha$ (as we discuss below, this conclusion changes after noise on $h$ and $J$ is included). On the other hand, we note that SA correctly predicts an enhanced isolated state at low $\alpha$ but does not predict the suppression at high $\alpha$ (see Appendix \ref{app:MoreNoiselessData}). This observation led us to consider classical models that interpolate between SSSV at high $\alpha$ and SA at low $\alpha$.  These models exploit the fact that 
in SA the qubits are replaced by fully incoherent, classical Ising spins, while in SSSV each qubit is replaced by a ``coherent" O(2) rotor.\footnote{Recall that SU(2) is (locally) isomorphic to SO(3), so a qubit can always be mapped to an SO(3) rotor. [Strictly, SO(3) is isomorphic to SU(2)/Z$_2$.] The restriction to O(2) rotors is heuristically justified by SSSV via the observation that the QA Hamiltonian contains only $x$ and $z$ components.} Therefore a natural way to interpolate between SSSV and SA is to ``decohere" the O(2) rotors over an $\alpha$-dependent timescale $\tau_{\alpha}$, and two natural decoherence models we considered are discussed in Appendix \ref{sec:dec-SSSV}. However, these models do not reproduce the behavior of the ME.

\begin{figure}[t] 
   \centering
   \includegraphics[width=0.48\textwidth]{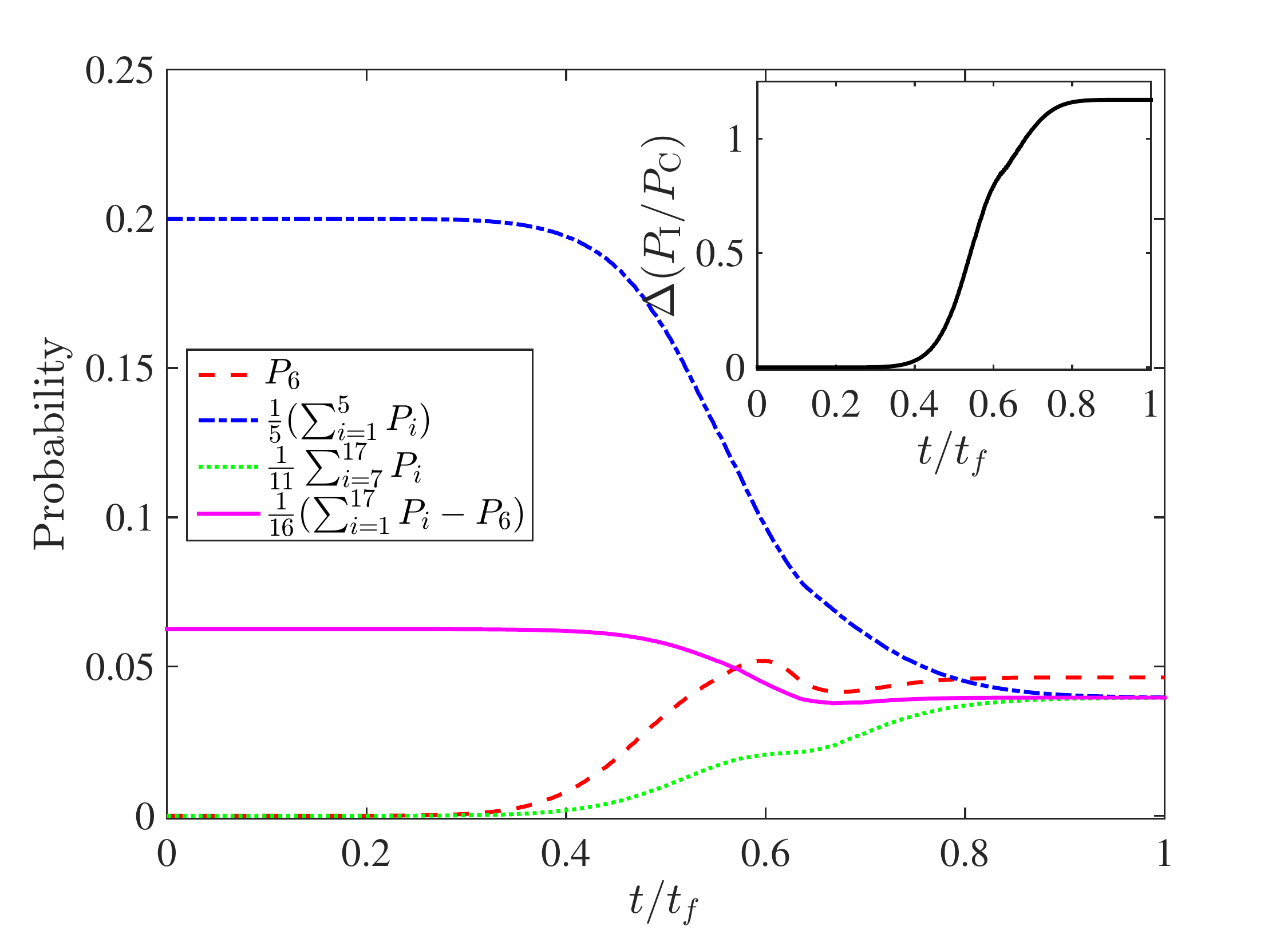} 
   \caption{ Time evolution (according to the ME) of energy eigenstate populations for $\alpha = 0.1$ and $\kappa = 8.9 \times 10^{-4}$ (this relatively large value was chosen here since it results in increased thermal excitation/relaxation).  $P_i$ denotes the population of the $i$th eigenstate, with $i = 1$ being the instantaneous ground state. The energy eigenstate that eventually becomes the isolated ground state is $i=6$ (dashed red line). This state acquires more population at the end of the evolution than the other $16$ eigenstates that eventually become the cluster (solid purple line). (Inset) The difference of the population ratio between the open system and the closed system evolution, $\Delta (P_\mathrm{I}/P_{\mathrm{C}}) = (P_\mathrm{I}/P_{\mathrm{C}})_{\mathrm{Open}} - (P_\mathrm{I}/P_{\mathrm{C}})_{\mathrm{Closed}}$. The deviation from closed system dynamics starts at $t/t_f \approx 0.4$, when the $i=6$ eigenstate becomes thermally populated at the expense of the lowest five eigenstates.}
   \label{fig:QA-CA}
\end{figure}

\begin{figure}[t]
\includegraphics[width=.98\columnwidth]{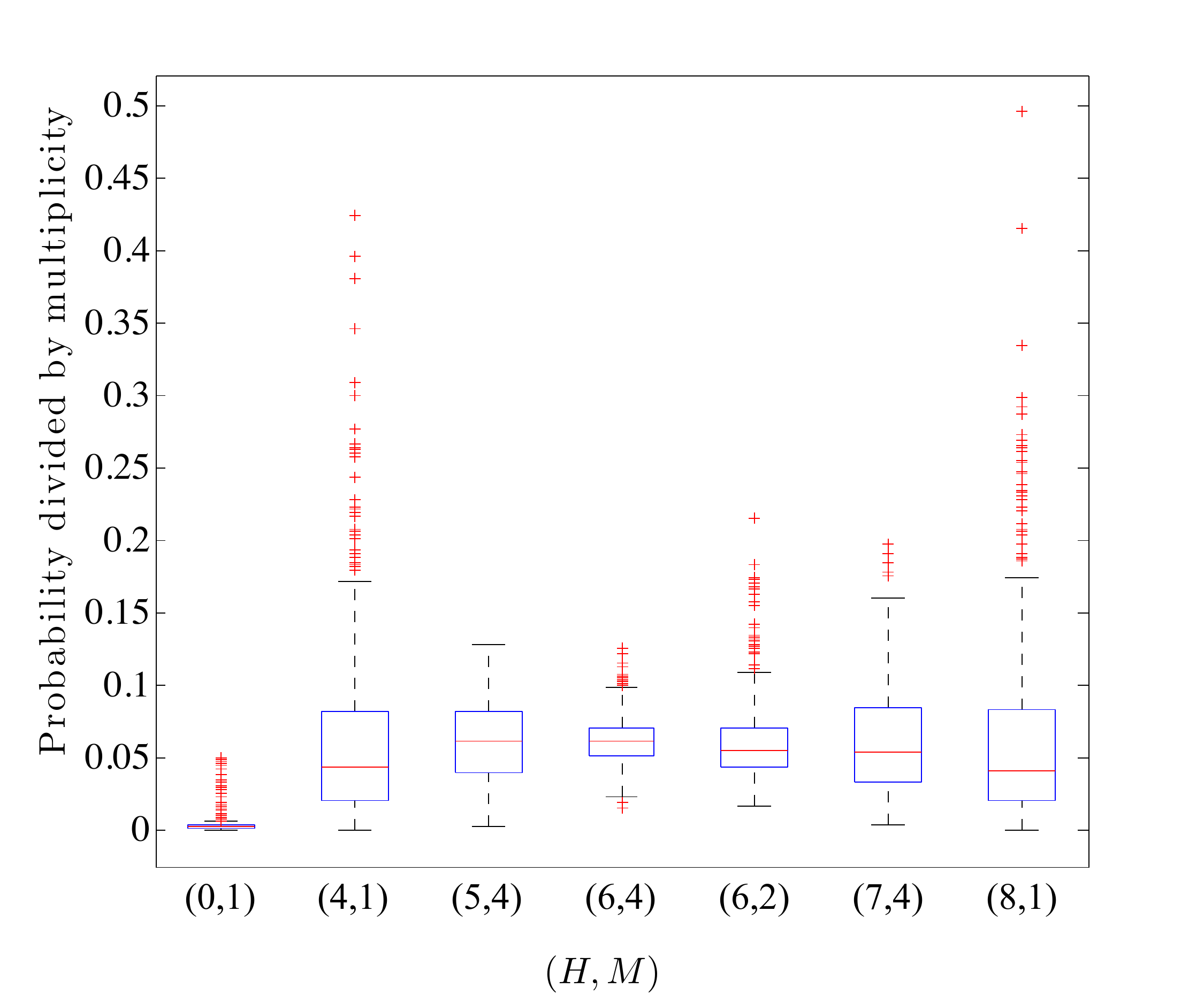}
\caption{ Statistical box plot of the probability divided by the multiplicity of being in a given state with Hamming distance $H$ and multiplicity $M$. Shown are the ME isolated state and cluster states for $N=8$, with $512$ noise realizations applied to the $h$'s and $J$'s [with distribution $\mathcal{N}(0,0.06)$] at $\alpha = 1$. The isolated state ($H=0,M=1$) is suppressed while the cluster states are, on average, equally populated. The red bar is the median, the blue box corresponds to the lower and upper quartiles, respectively, the segment contains most of the samples, and the $+$'s are outliers \cite{Frigge:1989vn}. The horizontal axis label indicates the Hamming distance from the isolated state and the multiplicity of the cluster-states at each value of $H$. States that are equivalent up to $90^{\circ}$ rotations are grouped together.  For example, there are four rotationally equivalent cluster-states that have two adjacent outer qubits pointing down, while the other two are pointing up. Only the $H=6$ case splits into two rotationally inequivalent sets.}
\label{fig:NoisyME}
\end{figure}

\begin{figure*}[t]
   \subfigure[\ DW2, $N=8$]{ \includegraphics[width=0.32\textwidth]{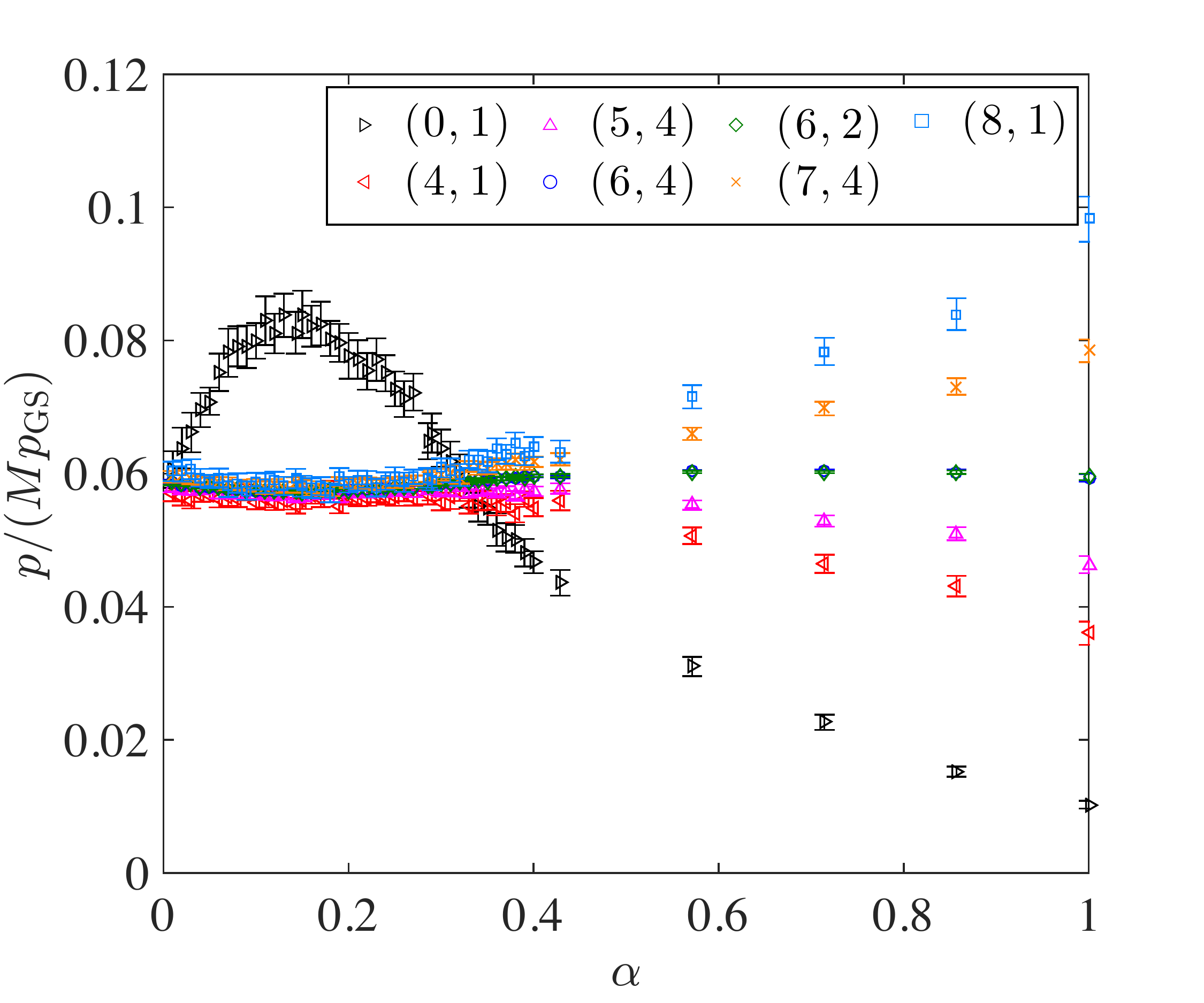}\label{fig:DW2-Uncalibrated2}} 
   \subfigure[\ DW2 \& ME, $\chi = 0.015$, $N=8$]{ \includegraphics[width=0.32\textwidth]{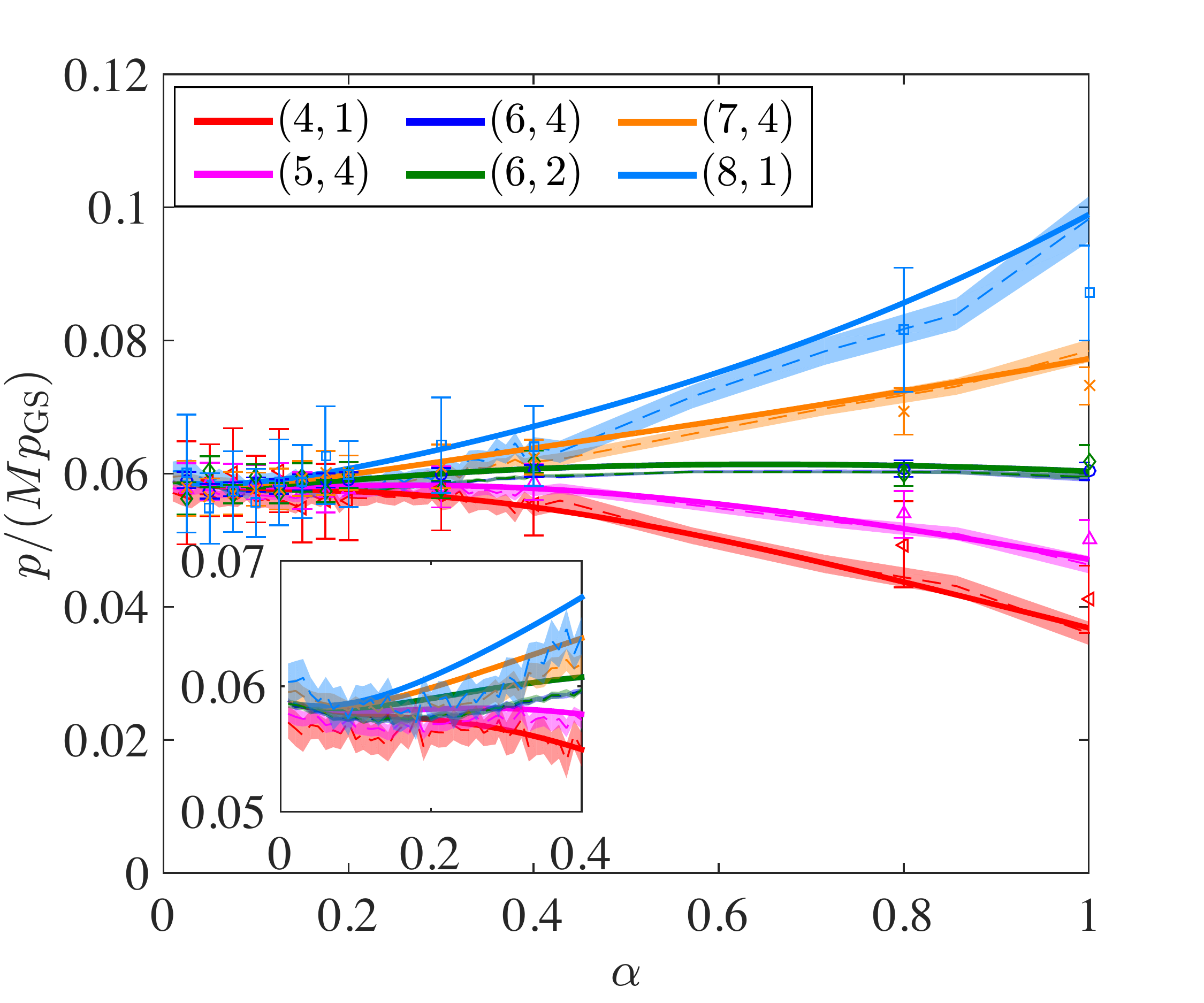}\label{fig:ME-Chi}} 
   \subfigure[\ DW2 \& SSSV, $\chi = 0.035$, $\sigma = 0.085$, $N=8$]{ \includegraphics[width=0.32\textwidth]{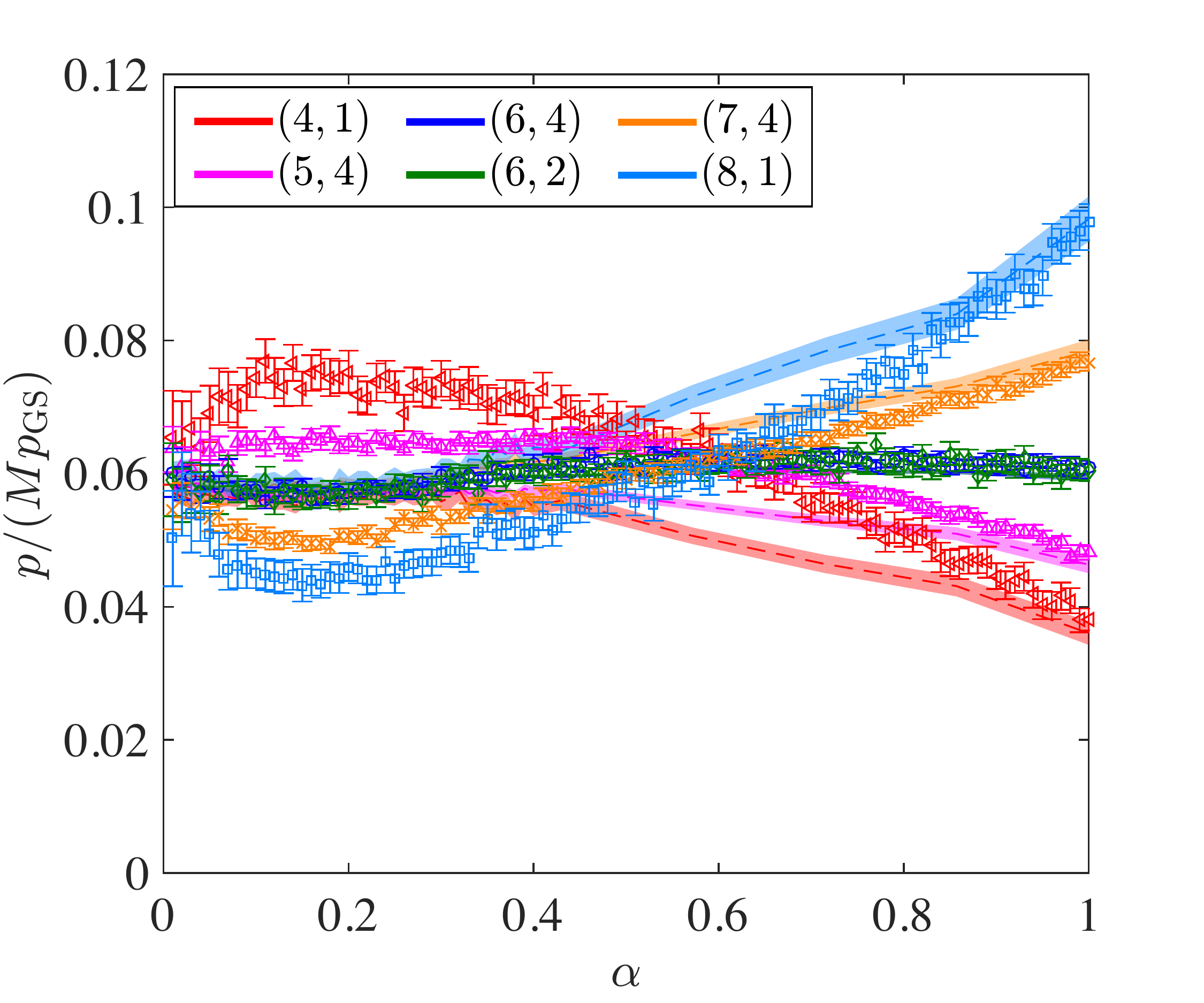} \label{fig:SSSV-Chi}} 
   \subfigure[\ DW2, ME ($\chi = 0.015$), SSSV, ($\chi = 0.035$, $\sigma = 0.085$), $N=8$]{ \includegraphics[width=0.32\textwidth]{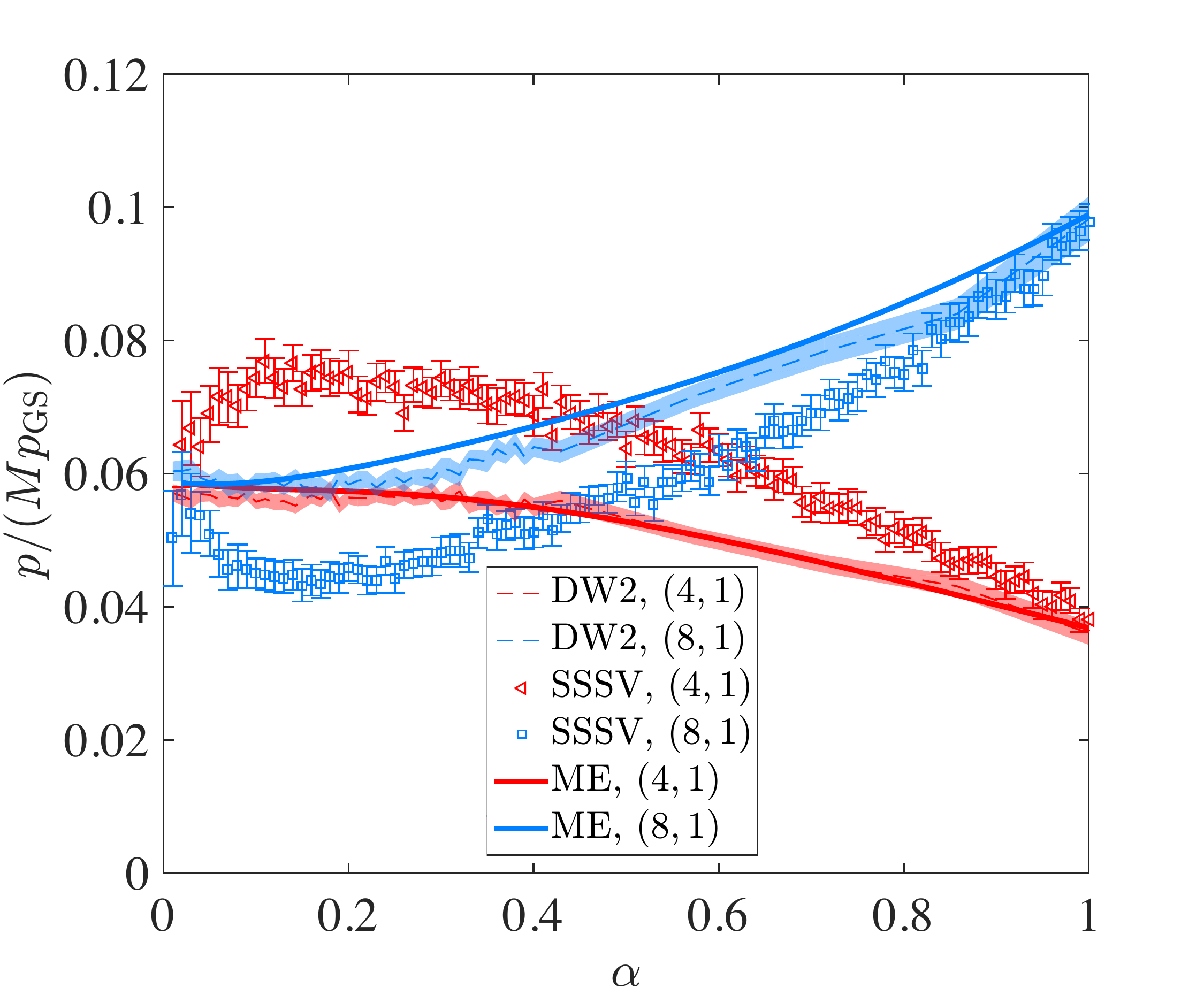}\label{fig:new1}} 
   \subfigure[\ DW2 \& SSSV ($\chi = 0.035$, $\sigma = 0.085$), $N=20$]{ \includegraphics[width=0.32\textwidth]{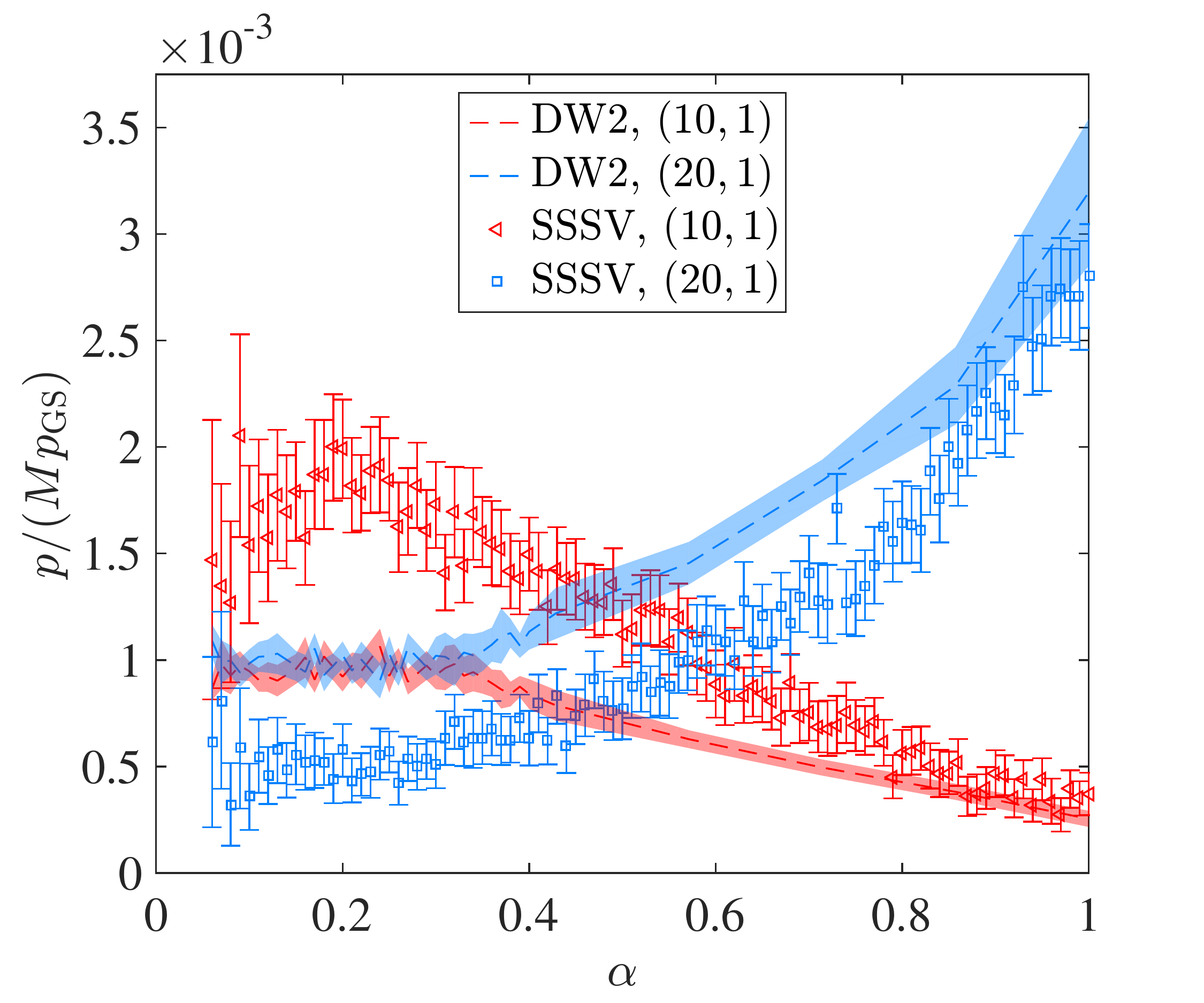} \label{fig:DW2-Chi-N=20}} 
   \subfigure[\ DW2, SSSV, ($\chi = 0.035$, $\sigma = 0.085$), ME ($\chi = 0.015$), and noisy ME ($\chi = 0.015$, $\sigma=0.025$), $N=8$]{ \includegraphics[width=0.32\textwidth]{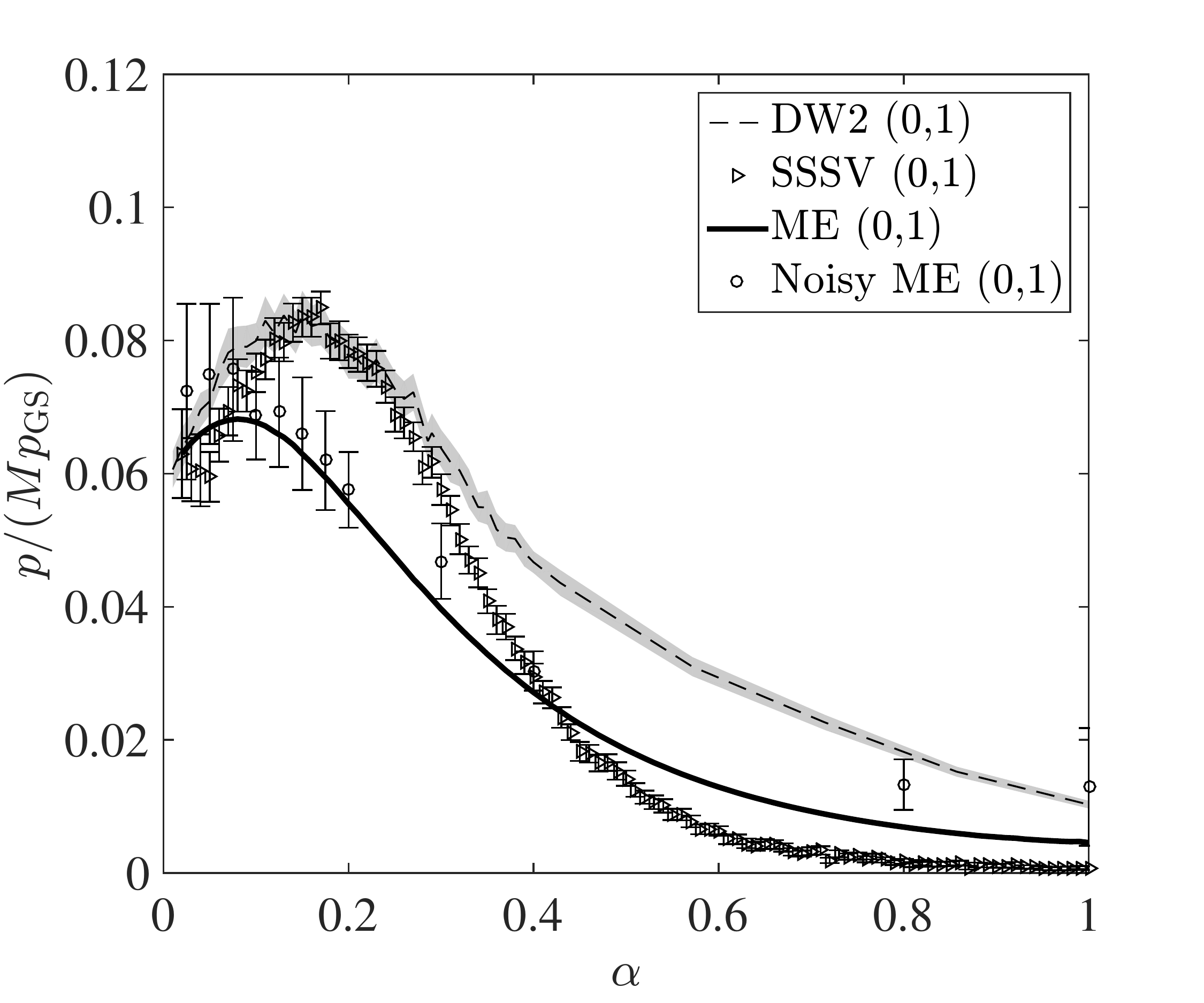} \label{fig:new2}} 
\caption{ (a) Ground state populations for DW2.  Legend: $(H,M)$, corresponding to Hamming distance from the isolated state and multiplicity respectively.  Error bars represent the 95\% confidence interval.  (b) Cluster state populations for the ME.  Solid lines correspond to the results with no noise on the $\{h_i, J_{ij}\}$'s, while the data points include Gaussian noise with mean 0 and standard deviation $\sigma = 0.025$ for 100 noise realizations.  The error bars represent the 95\% confidence interval.  The DW2 data from (a) is also plotted as the shaded region representing a 95\% confidence interval with the dashed lines corresponding to the mean.  The inset shows the behavior for the noiseless ME for small $\alpha$.  (c) Cluster state population for SSSV for $N=8$ with the DW2 data plotted as in (b).  In contrast to Fig.~\ref{fig:NoCalibration}, both the ME and the noisy SSSV model include the cross-talk correction, Eq.~\eqref{eq:cross-talk}, with $\chi$ chosen to optimize the fit for the cluster state populations at $\alpha=1$.  (d) Only the cluster states with Hamming distance 4 and 8 from the isolated state are shown for DW2, the ME, and SSSV from panels (b) and (c) in order to highlight their differences.  Panel (e) displays the same for $N=20$ (excluding the ME, which is too costly to simulate at this scale).   (f) The isolated state populations for DW2, SSSV, ME, and noisy ME, which highlights the qualitative agreement between the models and DW2. Experimental data were collected using the in-cell embeddings strategy described in Appendix \ref{app:experiment}. The embedding and gauge-averaging strategies are also discussed in Appendix \ref{app:experiment}.  The color coding of states is consistent across all panels.}
   \label{fig:cross-talk}
\end{figure*}

In order to understand what contributes to the increase in the isolated state population as $\alpha$ is lowered, it is useful to study the time evolution of the population in the lowest $17$ energy eigenstates according to the ME. An example is shown in Fig.~\ref{fig:QA-CA}, for $\alpha=0.1$, i.e., close to the peak of the isolated state population. This figure clearly shows how the relative ratio of the isolated state population to the mean cluster state population $P_\mathrm{I}/P_\mathrm{C}$ becomes $>1$. The sixth energy eigenstate (red line) evolves to become the isolated ground state, while the other $16$ eigenstates evolve to become the cluster (purple line). During the time evolution, the population in the sixth eigenstate grows slightly larger than that of the cluster (red curve ends up above the purple one), which explains why $P_\mathrm{I}/P_\mathrm{C}>1$. In more detail, we observe that (around $t/t_f = 0.4$) the sixth eigenstate acquires population (via thermal excitations) from the lowest five eigenstates (blue line).  Somewhat later (around $t/t_f = 0.6$) the sixth eigenstate loses some population due to thermal excitations, which is picked up in part by the highest $11$ eigenstates (green).  Finally, thermal relaxation returns some population to the $17$ eigenstates, but the sixth eigenstate gains more population than the other $16$ eigenstates since it is connected to a larger number of excited states.  During this relaxation phase, the system behaves like classical SA.  The inset shows that deviations from the closed system behavior occur around $t/t_f=0.4$, i.e., when the population of the sixth eigenstate first starts to grow (along with the highest $11$ eigenstates) due to excitations from the lowest five eigenstates.
%

\subsection{Cluster state populations}
The other important feature to note from Fig.~\ref{fig:NoCalibration} is that the population degeneracy of the cluster states is broken in the SSSV model, giving rise to a staircase pattern organized according to Hamming distance (HD) from the isolated state [Fig.~\ref{fig:SSSV-nonoise}], while the ME exhibits a uniform distribution over the cluster states [Fig.~\ref{fig:ME-Uncalibrated}].   Except for very small $\alpha$, the SSSV pattern remains fixed as $\alpha$ is decreased.  The preference for the $\ket{1111\,0000}$ (HD$=4$) state and the insensitivity of this feature to $\alpha$ in the SSSV model can be understood from the following simple argument.

Consider first the closed system (no thermal noise) case and note that for $t > 0.6 t_f$, the transverse field is almost completely turned off.  Therefore the cluster states' outer spins are free to rotate with no energy cost when the core spins are pinned at $M^z_{\textrm {c}} = \cos \theta_c = 1$, leading to $M^z_{\textrm {o}} = \cos \theta_0 = 0$ (the c and o subscripts stand for ``core" and ``outer", respectively).  However, in the open system case the core spins are not fixed at $M^z_{\textrm {c}} = 1$ due to thermal noise, and $M^z_{\textrm {o}} = -1$ becomes energetically favorable for the outer spins. To see why, consider the case of a single pair of core and outer spins.  In this case, the Ising potential is simply $V=-\alpha(h_{\textrm {o}} M^z_{\textrm {o}} + h_{\textrm {c}}  M^z_{\textrm {c}} +J_{\textrm {oc}} M^z_{\textrm {o}} M^z_{\textrm {c}}) = \alpha(M^z_{\textrm {o}} - M^z_{\textrm {c}} - M^z_{\textrm {o}} M^z_{\textrm {c}})$.
When $M^z_{\textrm {c}} = 1$, the dependence on $M^z_{\textrm {o}}$ vanishes so the outer spin is free to rotate, however when $M^z_{\textrm {c}} \neq 1$ (as happens when thermal noise is present), this Ising potential is minimized when $M^z_{\textrm {o}} = -1$. This explains why the SSSV model prefers the $\ket{1111\,0000}$ cluster state for all $\alpha$. 

Note that this argument depends on $|h|=|J|$; i.e., it will not necessarily apply when there is noise on $h$ and $J$.   As an example, if we add Gaussian noise $\Delta h_i,\Delta J_{ij} \sim \mathcal{N}(0,0.085)$ so that $h_i \mapsto h_i + \Delta h_i$ and $J_{ij} \mapsto J_{ij}+\Delta J_{ij}$ in Eq.~\eqref{eqt:SA-SD}, as first shown in Ref.~\cite{SSSV-comment}, the resulting ``noisy SSSV" model is able to reproduce the non-monotonic behavior of the isolated state observed for the ME, as shown in Fig.~\ref{fig:SSSV-Uncalibrated}.  However it maintains its preference and ordering of cluster state populations.

\section{Experimental results and numerical simulations including cross-talk}
\label{sec:calib-noise}
%

\subsection{The distribution of cluster states}

Having developed an understanding of the role of noise on the local fields and couplings on the ground state distribution, we now present experimental results for the DW2 in Fig~\ref{fig:DW2-Uncalibrated2}, which we believe to include cross-talk. We immediately observe a strong discrepancy between the DW2 and both the ME and SSSV results shown in Fig.~\ref{fig:NoCalibration}.  This discrepancy implies that both models require an adjustment.  We next introduce the cross-talk correction.  (An alternative model which gives rise to the breaking of the cluster state symmetry by detuning $|h|$ relative to $|J|$ is discussed in Appendix \ref{app:hvsJ}; this gives a less satisfactory fit to the experimental data.)  We fit the cross-talk magnitude $\chi$ at $\alpha = 1$ in order to force both the noisy SSSV model and the ME to reproduce the correct ordering of the cluster states at this value of $\alpha$. However, we refrain from excessively fine-tuning the models for additional values of $\alpha$. That is, we adhere to the idea that a good theoretical model should have predictive power after its free parameters are fit to the data once.

Noise on the local fields and couplings has no such effect on the ME. Indeed, we have checked that introducing noise on the couplings and the local fields does not, {on average}, break the population degeneracy of the ME cluster states, while it does break the degeneracy for any {given} noise realization.  This can be seen in Fig.~\ref{fig:NoisyME}. We have also checked that introducing different system-bath couplings for each qubit (by adding Gaussian noise to each coupling) does not break the population degeneracy of the cluster states.

\begin{figure*}[ht]
\subfigure[\ DW2, $N=8$]{ \includegraphics[width=0.32\textwidth]{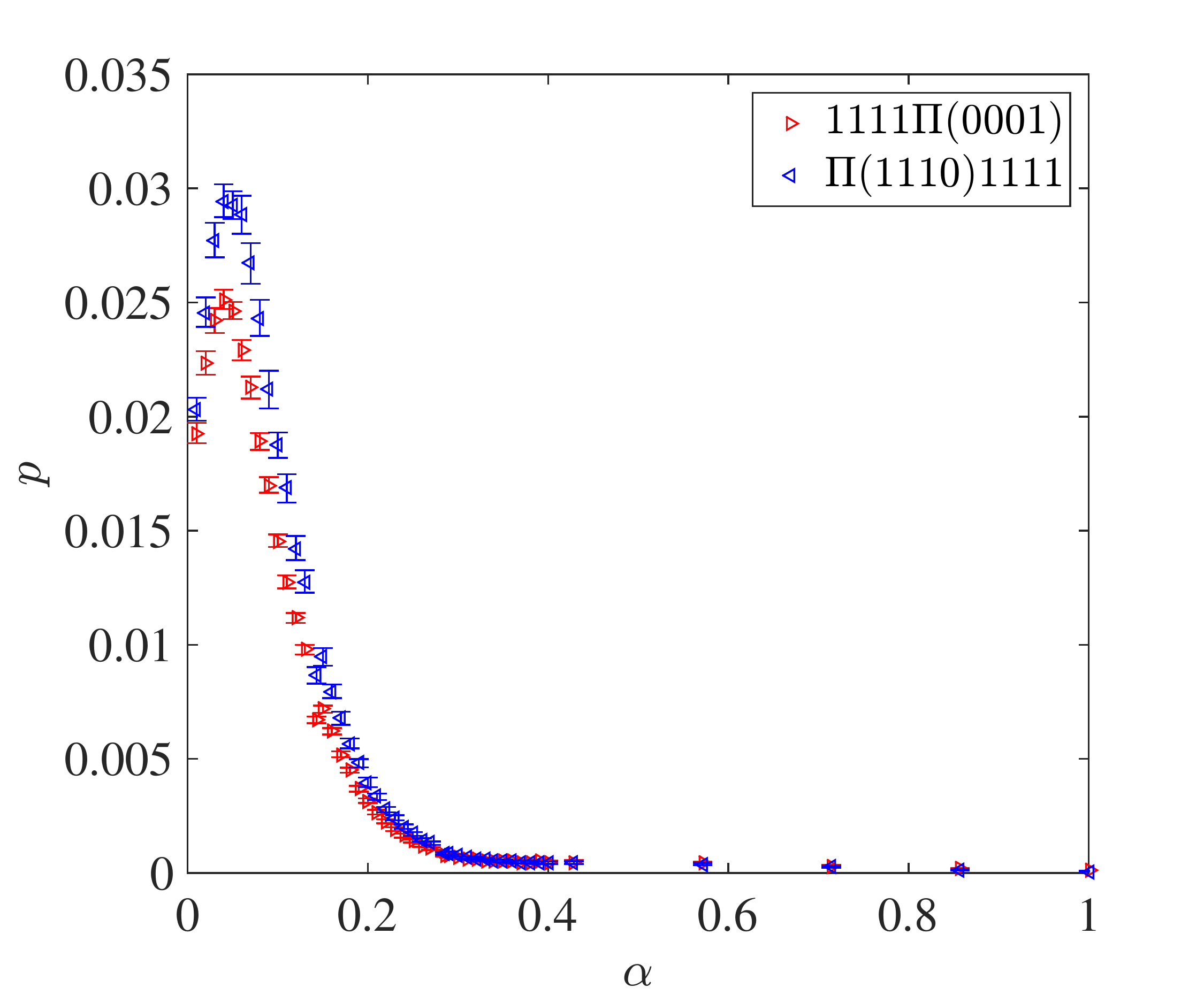}\label{fig:DW2-ES-nocalibration}} 
\subfigure[\ SSSV, $\chi = 0.035$, $\sigma = 0.085$, $N=8$]{ \includegraphics[width=0.32\textwidth]{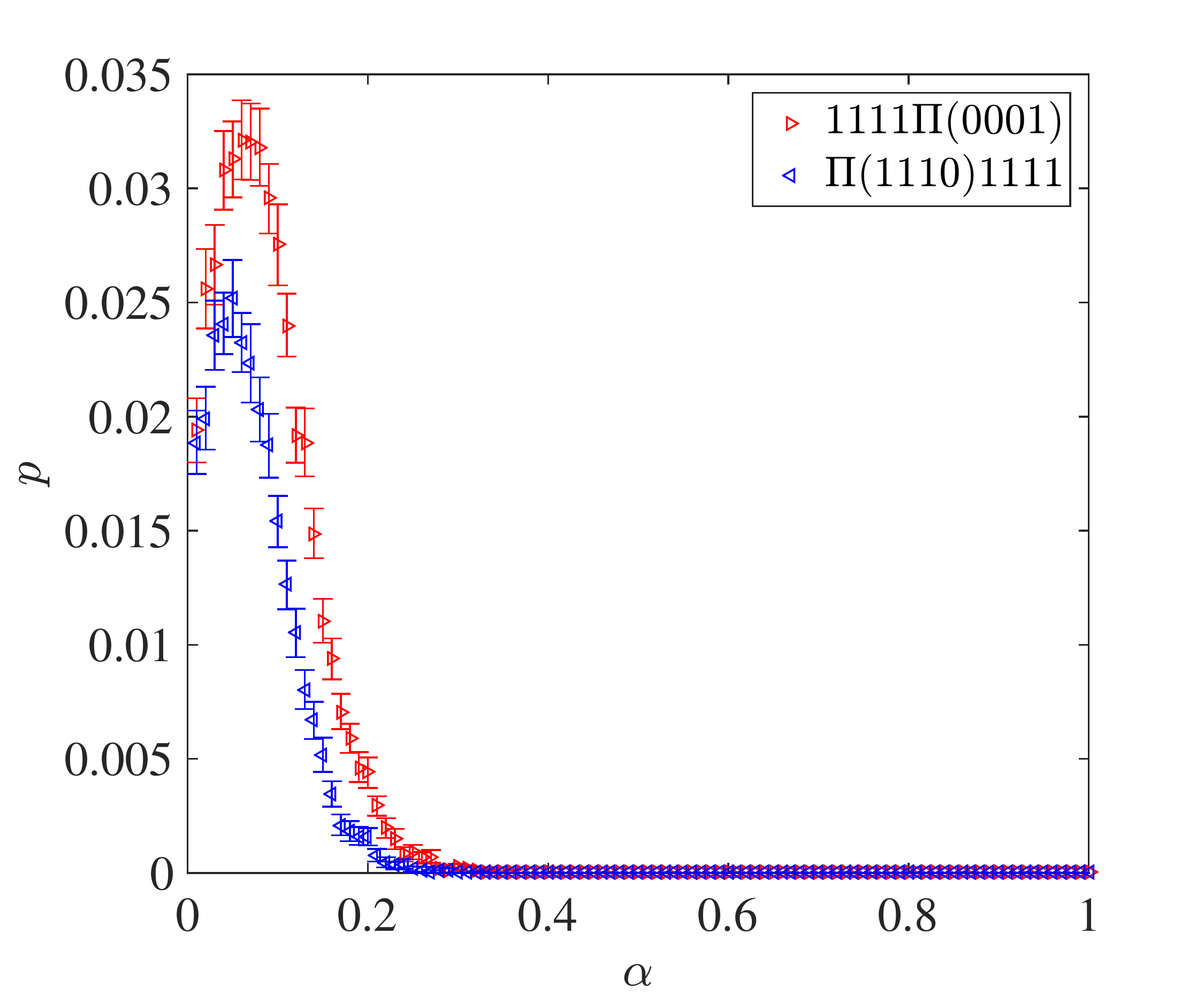} \label{fig:SSSV-ES-chi}} 
     \subfigure[\ ME, $\chi = 0.015$, $\sigma = 0.025$, $N=8$]{\includegraphics[width=0.32\textwidth]{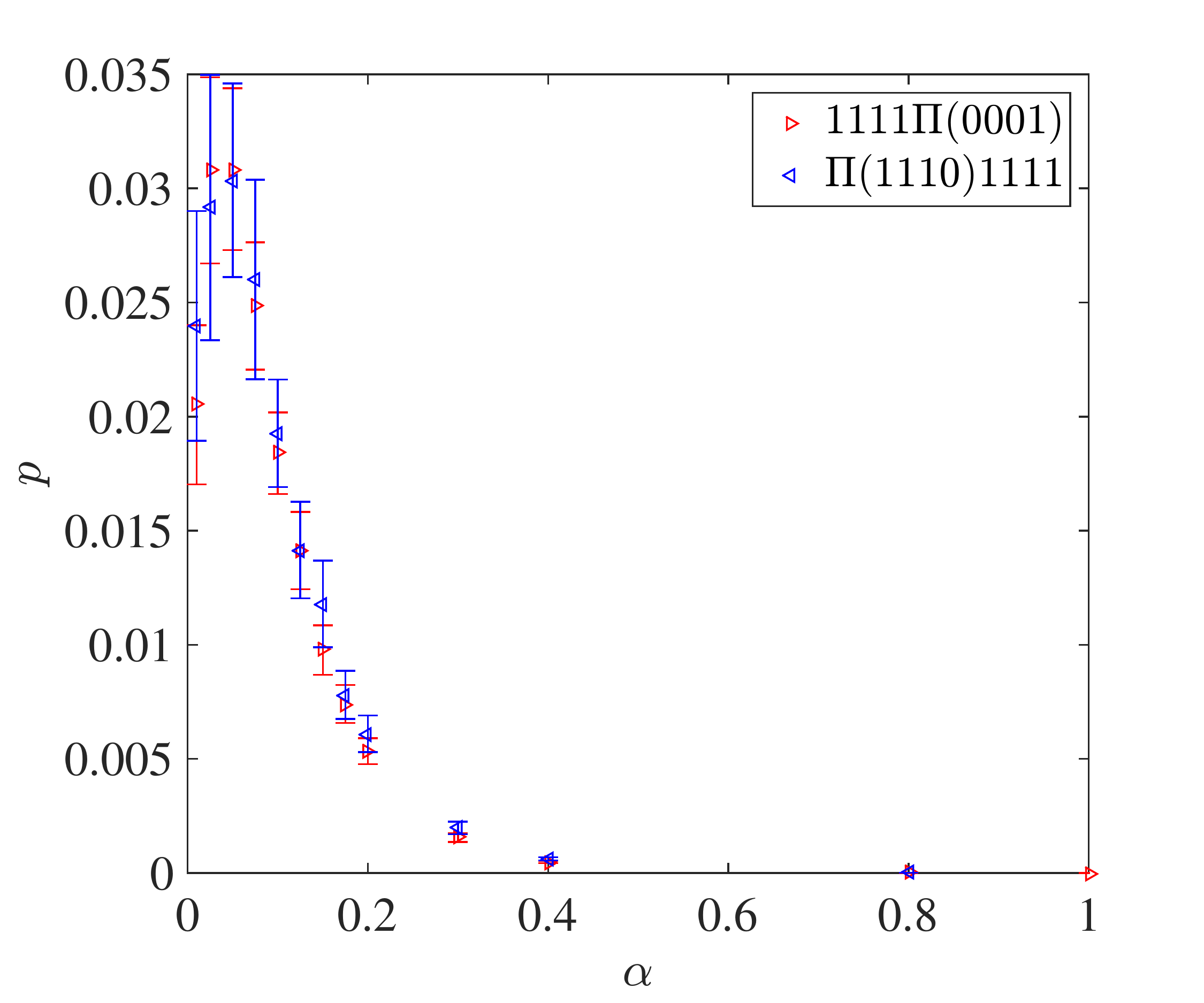}\label{fig:ME-ES}}
\subfigure[\ DW2, $N=20$]{ \includegraphics[width=0.32\textwidth]{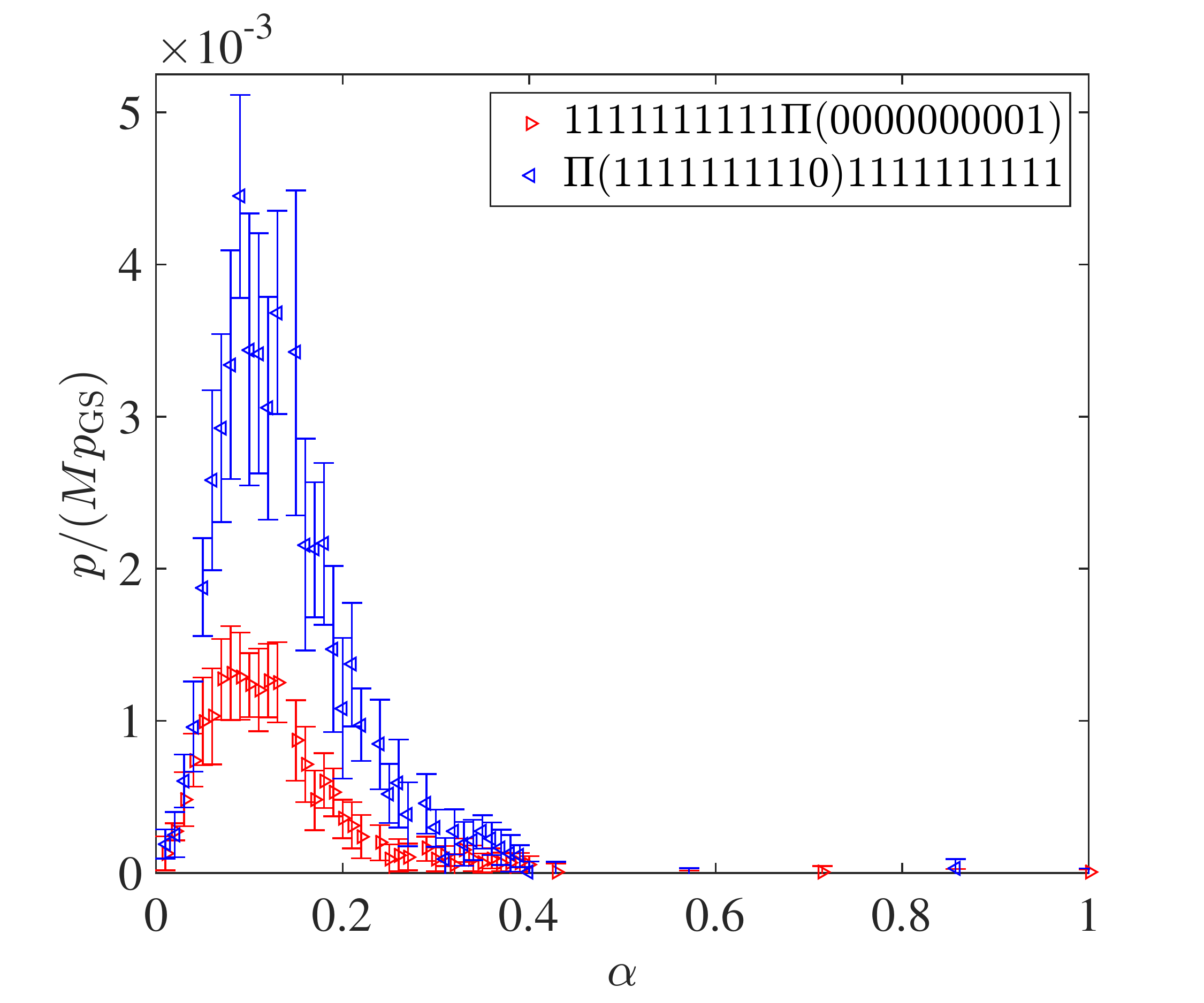}\label{fig:DW2-ES-nocalibration-N=20}} 
\subfigure[\ SSSV, $\chi = 0.035$, $\sigma = 0.085$, $N=20$]{ \includegraphics[width=0.32\textwidth]{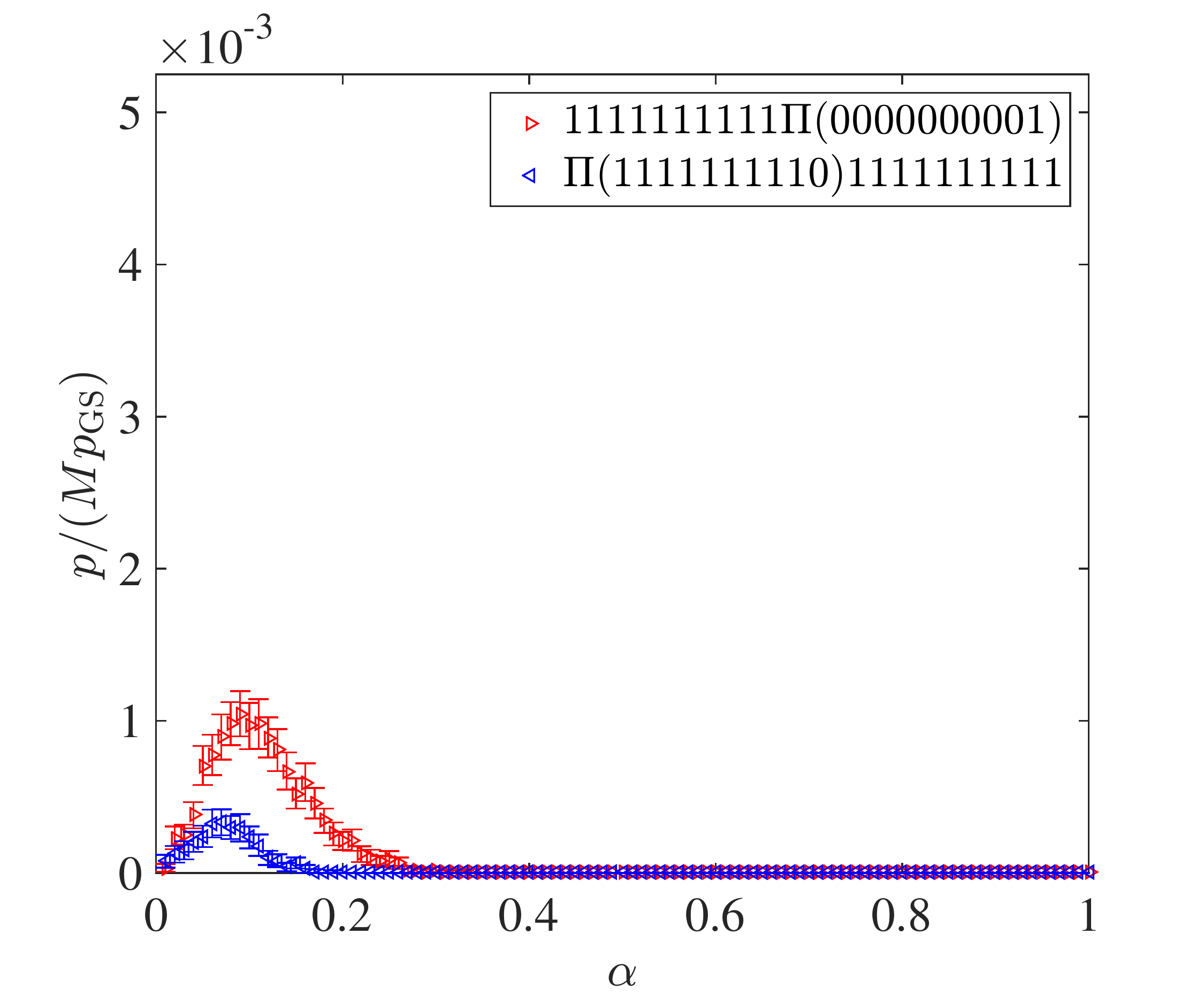} \label{fig:SSSV-ES-chi-N=20}}
\caption{ Subset of the first excited state populations for (a) DW2 for $N=8$; (b) SSSV for $N=8$; and (c) ME for $N=8$.  In (b) and (c), the simulations include qubit cross-talk correction with $\chi$ chosen as in Fig.~\ref{fig:cross-talk} to optimize the fit for the cluster state populations at $\alpha=1$. Panels (d) and (e) are for $N=20$. The $\Pi$ symbol denotes all permutations. The SSSV model does not reproduce the correct ordering.  The error bars represent the 95\% confidence interval.}
   \label{fig:cross-talk-ES}
\end{figure*}

The ME result [Fig.~\ref{fig:ME-Chi}] is now a significantly closer match to the DW2 cluster populations than before [Fig.~\ref{fig:ME-Uncalibrated}], over the entire range of $\alpha$ values.  The ME captures quantitatively the cluster state populations, while the noisy SSSV model [Fig.~\ref{fig:SSSV-Chi}], with $\chi$ and the noise variance optimized to match the DW2 results at $\alpha=1$, does not capture the cluster state populations correctly.  To highlight this difference, the same data are plotted in Fig.~\ref{fig:new1} for only two cluster states. The same conclusions apply for $N=20$ spins, as seen in Fig.~\ref{fig:DW2-Chi-N=20}, where we show only the two extremal of the $2^{10}$ cluster ground states.  

The ME's main discrepancy is in not capturing the full isolated state population, especially the strength of the peak at small $\alpha$, and it can only capture qualitatively the behavior of the isolated state as shown in Fig.~\ref{fig:new2}.  As illustrated by the SSSV results in Figs.~\ref{fig:SSSV-nonoise} and \ref{fig:SSSV-Uncalibrated}, the inclusion of noise on the local fields and couplings can have a dramatic effect on the small $\alpha$ behavior, while keeping the large $\alpha$ behavior mostly untouched.  Indeed, we have shown in Fig.~\ref{fig:NoisyME} that noise of a certain magnitude on the local fields and couplings does not significantly alter the cluster state distribution at $\alpha = 1$.  To study this effect over the entire range of $\alpha$ would require performing simulations for a large number of noise samples, which is unfeasible given the high computational cost of running the ME. However, even for a moderate number of small noise samples, we observe [Fig.~\ref{fig:new2}] an increase in the strength of the isolated state peak.  To increase the population at larger $\alpha$, increasing the system-bath coupling would increase the strength of thermal excitations, which would allow for the isolated state to be further populated.  We believe that an optimization over these parameters, albeit at a huge computational cost, could significantly improve the quantitative agreement between the ME and DW2.  However, our focus here has been to illustrate that the ME captures the behavior of the DW2 data remarkably well with no significant parameter fitting.

\subsection{The distribution of first excited states}

While the results presented in the previous subsection provide a clear quantitative discrepancy between the noisy SSSV model and the DW2 results, and demonstrate that the agreement with the ME is quantitatively better, 
it is important to provide a clearcut example of a qualitative discrepancy. To address this we now go beyond the ground subspace and consider an eight-dimensional subspace of the subspace of first excited states. We arrange these according to permutations of the core or outer qubits, i.e., we group the states as $\ket{1111 \, \Pi(0001)}$ and $\ket{\Pi(1110) \,1111}$, where $\Pi$ denotes a permutation. As shown in Fig.~\ref{fig:DW2-ES-nocalibration}, the DW2 prefers the set $\ket{\Pi(1110) \, 1111}$. However, the noisy SSSV model prefers the set $\ket{1111\, \Pi(0001)}$, as seen in Fig.~\ref{fig:SSSV-ES-chi}. This discrepancy becomes observable for $\alpha \lesssim 0.2$, where thermal excitations start to significantly populate the excited states. This also helps explain why $\alpha \approx 0.2$ played a threshold role in our ground state
analysis. 
This conclusion persists for $N=20$, as shown in Figs.~\ref{fig:DW2-ES-nocalibration-N=20} and \ref{fig:SSSV-ES-chi-N=20}.

In Fig.~\ref{fig:ME-ES}, we show similar results for the ME for $N=8$.  The results qualitatively match the DW2 ordering for $\alpha \gtrsim 0.15$. 
The error bars are large since it is computationally prohibitive to run a large number of noise instances, which also restricted us to a relatively low noise level ($\sigma = 0.025$). It is difficult to conclude much for $\alpha \lesssim 0.15$ because the high computational cost forces us to truncate the energy spectrum in our ME simulations, which predominantly degrades our ability to compute the excited state populations at low $\alpha$.

To summarize, we showed in the previous section that with the inclusion of the cross-talk terms in the Hamiltonian the ME captures the convergence of the cluster state populations for small $\alpha$ as well, while the noisy SSSV model predictions do not improve relative to the case without the cross-talk correction. The discrepancy between the noisy SSSV model and the experimental data is amplified when we consider the excited states, for which the former predicts the opposite population ordering from the one observed, as shown in Fig.~\ref{fig:cross-talk-ES}.

We note that varying $\alpha$ is not the only way in which a control parameter for thermal excitations can be introduced. In Appendix \ref{app:moreThermal} we discuss the similar effect of increasing the total annealing time or the number of spins, along with experimental results.

\begin{figure*}
   \centering
\subfigure[\ Closed system]{\includegraphics[width=0.48\textwidth]{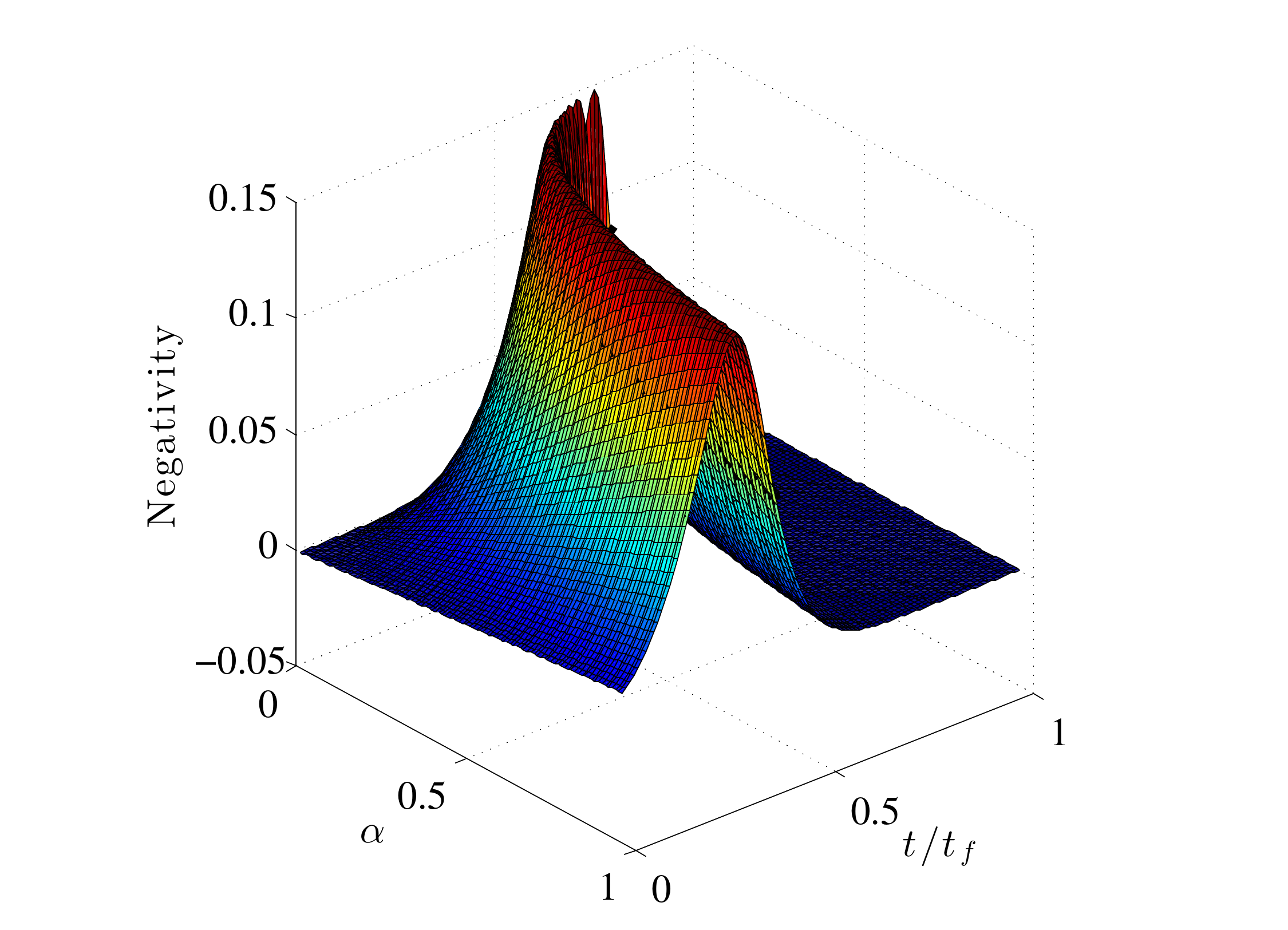} \label{fig:NegativityClosed}}
\subfigure[\ Open system]{\includegraphics[width=0.48\textwidth]{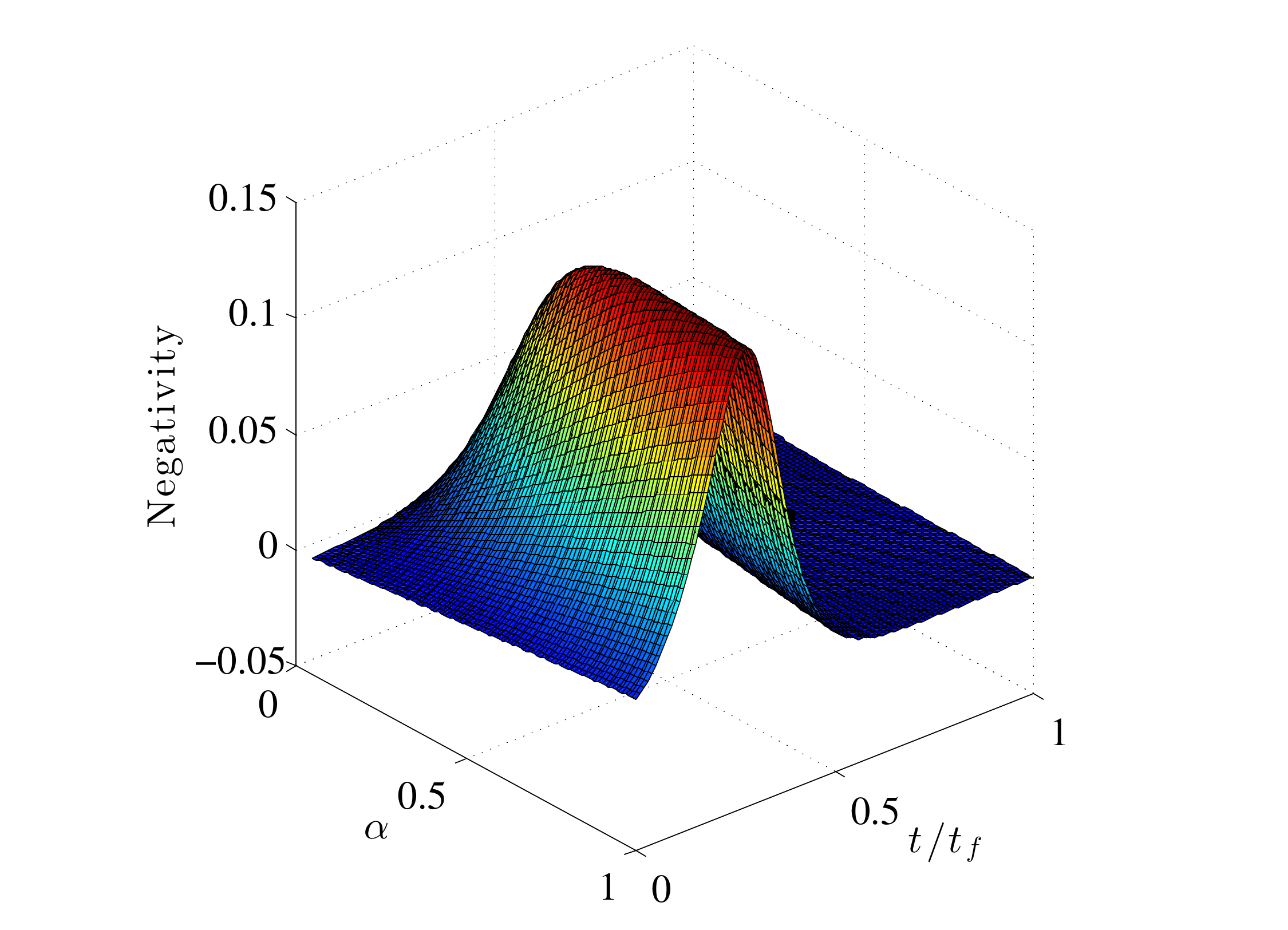}    \label{fig:NegativityOpen}}
   \caption{ Time-dependence of the negativity [Eq.~\eqref{eqt:negativity}] for (a) a closed system evolution and (b) an open system evolution of $N=8$ qubits (modeled via the ME with $\kappa= 1.27 \times 10^{-4}$), as a function of $\alpha$. The rapid decay of negativity for small $\alpha$ in the open system case signals a transition to classicality. However, for large $\alpha$ the closed and open system negativities are similar, suggesting that the system is quantum in this regime. The apparent jaggedness of the closed system plot near $\alpha = 0$ is due to our discretization of $\alpha$ in steps of 0.01.}
   \label{fig:Negativity}
\end{figure*}

\section{Ground state entanglement during the course of the annealing evolution}
\label{sec:ent}

Having established that the ME is, at this point, the only model consistent with the DW2 data, we are naturally led to ask whether the ME displays other quantifiable measures of quantum mechanical behavior. We thus use the ME to compute an entanglement measure for the time-evolved state. Ground state entanglement was already demonstrated experimentally in Ref.~\cite{DWave-entanglement} for a different Hamiltonian; here we are concerned with the time-dependent entanglement as a function of $\alpha$, and are relying on the good qualitative match between the ME and our experimental results to justify this as a proxy for the actual entanglement.  To this end we use the negativity (a standard measure of entanglement \cite{Vidal:02a})
\beq \label{eqt:negativity}
\mathcal{N}(\rho) = \frac{1}{2} \left( || \rho^{\Gamma_A}||_1 - 1 \right)\, ,
\eeq
where $ \rho^{\Gamma_A}$ denotes the partial transpose of $\rho$ with respect to a partition $A$. Figure~\ref{fig:Negativity} shows the numerically calculated  negativity as a function of $\alpha$ along the time evolution for a ``vertical" partition of the eight-qubit system, i.e., with an equal number of core and outer qubits on each side. Both the closed and open system evolution cases are shown. We observe that in the case of the closed system evolution there is always a peak in the negativity for all values of $\alpha \geq 0.01$ studied, with an $\alpha$-dependent position.  This is not surprising since as we change $\alpha$, we change the relative position of the fixed ratio value of $A(t)/ (\alpha B(t))$, and we expect the negativity peak to correspond to the position of the minimum gap of $H(t)$ \cite{WuSarandyLidar:04}.\footnote{While this peak position does not precisely match the position of the minimum gap, the result in Ref.~\cite{WuSarandyLidar:04} holds in the thermodynamic limit and predicts a strict singularity; a discrepancy is therefore excepted in the case of a finite system size.}  For the open system case, in contrast, the negativity peak drops when $\alpha$ is sufficiently small. This can be said to signal a transition to classicality. The reason for this drop is that as $\alpha$ decreases the system thermalizes more rapidly towards the Gibbs state, but the Gibbs state is also approaching the maximally mixed state, which has vanishing entanglement. However, for large $\alpha$ the peak position and value is similar to that of the closed system case, so that the simulated system exhibits quantum features and has not decohered into a classical evolution.  This can be interpreted as another reason for the failure of classical models to reproduce the experimental data.
%

\section{Discussion and Conclusions}
\label{sec:conc}
Motivated by the need to discern classical from quantum models of the D-Wave processor, in this work we examined three previously published classical models of the D-Wave device (SA, SD \cite{Smolin}, SSSV \cite{SSSV}).
We studied the dependence of the annealing process on the energy scale of the final ``quantum signature" Hamiltonian. Lowering this energy scale acts as an effective temperature increase and thus enhances the effects of  thermal fluctuations. 
While this strategy might appear counterproductive as a means to rule out classical models since it promotes a transition to the classical regime, it in fact presents a challenge for classical models that must now accurately describe not only the ground subspace but also the excited state spectrum of a quantized system.  

We found that all of the classical models we studied are inconsistent with the experimental data for our quantum signature Hamiltonian, covering the range of $8$ to $20$ qubits (thus extending beyond the $8$-qubit unit cell of the DW2 device), in a ``black-box" setting of a study of the input-output distribution of the device. 
The SA and SD models were already rejected based on such inconsistency in earlier work \cite{q-sig,q108,comment-SS} and the present evidence supports and strengthens these conclusions. The SSSV model was of particular interest since it matches the ground state success probabilities of random Ising model experiments on the DW1 device \cite{SSSV}. While it is possible that with additional fine-tuning a better match can be achieved with a classical model, an adiabatic quantum ME \cite{ABLZ:12-SI} which we have examined is capable of reproducing most of the key experimental features with only one free parameter (the effective system-bath coupling $\kappa$).
Our most complete and accurate model for the D-Wave device accounts for qubit cross-talk and local field and coupling noise, where we demonstrated that the ME captures all the features in the experimental data, in contrast to the noisy SSSV model (Fig.~\ref{fig:cross-talk}). We have thoroughly analyzed and explained these findings.

It is important to stress that the ME exhibits decoherence not in the computational basis but in the energy eigenbasis. Such decoherence is not necessarily a detriment to QA since it is consistent with maintaining computational basis coherence in the ground state.

How can the rejection of the classical SSSV model by our experimental data on quantum signature Hamiltonian problem instances of up to $20$ qubits 
be reconciled with the conclusions of Ref.~\cite{SSSV}, which demonstrated a strong correlation between success probabilities of the SSSV model and the DW1 device for random Ising problem instances of $108$ qubits? One obvious consideration is problem size, though we have found no evidence to
suggest that the agreement with experiment improves for the SSSV model as the number of qubits increases. More pertinent seems to be the fact that the quantum signature Hamiltonian experiment probes different aspects of the QA dynamics than the random Ising problem instances experiment. The former is, by design, highly sensitive to the detailed structure of the ground state degeneracy and the manner in which this degeneracy is dynamically generated, and these aspects are different for quantum and classical models. In this sense, it is a more sensitive probe than the random Ising experiment \cite{q108}, which did not attempt to resolve the ground state degeneracy structure. While Ref.~\cite{SSSV} established that the SSSV model correlates very well with the experimental success probability distribution for random Ising instances, and even better with SQA, our results suggest the possibility that a closer examination would reveal important differences between the SSSV model and QA also for the random Ising experiment. 
For example, Ref.~\cite{q108} presented additional evidence for QA by also considering excited states and correlations between hardness and avoided level crossings with small gaps. 
Specifically, we conjecture that a detailed study of the ground state degeneracy for random Ising instances would determine the suitability of the SSSV model as a classical model for QA in this setting as well. 
Such a study might also circumvent an important limitation of our quantum signature Hamiltonian approach: the exponential degeneracy of the cluster states ($2^{N/2}$) makes gathering statistically significant data prohibitively time-consuming for $N\gtrsim 20$. 

Clearly, ruling out any finite number of classical models still leaves open the possibility that a new classical model can be found that explains the experimental data. Nevertheless, in the absence of a strict no-go test such as a Bell inequality violation, ruling out physically reasonable  classical models while establishing close agreement with a quantum model (the adiabatic ME), is a strategy that should bolster our confidence in the role played by quantum effects, even if it falls short of a proof that all classical models are inconsistent with the experiment. 
 
Finally, we stress that the results reported here do not address the scaling of the performance of the D-Wave devices against state-of-the-art classical solvers, or whether this scaling benefits from a quantum speedup \cite{speedup}. Recent work has highlighted the importance of the choice of the benchmark problems \cite{2014Katzgraber}. Moreover, a careful estimate of the scaling performance of the D-Wave devices must take into account the effects of limited connectivity and precision in setting the intended problem \cite{2014arXiv1406.7553V,2014arXiv1406.7601P}.

The presence of quantum speedup is possible only if the device displays relevant quantum features and defies a classical description. Our work rules out plausible classical models while at the same time showing consistency with an open quantum system description. For small values of $\alpha$ our ME predicts that entanglement rapidly vanishes, signaling a transition to classicality as the effective temperature becomes high enough, though a quantized energy spectrum persists. 
This observation can be of practical importance in the case of optimization problems where one expects that classical annealing can be more efficient than QA \cite{2014Katzgraber,speedup}. In this case one might obtain a performance improvement by allowing the device to work in the classical, thermal region. The possibility of such ``thermally assisted" QA has been indeed demonstrated experimentally  in~\cite{DWave-16q}, in the case of a specific toy problem. It is interesting to more generally characterize the potentially beneficial role played by thermal effects in affecting the performance of QA and adiabatic quantum computing \cite{TAQC,Vega:2010fk}. Apart from being a practical issue for the D-Wave device, thermal excitations present a fundamental obstacle for any adiabatic algorithm \cite{childs_robustness_2001,PhysRevLett.95.250503,PhysRevA.79.022107}. This issue must be addressed by  adding error correction to QA \cite{PAL:13,Young:2013fk}, or by exploiting thermal noise as a computational resource \cite{2013arXiv1307.1114V}.
Future work shall revisit these questions using new tests and larger system sizes.

\acknowledgments

We would like to thank Gabriel Aeppli, Andrew Fisher, Andrew Green, Seung-Woo Shin, Matthias Troyer, and Umesh Vazirani for valuable discussions. We thank Simone Severini for valuable discussions and a thorough and critical reading of an early version of the manuscript. Part of the computing resources were provided by the USC Center for High Performance Computing and Communications.  This research used resources of the Oak Ridge Leadership Computing Facility at the Oak Ridge National Laboratory, which is supported by the Office of Science of the U.S. Department of Energy under Contract No. DE-AC05-00OR22725. The work of W.V. and P.A.W. has been done within a Global Engagement for Global Impact programme funded by EPSRC Grant No. EP/K004506/1. The work of A.M., T.A. and D.A.L. was supported under ARO MURI Grant No. W911NF-11-1-0268, ARO Grant No. W911NF-12-1-0523, and the Lockheed Martin Corporation. A.M. was also supported by the USC Provost's Ph.D. fellowship.

%

\appendix
\section{The D-Wave Two device}
\label{app:device}
\begin{figure}[t]
\centering
\includegraphics[width=\columnwidth]{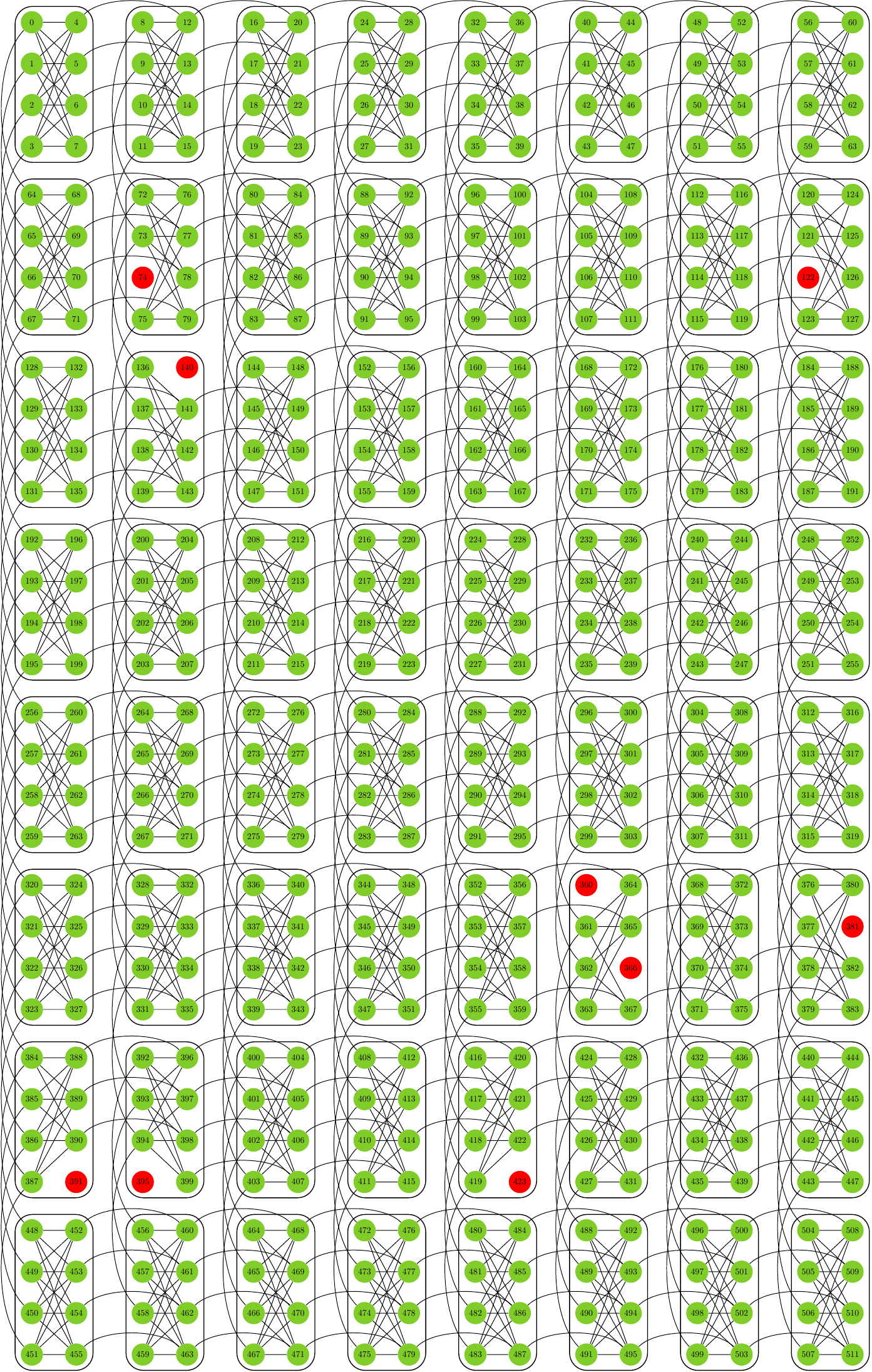}
\caption{ {Qubits and couplers in the DW2 device.} The DW2 ``Vesuvius'' chip consists of an $8\times8$ two-dimensional square lattice of eight-qubit unit cells, with open boundary conditions. The qubits are each denoted by circles, connected by programmable inductive couplers as shown by the lines between the qubits. Of the $512$ qubits of the device located at the University of Southern California used in this work, the $503$ qubits marked in green and the couplers connecting them are functional.}
\label{fig:DW2schematic}
\end{figure}

All our experiments were performed on the DW2 ``Vesuvius''  processor located at the Information Sciences Institute of the University of Southern California.  Details of the device have been given elsewhere \cite{Berkley:2010zr,Johnson:2010ys,Harris:2010kx}, and we only provide a brief overview here.  As shown Fig.~\ref{fig:DW2schematic}, the device is organized into an $8 \times 8$ grid of unit cells, each comprising eight qubits arranged in a $K_{4,4}$ bipartite graph, which together form the ``Chimera'' connectivity graph \cite{Choi2} of the entire device. Of the $512$ qubits $503$ were calibrated to within acceptable working margins in the DW2 processor used in our experiments. Figure~\ref{fig:DW2schematic} also gives a schematic representation of the most general problem Hamiltonian [as specified in Eq.~\eqref{eq:problem}] that can be implemented in the device. 

\section{Enhancement \textit{vs} suppression of the isolated state in SA \textit{vs} QA}
\label{app:generalized}
Here we review and generalize the detailed argument given in Ref.~\cite{q-sig} for the enhancement of the isolated state in SA {vs} its suppression in QA. 

\subsection{Classical master equation explanation for the enhancement of the isolated state for general $N$}
\label{app:B1}
We first explain why SA predicts an enhancement of the isolated state for general (even) $N$.
To do so we closely follow the arguments from Ref.~\cite{q-sig} concerning the $N=8$ case. Consider a signature Hamiltonian with $N=2n$ qubits. As depicted in Fig.~\ref{fig:12-16-20spinIsing}, $n$ of these are the ferromagnetically coupled ``core" qubits ($J_{ij}=1$), while the other $n$ ``outer" qubits are each ferromagnetically coupled to a single core qubit ($J_{ij}=1$). The local fields applied to the core qubits are $h_i=1$, while $h_i=-1$ for the outer qubits. \\

Under our classical annealing protocol the system evolves via single spin flips. That is, at each step of the evolution a state can transfer its population only to those states which are connected to it by single spin flip. Thus the rate of population change in a state depends only on the number of states it is connected to via single spin flips. Let the index $j$ run over all the states connected to state $a$; the Pauli master equation for the populations can then be written as
\begin{equation}
\label{eq:ThermalRate}
\dot{p}_a = \sum_{j} f(E_a - E_j)p_{j} - f(E_j-E_a)p_a ,
\end{equation}
where we have assumed that the transfer function $f(\Delta E)$ does not depend on $j$ and satisfies the detailed balance condition. For a derivation starting from the quantum ME, see Ref.~\cite{q-sig} (Supplementary Information).

We now derive a classical rate equation for the generalization to $N$ spins of the clustered and isolated states given in Eq.~\eqref{eq:C-I}. Let $P_\textrm{I}$ denote the population in the isolated state $\ket{\underbrace{11 \cdots 1 }_{\text{$n$ outer}}\, \underbrace{11\cdots 1}_{\text{$n$ core}}}$, and let $P_\textrm{C} = 2^{-n} \sum_c P_c$ denote the average population of the cluster-states $\{\ket{00\cdots 0\,00\cdots 0},\dots,\ket{11\cdots 1\,00\cdots 0}\}$. For the isolated state, flipping either a core spin or an outer spin creates an excited state. An outer-spin flip changes the core-outer spin-pair from $\ket{11}$ to $\ket{10}$. This flip has an associated cost of $4$ units of energy, and there are $n$ such cases. A flip of one of the core qubits changes the core-outer spin-pair from $\ket{11}$ to $\ket{01}$. This results in two unsatisfied links in the core ring, raising the energy by $4$ units. However, the flip leaves the energy of the given core-outer spin-pair unchanged. There are again $n$ such cases. Thus, for the isolated state, the rate equation is
\begin{align}
\dot{P_\textrm{I}} = 2n[f(-4)P_4 - f(4)P_\textrm{I}]
\end{align}
where $P_4$ is the population in the excited states which are $4$ units of energy higher than the ground states.

The derivation of the rate equation for the cluster-states is somewhat more involved. We note first that a flip of any of the outer spins involves no energy cost, so all the excited states created from the cluster-states arise from flipping a core spin. Depending on the state of the outer spin when the core qubit is flipped, we have two different cases. 
\begin{itemize}
\item If the outer is spin $\ket{0}$, the configuration of the core-outer pair changes from $\ket{00}$ to $\ket{10}$. This transition involves a change of 4 units of energy. Moreover, this creates a pair of unsatisfied links in the core ring, at a cost of another $4$ energy units. Overall, it takes $8$ units of energy to accomplish this flip. To count the total number of such excited states connected to all cluster-states, let us consider a cluster-state with $l$ outer spins in $\ket{0}$ and $n-l$ outer spins in $\ket{1}$. There are $\binom{n}{l}$ such cluster-states. In each of these states, we can choose any of the $l$ core spins to flip. Thus, the overall number of all such possible excited states connected to cluster-states is $\sum_{l=0}^{n} l \binom{n}{l} = n 2^{n-1}$.  
\item If the outer is spin $\ket{1}$, the configuration of the core-outer pair changes from $\ket{01}$ to $\ket{11}$. This core-outer transition involves no change of energy. However, this creates a pair of unsatisfied links in the core-ring, at a cost of $4$ energy units. The counting argument for number of these excited states is same as in the previous case. Thus, the number of all such possible excited states connected to cluster-states is again $n 2^{n-1}$.
\end{itemize}
We assume that all cluster-states have the same population, equal to the average population. The rate equation of the average cluster-states population is then
\bes
\begin{align}
\dot{P_\textrm{C}} &= {n 2^{n-1} \over 2^n} ([f(-8)P_8 - f(8)P_\textrm{C}] + [f(-4)P_4 - f(4)P_\textrm{C}])   \\
          &= {n \over 2}  \left[f(-8)P_8 - f(8)P_\textrm{C} + f(-4)P_4 - f(4)P_\textrm{C} \right]  ,
\end{align}
\ees
where $P_8$ is the population in the excited states that are $8$ units of energy above the ground states.   

For most temperatures of interest, relative to the energy scale of the
Ising Hamiltonian, the dominant transitions are those between the cluster and states with energy $-4$.  Transitions to energy $0$ states are suppressed by the high energy cost, and transitions from energy $0$ states to the cluster-states are suppressed by the low occupancy of the $0$ energy states.
\begin{equation}
\dot{P_\textrm{C}} \approx {n \over 2} [f(-4)P_4 - f(4)P_\textrm{C}]
\label{eq:coreRate}
\end{equation}

In classical annealing at constant low temperature starting from
arbitrary states (that is, the high energy distribution), probability
flows approximately $\dot P_\textrm{I} / \dot P_\textrm{C} \approx 4$ times faster into the
isolated state initially, and it gets trapped there by the high energy
barrier. To show that $\dot P_\textrm{I} \ge \dot P_\textrm{C}$ for slow cooling schedules, assume that this is indeed the case initially. Then, in order for $P_\textrm{C}$ to become larger than $P_\textrm{I}$, they must first become equal at some inverse annealing temperature $\upbeta'$:  $P_\textrm{I} (\upbeta')= P_\textrm{C} (\upbeta') \equiv P_g$, and it suffices to check that this implies that $P_\textrm{I}$ grows faster than $P_\textrm{C}$. Subtracting the two rate equations at this temperature yields
\bes
\begin{align}
 \dot P_\textrm{I}- \dot P_\textrm{C} &= \frac{3n}{2} ( f(-4)\, P_4 - f(4) \, P_g)  \\
 &= \frac{3n}{2} f(-4) P_g \left( \frac{P_4}{P_g} - \frac{P(g \to 4)} {P(4 \to g)}\right)\;,
\end{align}
\ees
where in the second line we used the detailed balance condition, and $P(4 \to g)$ denotes the probability of a transition from the excited states with energy $4$ units above the ground state to the ground state $g$. Now, because the dynamical SA process we are considering proceeds via cooling, the ratio between the non-equilibrium excited state and
 the ground state probabilities will not be lower than the corresponding thermal equilibrium transition ratio, i.e., $\frac{P_4}{P_g} \geq  \frac{P(g \to 4)} {P(4 \to g)} = e^{-4\upbeta'}$. Therefore, as we set out to show,
 \begin{align}
     \dot P_\textrm{I} - \dot P_\textrm{C} \ge 0\;,
 \end{align}
 implying that at all times $P_\textrm{I} \ge P_\textrm{C}$. 

\subsection{Perturbation theory argument for the suppression of the isolated state in QA for general $N$}
\label{app:pert-theory}
We consider the breaking of the degeneracy of the ground state of our $N$ spin benchmark Ising Hamiltonian by treating the transverse field $H_X =  -\sum_{i=1}^N \sigma_i^x $ as a perturbation of the Ising Hamiltonian (thus treating the QA evolution as that of a closed system evolving backward in time). As pointed out in the main text, the ground state is $2^{N/2} + 1$-fold degenerate.  According to standard first order degenerate perturbation theory, the
perturbation $\hat{P}_g$ of the ground subspace is given by the spectrum of the
projection of the perturbation $H_X$ onto the ground subspace.  \beq
\Pi_0 = \left(\ket{1}\bra{1}\right)^{\otimes N} + \left(\ket{0}\bra{0} \right)^{\otimes N/2}  \left(\ket{+}\bra{+} \right)^{\otimes N/2} \ ,
\eeq
where the first term projects onto the isolated state, and we have written the state of the outer qubits of the cluster in terms of $\ket{+} = (\ket{0}+\ket{1})/\sqrt{2}$. 
We therefore wish to understand the
spectrum of the operator
\beq
\hat{P}_g =  \Pi_0 \left(-\sum_{j=1}^N \sigma_j^x \right)\Pi_0 \;.
\eeq
The isolated state is unconnected via single spin flips to any other state in the ground subspace, so we can write $\hat{P}_g$ as a direct sum of the $0$ operator acting on the isolated state and the projection onto the space $\Pi_0' = \Pi_0 - (\ket{1}\bra{1})^{\otimes N} = \left(\ket{0}\bra{0} \right)^{\otimes N/2}  \left(\ket{+}\bra{+} \right)^{\otimes N/2}$ of the cluster
\bes
\begin{align}
\hat{P}_g &=  -0 \oplus \Pi_0' \left(-\sum_{j=1}^N \sigma_j^x \right)\Pi_0' \\
& =  - 0 \oplus \left(-\sum_{j=N/2+1}^N \sigma_j^x\right)\;,
\end{align}
\ees
where the sum is over the outer qubits.

This perturbation 
splits the ground space of $H_\textrm{I}$, lowering the energy of
$\ket{00\cdots 0 ++\cdots +}$, and the $N/2$ permutations of $\ket{-} = (\ket{0}-\ket{1})/\sqrt{2}$ in the outer qubits of $\ket{00\cdots 0++\cdots +-}$. None of these states
overlaps with the isolated ground state, which is therefore not a ground
state of the perturbed Hamiltonian. Furthermore, after the
perturbation, only a higher (the sixth for $N=8$) excited state overlaps with the isolated state. The isolated state becomes a ground state only at the very end of the evolution (with time going forward), when the perturbation has vanished. This explains why the isolated state is suppressed in a closed system model.  A numerical solution of the ME agrees with this prediction for sufficiently large values of the problem energy scale $\alpha$.

\section{Experimental data collection methodology}\label{app:experiment}

Our data collection strategy was designed to reduce the effects of various control errors. In this appendix we explain the main sources of such errors and our methods for reducing them. These methods are distinct from, and complementary to other error correction methods \cite{Lidar-Brun:book}, inspired by stabilizer codes, that have been previously proposed and implemented \cite{PAL:13,Young:2013fk}.

Each time (programming cycle) a problem Hamiltonian is implemented on the DW2 device, the values of the local fields and couplings $\{h_i,\,J_{ij}\}$ are set with a Gaussian distribution centered on the intended value, and with standard error of about $5\%$~\cite{Trevor}. To average out these random errors we ran several different programming cycles for the same problem Hamiltonian as described in Sec.~\ref{app:data-collection} on this appendix. Differences among the individual superconducting flux qubits can contribute to systematic errors. To average out these local biases, we embedded our Hamiltonian multiple times in parallel on the device using different flux qubits, as also explained in Sec.~\ref{app:data-collection} of this appendix.  Furthermore, we implemented different ``gauges", a technique introduced in Ref.~\cite{q-sig}. A gauge is a given choice of $\{h_i,\,J_{ij}\}$; a new gauge is realized by randomly selecting $a_i =\pm 1$ and performing the substitution $h_i \mapsto a_i h_i$ and $J_{ij} \mapsto a_i a_j J_{ij}$. Provided we also perform the substitution $\sigma_i^z \mapsto a_i \sigma_i^z$, we map the original Hamiltonian to a gauge-transformed Hamiltonian with the same energy spectrum but where the identity of each energy eigenstates is relabeled accordingly.  In total, there are $2^N$ different gauges for an $N$-spin problem. We averaged our data using different programming cycles, embeddings, and gauges, as explained in Sec.~\ref{app:data-analysis} of this appendix. In addition we checked for errors due to correlations between successive runs (Sec.~\ref{app:autoCorr} of this appendix) and found these to be negligible. Finally, we describe a new method for correcting control errors that assumes that the degenerate cluster states should ideally have the same population (Appendix~\ref{sec:control-errs}).

\begin{figure}[t] 
\includegraphics[width=\columnwidth]{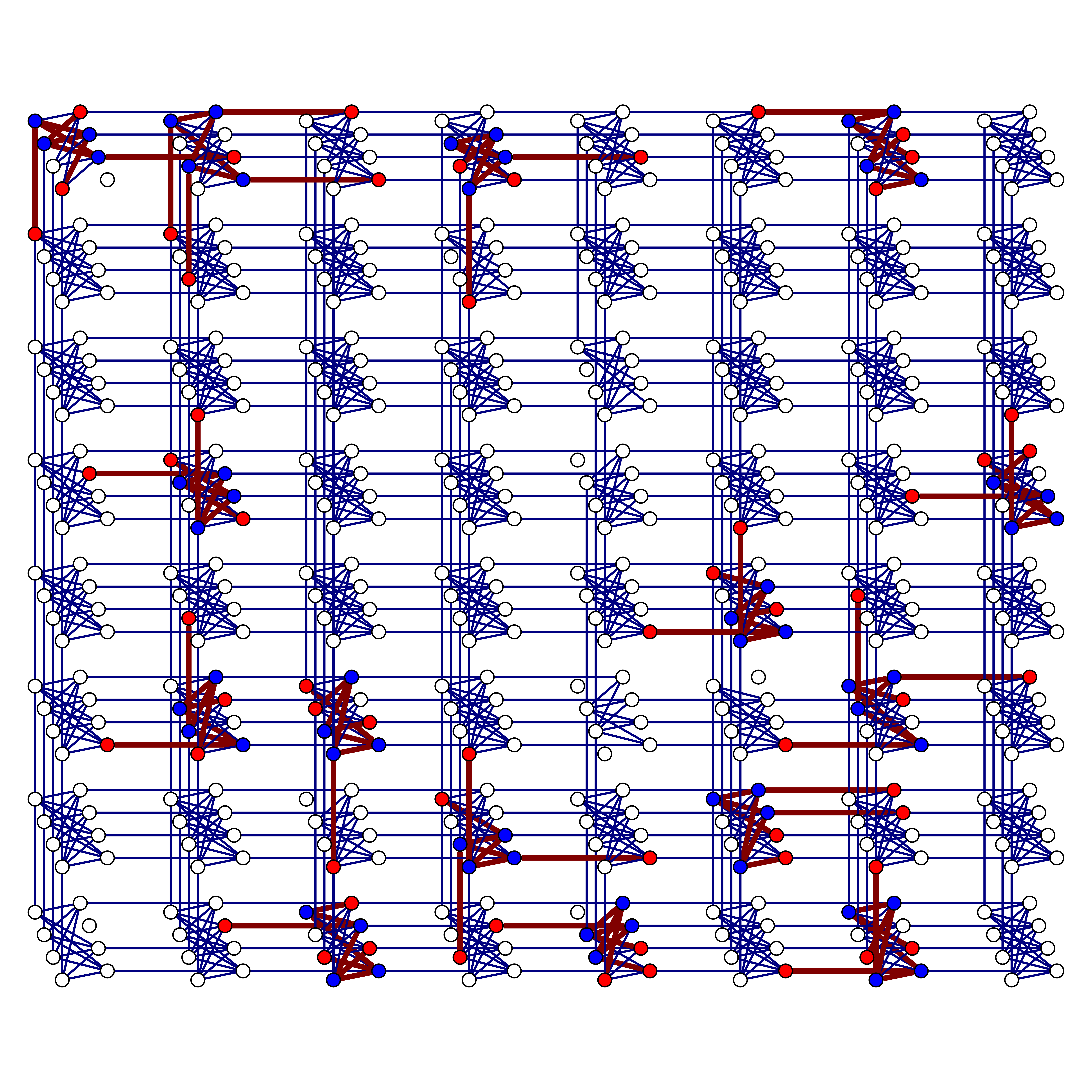}
\caption{Embedding according to the random parallel embeddings strategy for eight spins. An example of $15$ randomly generated different parallel copies of the eight-spin Hamiltonian. Our data collection used a similar embedding with $50$ different copies.}
\label{fig:8q-embed-15}
\end{figure}
\begin{figure}[t] 
\includegraphics[width=\columnwidth]{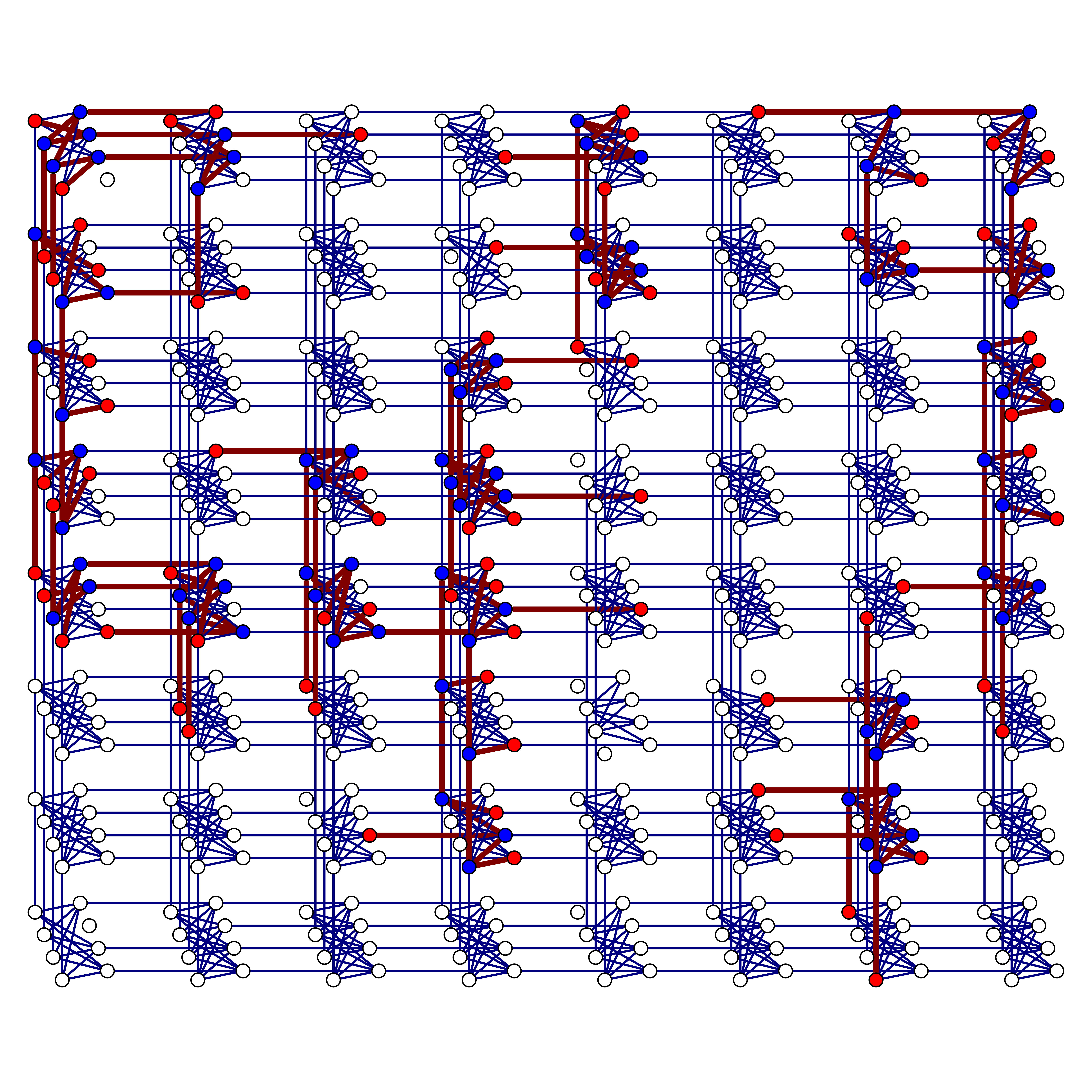} 
\caption{Embedding according to the random parallel embeddings strategy for $16$ spins. An example of $10$ randomly generated different parallel copies of the $16$-spin Hamiltonian. Our data collection used a similar embedding with $93$ different copies.}
\label{fig:randembedding16}
\end{figure}
\begin{figure}[t] 
\includegraphics[width=0.325\columnwidth]{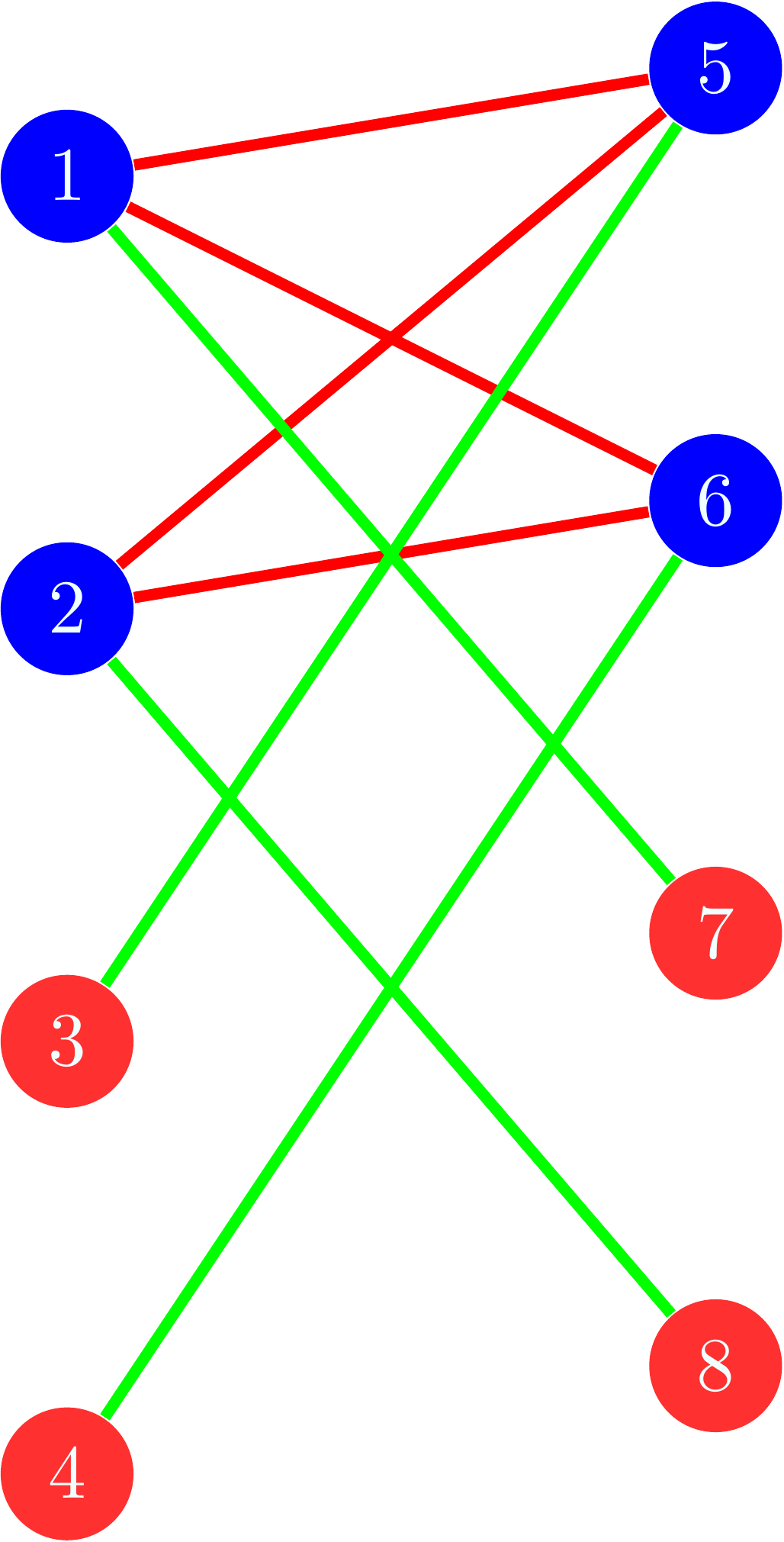} 
\caption{Embedding according to the ``in-cell embeddings" strategy. An example of a randomly generated in-cell embedding.}
\label{fig:8q-embed}
\end{figure}
\begin{figure*}[t]
\subfigure[\ $\alpha = 0.1$]{\includegraphics[width=0.33\textwidth]{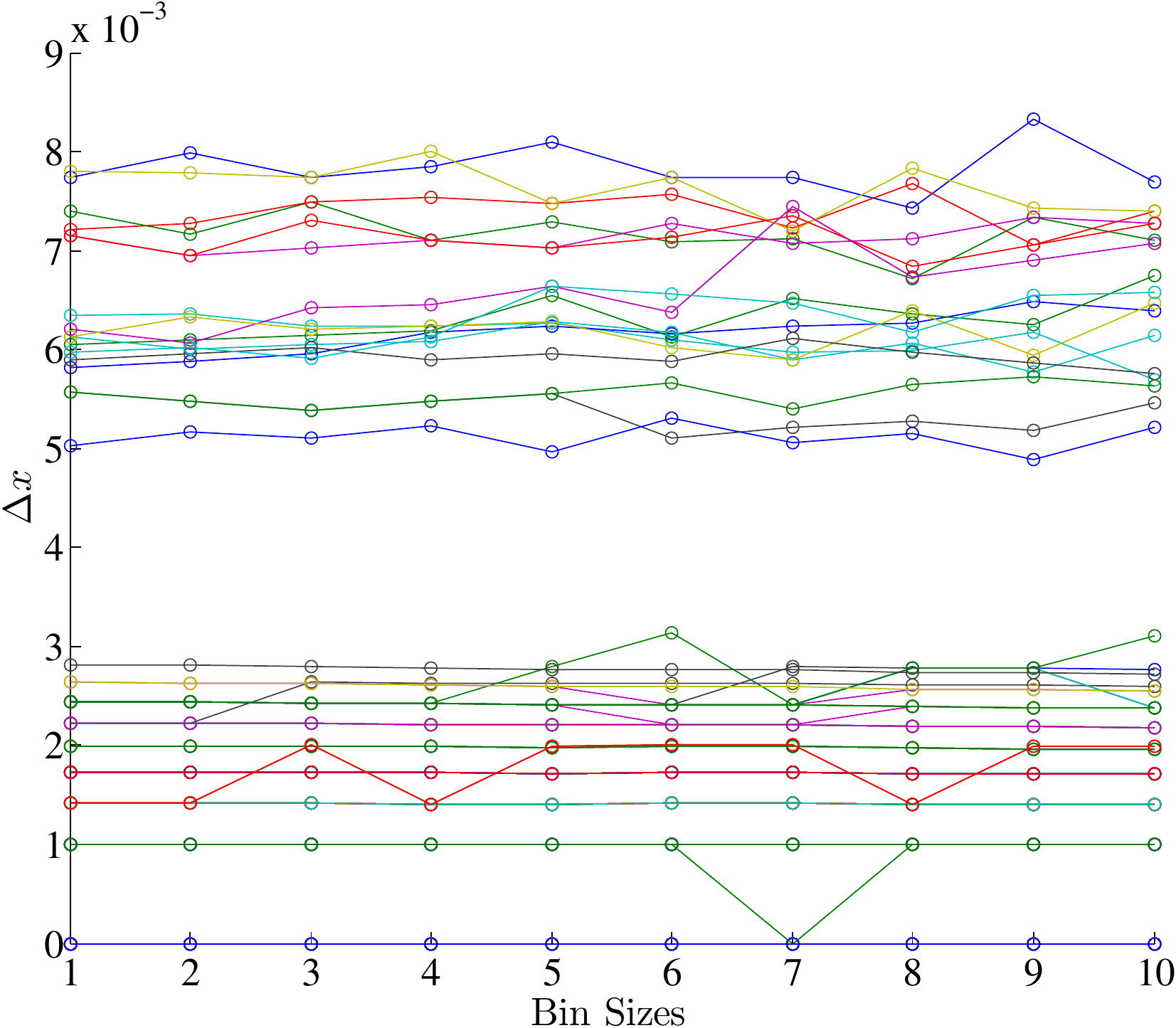}} 
\subfigure[\ $\alpha = 0.35$]{\includegraphics[width=0.33\textwidth]{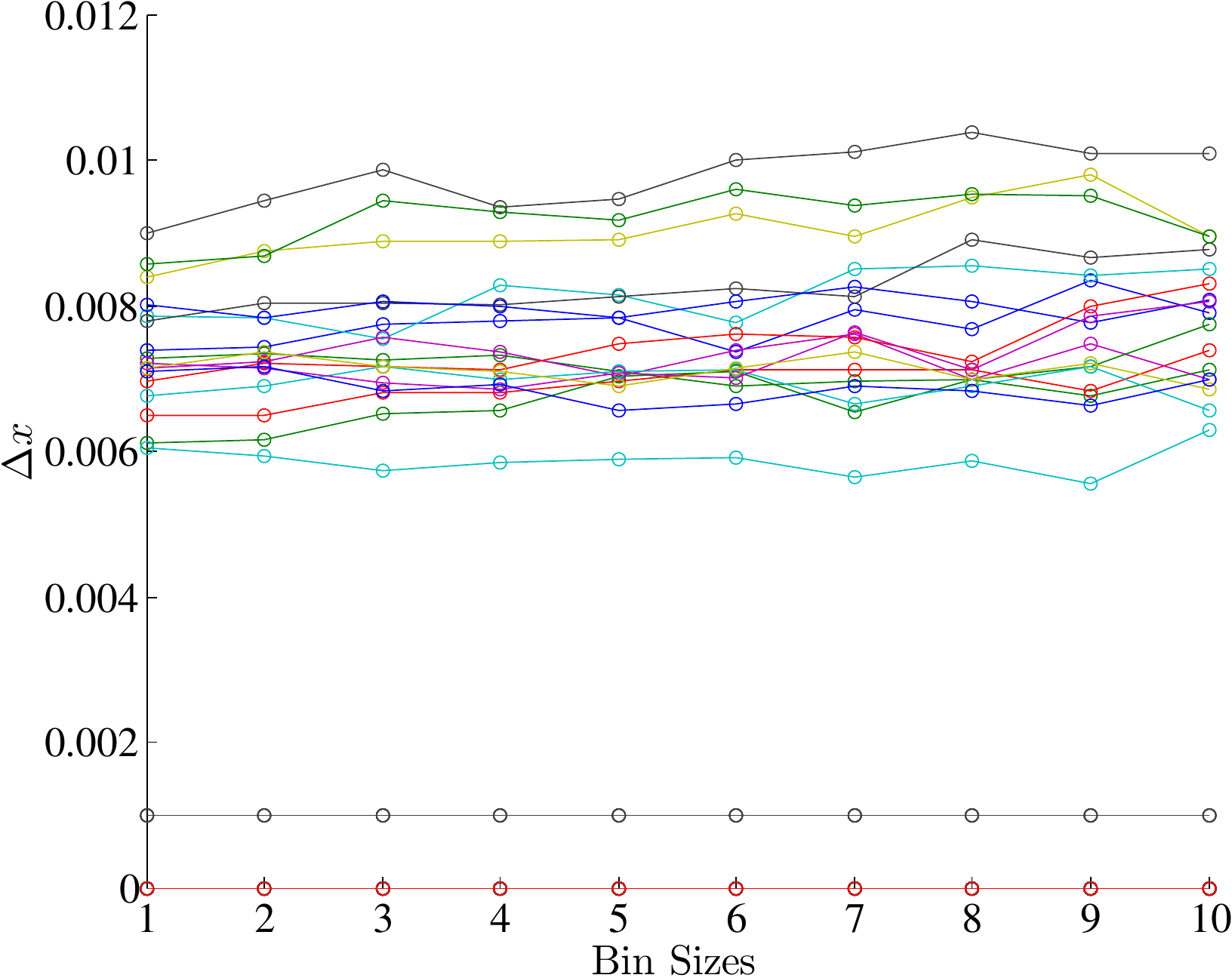}} 
\subfigure[\ $\alpha = 1$]{\includegraphics[width=0.33\textwidth]{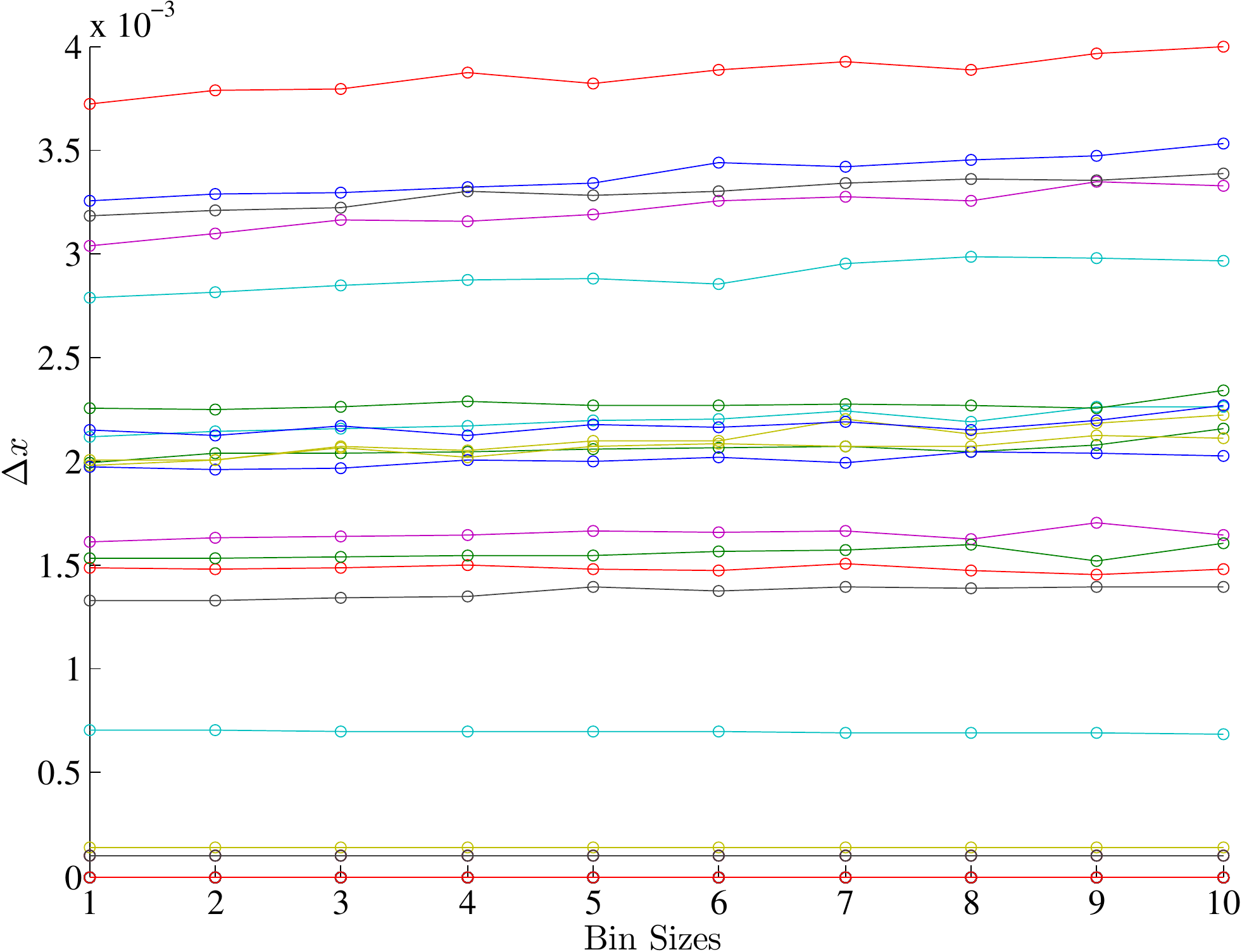}} 
\caption{\protect Some representative autocorrelation tests at $t_f=20\mu s$ showing the standard error of the mean $\Delta x$ as a function of binning size, for different values of $\alpha$: (a) $\alpha = 0.1$, (b) $\alpha = 0.35$, (c) $\alpha = 1$. Each curve is the result of the binning test for a different state. The relatively flat lines for all states suggest that there are no significant autocorrelations in the data.}
\label{fig:autocorrelation}
\end{figure*}


\subsection{Data collection strategies}
\label{app:data-collection}
We used two different data collection strategies that resulted in perfectly consistent results.

\emph{Strategy A: Random parallel embedding}.  As illustrated in Fig.~\ref{fig:8q-embed-15}, $50$ different parallel embeddings of the $8$-spin problem Hamiltonian were generated randomly in such a way that an embedding is not necessarily limited to a unit cell.  We thus solved $50$ different copies of the same $8$-spin problem in parallel during each programming cycle. We generated two such parallel embeddings containing $50$ copies each.  For each embedding,  we performed $100$ programming cycles and $1000$ readouts for the runs with annealing time $t_f = 20\mu$s, $200\mu$s; $200$ programming cycles and $498$ readouts for  the runs with annealing time $t_f = 2000\mu$s; $500$ programming cycles and $48$ readouts for  the runs with annealing time $t_f = 20000\mu$s. An example set of randomly generated embeddings  for the $16$ spin Hamiltonian is shown in Fig.~\ref{fig:randembedding16}. 

\emph{Strategy B: In-cell embeddings}. We utilized $448$ qubits to program $56$ parallel copies of the eight-spin problem Hamiltonian, with an identical gauge for all copies, with one copy per unit cell. All possible $256$ gauges were applied sequentially. For a given annealing time $t_f$, the number of readouts was $\min (1000,\lfloor 5\times10^5/t_f \rfloor)$. For example, $1000$ readouts were done for $ t_f=20\mu$s and $100$ readouts for $t_f=5000\mu$s. One such copy of an in-cell embedding is shown in Fig.~\ref{fig:8q-embed}. No in-cell embeddings are possible for problems involving $N>8$ spins.

\emph{Strategy C: Designed parallel embedding}. For $N=40$, we utilized $320$ qubits to program $8$ parallel copies of the $40$-spin problem Hamiltonian. The $40$-qubit Hamiltonian was designed with three different embeddings spread across the Chimera graph. Each embedding occupied six different unit cells. One hundred random gauges were chosen out of the $2^{40}$ possible gauges and were identically applied to all eight copies. To collect significant statistics, we performed $200$ programming cycles with $10000$ readouts for every gauge. The annealing time was $t_f=20\mu$s.
\subsection{Data analysis method}
\label{app:data-analysis}
The following method was used to analyze the data.  Let us denote the number of gauges by $N_{\mathrm{G}}$ and the number of embeddings by $N_{\mathrm{E}}$.  For a given embedding $a$ and gauge $g$, the number of total readouts (number of readouts times the number of programming cycles) for the $i$th computational state is used to determine the probability $p_{a,g} (i)$ of that computational state.  The gauge-averaged probabilities for the $i$th computational state $p_a^{\mathrm{GA}}(i)$ of the $a$th embedding are determined by averaging over the gauges for a fixed embedding:
\beq
p^{\mathrm{GA}}_a(i) = \frac{1}{N_{\mathrm{G}}} \sum_{g=1}^{N_{\mathrm{G}}} p_{a,g} (i) \ , \quad a = 1, \dots, N_{\mathrm{E}}\, .
\eeq
Let us now consider a function of interest $\mathcal{F}$, for example $P_{\mathrm{I}}/P_{\mathrm{C}}$ or the trace-norm distance $\mathcal{D}(\rho_{\mathrm{DW2}},\rho_{\mathrm{Gibbs}})$.  Using the raw probabilities $p_{a,g}(i)$, we can calculate $\mathcal{F}_{a,g}$.  For example, if $\mathcal{F} = P_{\mathrm{I}}/P_{\mathrm{C}}$, we have:
\beq
\mathcal{F}_{a,g} = \left( \frac{P_{\mathrm{I}}}{P_{\mathrm{C}}} \right)_{a,g} = \frac{16 p_{a,g}(\mathrm{I})}{\sum_{i=1, i \in \mathrm{C}}^{16} p_{a,g}(i)}  \ .
\eeq
For a fixed embedding $a$, we calculate the standard deviation $\sigma^{\mathrm{G}}_a$ associated with the distribution of $\mathcal{F}$ using the raw probabilities values over the $N_{\mathrm{G}}$ gauges, i.e.
\beq
\sigma^{\mathrm{G}}_a = \mathrm{std} \left[ \left\{ \mathcal{F}_{a,g} \right\}_{g=1}^{N_{\mathrm{G}}} \right] \ .
\eeq
 For each embedding, we also calculate $\mathcal{F}_a^{\mathrm{GA}}$ using the gauge-averaged probabilities, e.g.,
 \beq
\mathcal{F}_{a}^{\mathrm{GA}} = \left( \frac{P_{\mathrm{I}}}{P_{\mathrm{C}}} \right)_{a}^{\mathrm{GA}} = \frac{16 p_{a}^{\mathrm{GA}}(\mathrm{I})}{\sum_{i=1, i \in \mathrm{C}}^{16} p_{a}^{\mathrm{GA}}(i)} \ .
\eeq
 Therefore, for each embedding, we now have the following sets of data $\{(\mathcal{F}_a^{\mathrm{GA}}, \sigma_a^{\mathrm{G}})\}_{a=1}^{N_{\mathrm{E}}}$.  We refer to this as the gauge-averaged data, of which we have $N_{\mathrm{E}}$ data points.  We then drew $1000$ bootstrap \cite{Efron:1994qp} data samples from the gauge-averaged data (giving us a total of $1000 \times N_{\mathrm{E}}$ data points), which we denote by $\mathcal{F}_{a,b}^{\mathrm{GA}}$ where $a = 1, \dots , N_{\mathrm{E}} , \ b = 1, \dots , 1000$.  In order to account for the fluctuations in the gauge data, for a fixed bootstrap sample $b$, we add noise (normally distributed with the standard deviation $\sigma_a^{\mathrm{G}}$) to every $\mathcal{F}_{a,b}^{\mathrm{GA}}$ in the bootstrap sample.  For each of the $1000$ bootstrap data samples, we calculated the mean:
 \beq
\bar{\mathcal{F}}_b^{\mathrm{GA}} = \frac{1}{N_\mathrm{E}} \sum_{a=1}^{N_{\mathrm{E}}} \mathcal{F}_{a,b}^{\mathrm{GA}} \ , \quad b = 1, \dots, 1000 \ .
\eeq
Therefore we now have a distribution of 1000 means.  The mean of the 1000 means corresponds to the data points in our plots, and twice the standard deviation of the $1000$ means is the error bar used in the plots in the main text.

\subsection{Autocorrelation Tests}  
\label{app:autoCorr}
%
Correlations between the outputs of different runs on the device could be a result of errors on the device (such correlations were reported in \cite{q108}) in that the results of each run are not completely independent.  This can in turn affect the ground state populations by preferentially picking the first state observed.
To test for this possibility, we use a binning test, which is a simple method to test for autocorrelations in statistical data \cite{Wolff:2004nr}.  Consider a list of $n$ uncorrelated binary numbers $\{x_i\}$ with $P(x_i=1)=p$. The standard error of the mean for this dataset is $\Delta x \equiv \sqrt{\Var[x]/n}=\sqrt{p(1-p)/n}$.  We bin together the average of consecutive pairs in this list to produce a new list ${y_i}$ of $n/2$ numbers such that $y_i \in \{0,0.5,1\}$.  Since $P(y_i=1)=p^2$, $P(y_i=0.5)=2p(1-p)$ and $P(y_i=0)=(1-p)^2$, the error in the mean of this derived list is $\Delta y=\sqrt{p(1-p)/n}=\Delta x$.  If however, the list were correlated such that $P(x_{i+1}=1 \mid x_i=1)=q \not= p$, then $\Delta y = \sqrt{(p+q-p^2-q^2)/(2n)} \not= \Delta x$. The idea of the binning test easily follows: keep on binning data with larger bin sizes until the error in the means converges to a constant value. The minimal bin size where this occurs is the autocorrelation length, $\xi$. 

We used the binning test on all $256$ different states for the $N=8$ problem. For each state, we generated a list of $1000$ binary numbers $\{x_i\}$ such that $x_i=1$ when that state was read from the D-Wave device and $x_i=0$ otherwise. The probability of occurrence of the state is denoted by $\bar{x}$ and $\Delta x$ is its error. We found that the error in the mean does not change appreciably with the size of bins used. This indicates that the autocorrelation length for any state in our system is zero, and there are no significant autocorrelations in our data.  Figure \ref{fig:autocorrelation} shows a few representative cases from the data collected using the ``in-cell embeddings" strategy for various choices of $\alpha$ and random gauge choices. 

The \textit{Wald-Wolfowitz runs test} is a standard statistical test for autocorrelations. We tested the null hypothesis, $H_0$, that the sequence in consideration was generated in an unbiased manner. The Wald-Wolfowitz test relies on comparing the number of ``runs" in the dataset to a normal distribution of runs. A run is defined as consecutive appearance of same state. In our dataset of binary valued sequences, a run occurs every time there is a series of either $0$'s or $1$'s. For example, the sequence $011100100111000110100$ contains 6 runs of $0$ and 5 runs of $1$. The total number of runs is 11. Let $R$ be the number of runs in the sequence, $N_1$ be the number of times value $1$ occurs and $N_0$ be the number of times value $0$ occurs. (In our example, $R=11$, $N_1 = 10$ and $N_0=11$.) It can be shown that if the sequence were unbiased, the average and the standard deviation of the number of runs would be given by \cite{bradley1968distribution}
\begin{eqnarray}
\bar{R} &=& {2 N_1 N_0 \over N_1+N_0}+1 ,\\
\sigma_R^2 &=& {2 N_0 N_1(2 N_1 N_0 - N_1 - N_0) \over (N_1+N_0)^2(N_1+N_0-1)}
\end{eqnarray}     
The test statistic is $Z = {R - \bar{R} \over \sigma_R}$. At 5\% significance level, the test would reject the null hypothesis if $\abs{Z} > 1.96$ (in this case the obtained value of the number of runs differs significantly from the number of runs predicted by null hypothesis).

We applied the Wald-Wolfowitz test to the binary sequences used in the binning test. We found that for each value of $\alpha$, fewer than 0.01\% of the sequences failed the Wald-Wolfowitz test. For example, $112$ sequences for $\alpha=1$, $343$ sequences for $\alpha=0.35$ and $195$ sequences for $\alpha=0.1$ failed the test. The total number of such sequences tested for each value of $\alpha$ were $256\times 56 \times 256 \approx 3.6\times 10^6$. This suggests once more that autocorrelations do not affect our dataset significantly. 
%

\section{Kinks in the time dependence of the gap}
\label{app:kink}
%

\begin{figure*}[t] 
\centering
\subfigure[]{\includegraphics[width=3.3in]{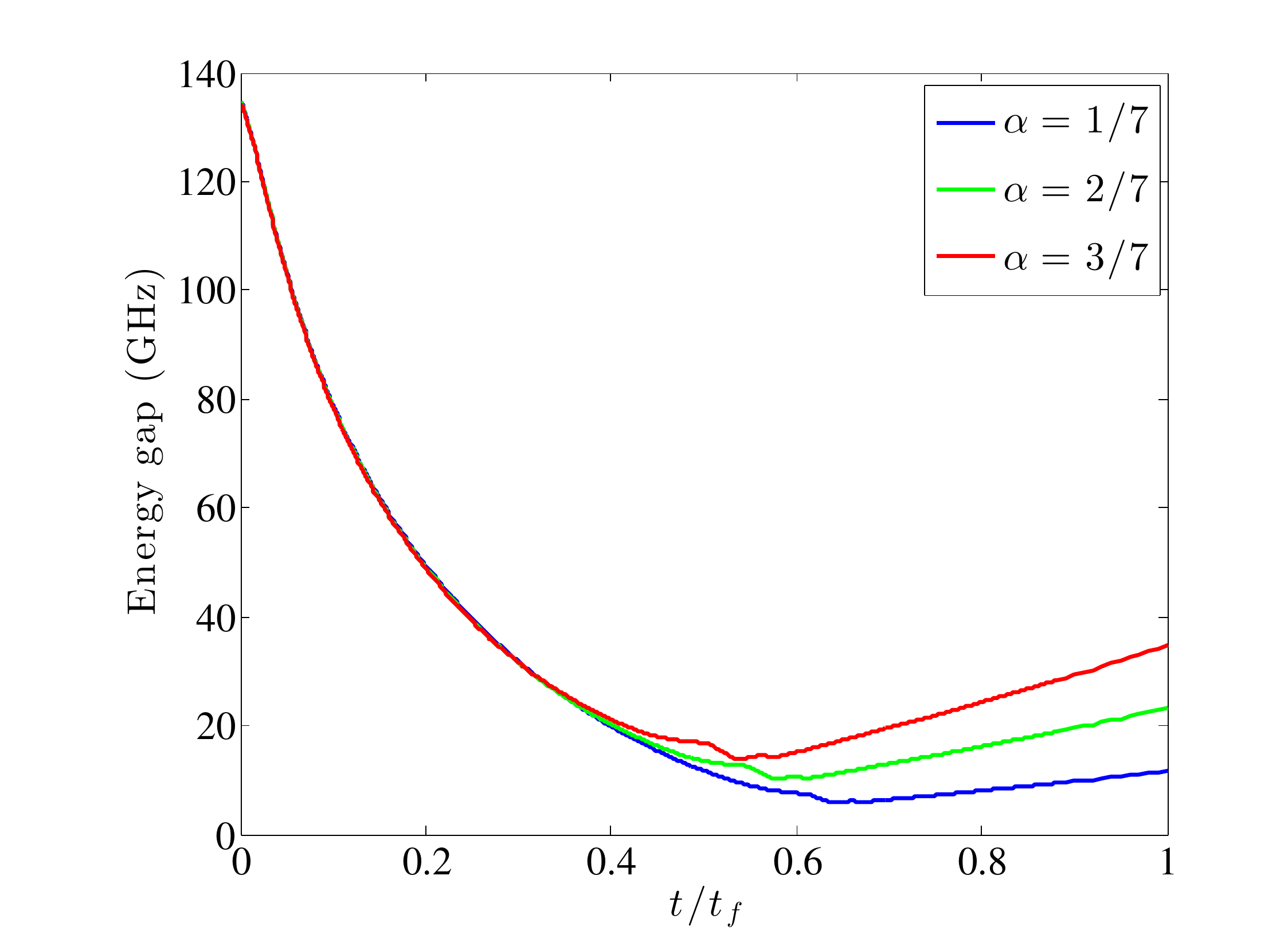} \label{fig:GapVsTime-SM}} 
\subfigure[\ $\alpha = 1/7$]{\includegraphics[width=3.3in]{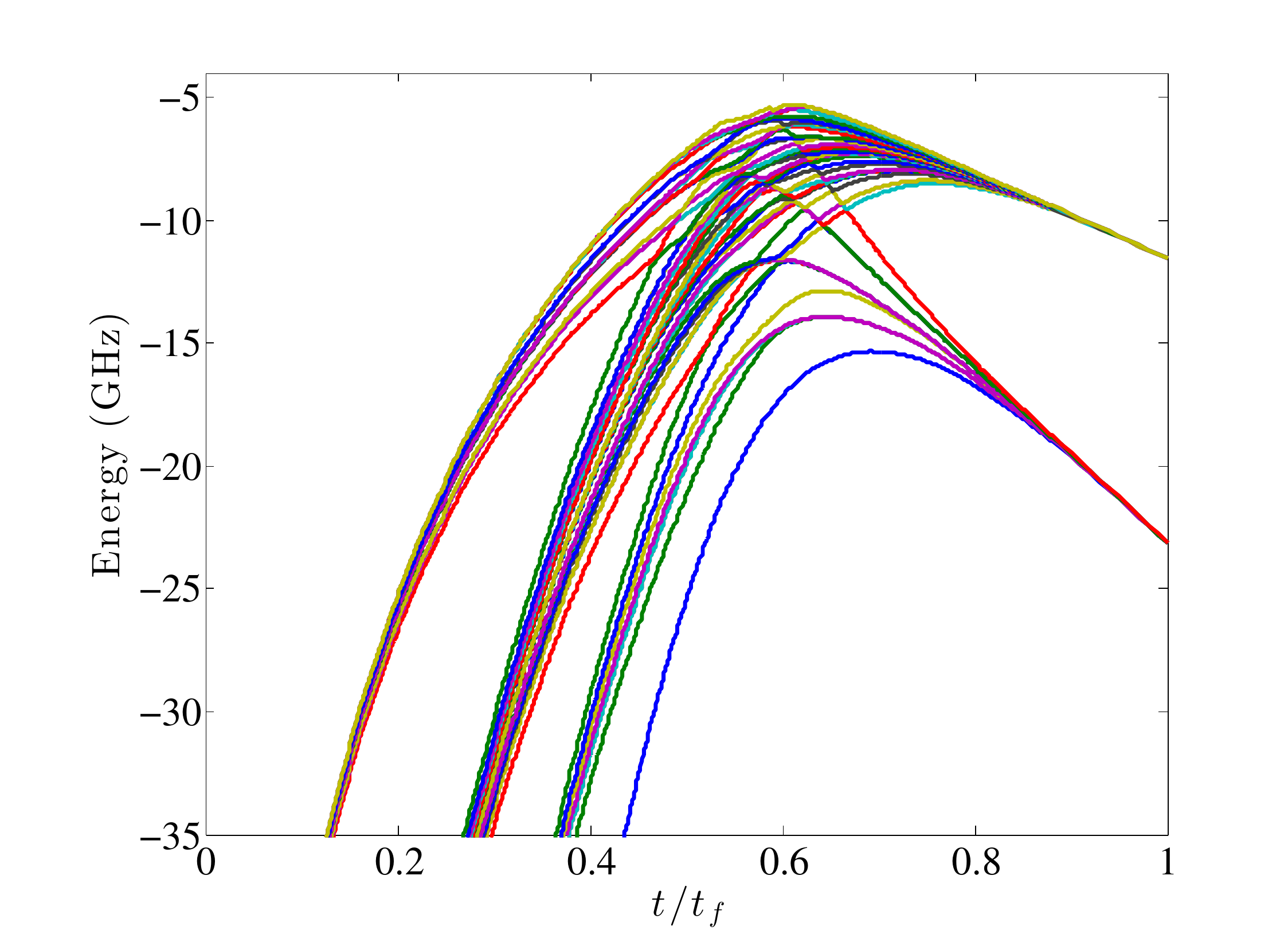} } 
\subfigure[\ $\alpha = 2/7$]{\includegraphics[width=3.3in]{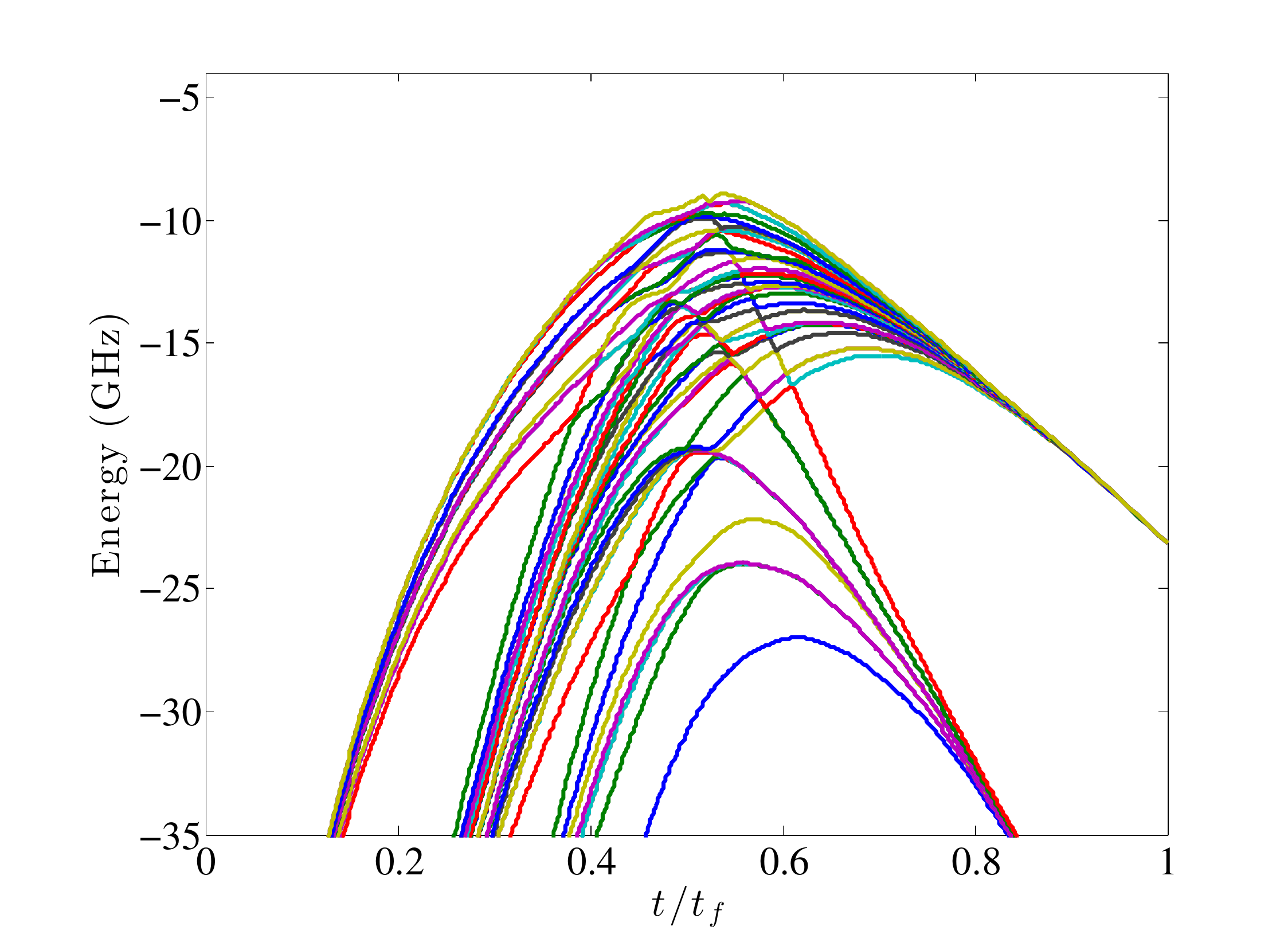} }    
\subfigure[\ $\alpha = 3/7$]{\includegraphics[width=3.3in]{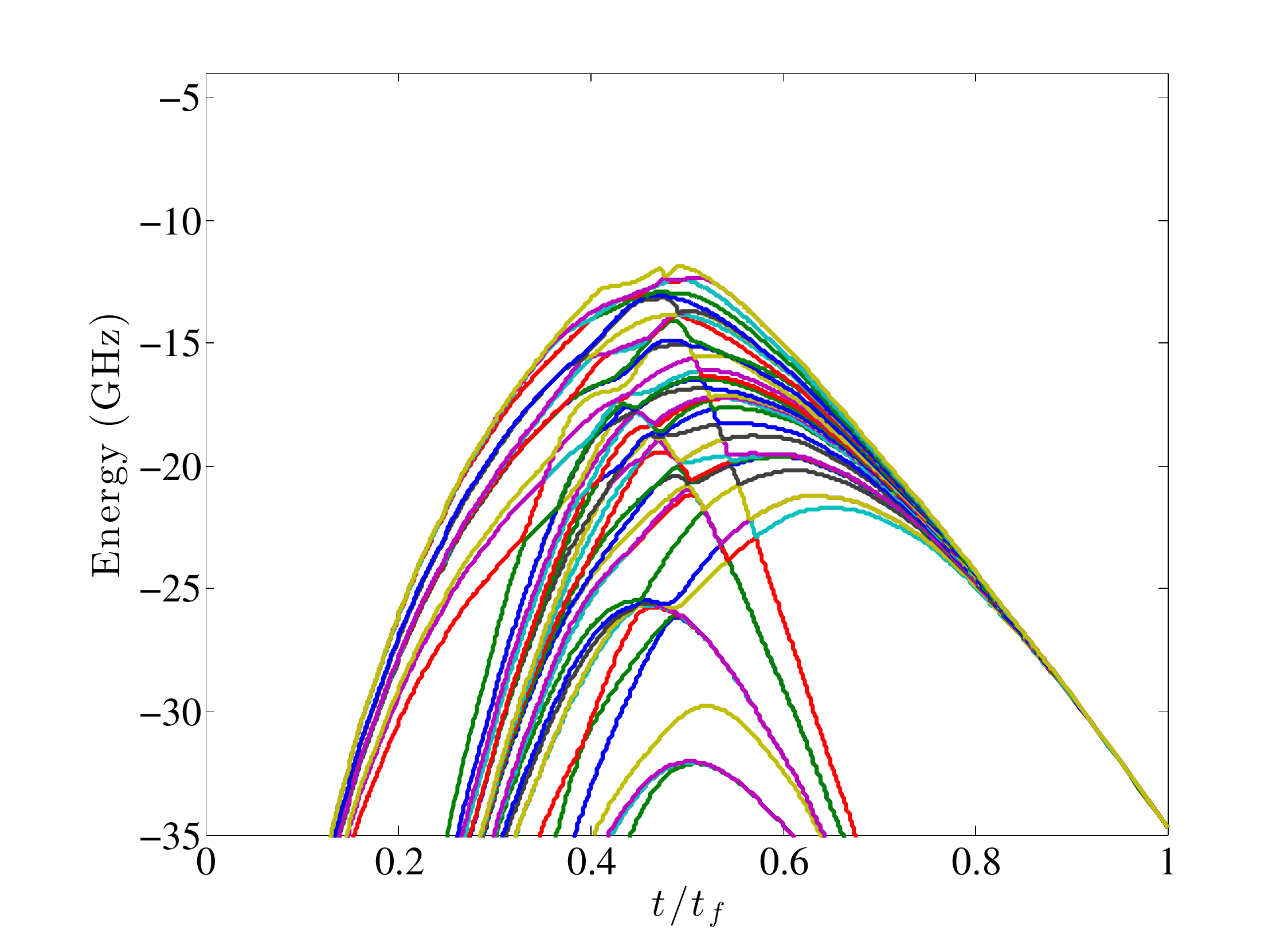} } 
\caption{(a) Time-dependence of the gap between the $18$th excited state and the instantaneous ground state, for different values of the energy scale factor $\alpha$. (b)-(d) Time-dependence of the lowest $56$ energy eigenvalues for different values of $\alpha$ [(b)$\alpha = 1/7$, (c) $\alpha = 2/7$, (d) $\alpha = 3/7$].  Note that the identity of the lowest $17$ energy eigenvalues  changes over the course of the evolution.}
\label{fig:energy}
\end{figure*}

Here we explain the origin of the kinks in the time dependence of the gap seen in Fig.~\ref{fig:GapVsTime}.  First, just as in Fig.~\ref{fig:GapVsTime} but for different values of $\alpha$, we show in Fig.~\ref{fig:energy}(a) how as $\alpha$ is decreased, the minimal gap occurs at a later time in the evolution and decreases in magnitude. The kinks that appear in both Fig.~\ref{fig:GapVsTime} and Fig.~\ref{fig:energy}(a) are a consequence of energy level crossings apparent in the evolution of the spectrum, as shown in Figs.~\ref{fig:energy}(b)-\ref{fig:energy}(d) for the same values of $\alpha$ as in Fig.~\ref{fig:energy}(a). There are energy eigenstates that become part of the $17$ degenerate ground states that ``cut'' through other energy eigenstates.

\section{Simulation details}
\label{app:sims}

\subsection{Simulated Annealing} 
\label{app:SA}
\begin{figure}[t]
\includegraphics[width=\columnwidth]{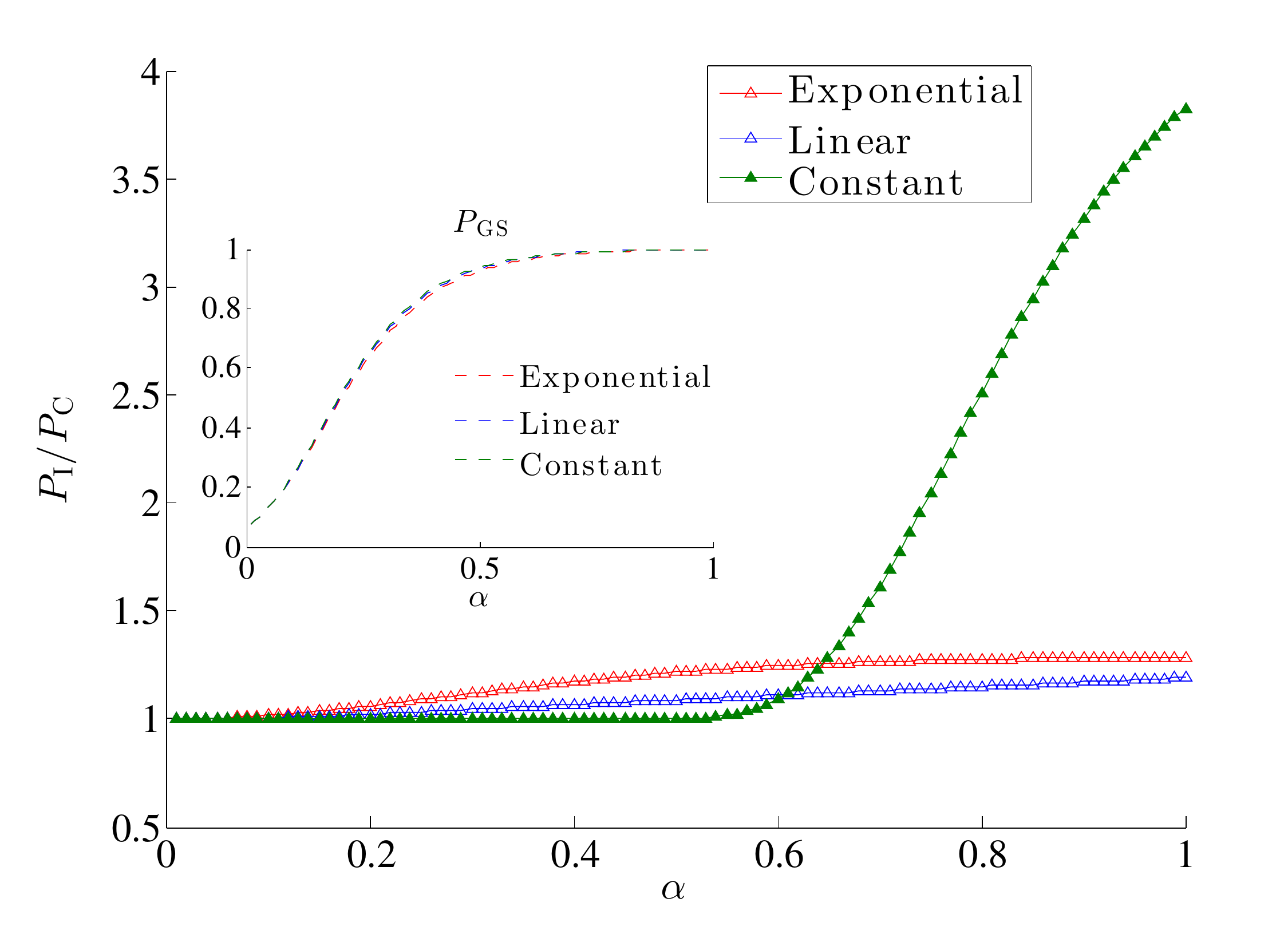}
\caption{Simulated annealing shows quantitatively similar behavior for various annealing schedule. The schedules are: exponential $T(k) = T(0)({T(K)/T(0)})^{k/K}$, linear $T(k) = T(0) + {k \over K}(T(K)-T(0))$, and constant $T(k)=T(K)$, with $K=1000$, with $k_B T(0) /\hbar = 8$ GHz and $k_B T(K) /\hbar = 0.5$ GHz.}
   \label{fig:SAAnnealing}
\end{figure}

We describe here our implementation of classical SA.  As the state of system at any given step is a classical probability distribution, we can represent it by a state vector $\vec{p}$, where the component $p_i$ of the vector denotes the probability of finding the system in the $i$th state with energy $E_i$.  We initialize in the maximally mixed state (infinite temperature distribution), i.e., $p_i = 1/2^N \ \forall i$. Note that the initial Gibbs state of the quantum annealer also has a uniform probability distribution over all computational states, so this choice for the classical initial state is well motivated.  The system then evolves via single spin flips.  The transition probability between two states with energy difference $\Delta E$ is given by $\frac{1}{N} \min(1,\exp(-\upbeta \Delta E))$, the Metropolis update rule \cite{1953JChPh..21.1087M}. 
The transition matrix has elements
\beq
\mathrm{T}(i \rightarrow j) = {\frac{1}{N}} \min(1,\exp(-\upbeta (E_j-E_i))) \ .
\label{eq:Markov-matrix}
\eeq
The system is evolved for $1000$ steps by acting with the transition matrix on the state vector $\vec{p}$.  At each step, the temperature is adjusted so as to reduce thermal excitations. If the temperature is reduced slowly enough and to low enough energies, SA can find an optimal solution. The choice of temperature schedule to follow for SA is often motivated by experimental circumstances, and in the main text we used $\upbeta^{-1} B(t)$ (as shown in Fig.~\ref{fig:AnnealingSchedule}) as the schedule with $\upbeta^{-1}/\hbar = 2.226$ GHz.  Here we tested three other different temperature schedules.  As shown in Fig.~\ref{fig:SAAnnealing}, we find that the qualitative features of the simulation results do not depend on a particular choice of temperature schedule.  While the numerical values of the ratio of the isolated state and cluster populations changes, the ratio is always greater than unity. The ground state population curves are indiscernible regardless of the temperature schedule used.

\subsection{Spin Dynamics}
\label{app:SD}
In the O(3) SD model, qubits are replaced by classical spins $\vec{M}_i = (\sin\theta_i\cos\phi_i,\sin\theta_i\sin\phi_i,\cos\theta_i)$.  This is a natural semi-classical model since it amounts to the saddle-point approximation of the path integral for the spin system (the derivation is presented in Appendix~\ref{app:derivation}) and can be interpreted as describing coherent single qubits interacting classically. This model is closely related to one that was proposed and analyzed by Smolin and Smith \cite{Smolin} in its planar, O(2) version, i.e., $\vec{M}_i=(\sin\theta_i,0,\cos\theta_i)$ (a spin in the $x-z$ plane).  While the SD model was  already shown to be inconsistent with the experimental data in the context of correlations with DW1 spin glass benchmarks in Refs.~\cite{q108,comment-SS}, the SD model was shown in Ref.~\cite{Smolin} to give the same suppression of the isolated state prediction as QA for $\alpha=1$, and hence the evidence in Ref.~\cite{q-sig} alone does not suffice to rule out the SD model as a classical description of the D-Wave device.
In this section we demonstrate that similarly to SA, the SD model is also inconsistent with the ME results we obtain when we tune the energy scale factor $\alpha$ of the quantum signature Hamiltonian. 


As shown in section~\ref{app:O3-open}, the SD model with thermal fluctuations is described by a (Markovian) spin-Langevin equation \cite{Jayannavar:1990,PTPS.46.210} with a Landau-Lifshitz friction term \cite{LL:1935,PTPS.46.210},
\beq
 \frac{d}{dt} \vec{M}_i  = -\left( \vec{H}_i + \vec{\xi}(t) + \chi  \vec{H}_i \times \vec{M}_i  \right) \times \vec{M}_i \ ,
 \label{eq:spin-Langevin}
\eeq
with the Gaussian noise $\vec{\xi} = \{\xi_i\}$ satisfying
\beq
\braket{\xi_i(t)} = 0 \ , \quad \braket{\xi_i(t) \xi_j(t')} = 2 k_B T \chi \delta_{ij} \delta(t- t') \ ,
\eeq
and 
\beq
\vec{H}_i = 2 A(t) \hat{x} +2 B(t) \left(h_i  + \sum_{j \neq i} J_{ij}  \vec{M}_j \cdot \hat{z}  \right) \hat{z}  \ ,
\eeq
where $\hat{x}$ and $\hat{z}$ are unit vectors. For the $n$th run out of a total of $N_r$ runs we obtain a set of angles $\{ \theta_j^{(n)} \}$, which are interpreted in terms of a state in the computational basis by defining the probability of the $\ket{0}$ state for the $j$th spin (out of $N$) as $\cos^2 (\theta_j^{(n)}/2)$.  Therefore, we define:
\bes
\begin{align}
P_{\mathrm{C}} =&  \frac{1}{16 N_r} \sum_{n=1}^{N_r} \prod_{j=N/2+1}^N \cos^2 \left(\theta_j^{(n)}/2 \right)\\
P_{\mathrm{I}} =&  \frac{1}{N_r} \sum_{n=1}^{N_r} \prod_{j=1}^N \sin^2 \left(\theta_j^{(n)}/2 \right)\, ,
\end{align}
\ees
where the product over the last $N/2$ spins in $P_{\mathrm{C}}$ is $1$ if and only if all the core spins are in the $\ket{0}$ state, i.e., a cluster state, and likewise the product over all $N$ spins in $P_{\mathrm{I}}$ is $1$ if and only if all the spins are in the $\ket{1}$ state, i.e., the isolated state.  To incorporate the $\alpha$-dependence we simply rescale $B(t)$ to $\alpha B(t)$.

\subsection{Master Equation}
\label{app:ME}
%
We used an adiabatic Markovian ME in order to simulate the DW2 as an open quantum system. Details of the derivation of the ME can be found in Ref.~\cite{ABLZ:12-SI}.  The derivation assumes a system-bath Hamiltonian of the form:
\beq
H = H_S(t) + H_B + g \sum_{\alpha} A_{\alpha} \otimes B_{\alpha} \ ,
\eeq
where $A_\alpha$ is a Hermitian system operator acting on the $\alpha$th qubit and $B_{\alpha}$ is a Hermitian bath operator.  We restrict ourselves to a model of independent baths of harmonic oscillators, i.e., each qubit experiences its own thermal bath, with a dephasing system-bath interaction,
\beq
A_{\alpha}  = \sigma_{\alpha}^z \, ; \quad B_\alpha =   \sum_k \left( b_{k,\alpha} + b_{k,\alpha}^{\dagger} \right) \ , 
\eeq
where $b_{k,\alpha}$ and $b_{k,\alpha}^{\dagger}$ are lowering and raising operators and $k$ is a mode index.  We use the double-sided adiabatic ME \emph{without} the rotating wave approximation \cite{ABLZ:12-SI}:
\begin{align}
\frac{d}{dt}{\rho }_{S}(t)&=-i\left[ H_{S}(t),\rho _{S}(t)\right]\notag \\
&+g^{2}\sum_{\alpha \beta }\sum_{ab}\Gamma _{\alpha \beta }(\omega _{ba}(t))%
\left[ L_{ab,\beta }(t)\rho _{S}(t),A_{\alpha }\right] \notag \\
&+\mathrm{h.c.}\ ,
\label{eqt:NRWA}
\end{align}
where $\omega_{ba} = \eps_b(t) - \eps_a(t)$ are differences of instantaneous energy eigenvalues given by $H_S(t) \ket{\eps_a(t)} = \eps_a(t) \ket{\eps_a(t)}$ and
\bes
\begin{align}
L_{ab,\alpha }(t) & =  \bra{\eps_a(t)}A_{\alpha} \ket{\eps_b(t)} \ket{\varepsilon_a(t)}\! \bra{\varepsilon_b(t)} = L^\dag_{ba,\alpha }(t)\ , \\
\Gamma_{\alpha \beta}(\omega) & = \int_{0}^{\infty} e^{i \omega t} \langle e^{-i H_B t} B_{\alpha} e^{i H_B t} B_{\beta} \rangle \, dt \notag \\
&=\frac{1}{2} \gamma(\omega) + i S(\omega) \ , \\
\gamma_{\alpha \beta}(\omega) & = \int_{-\infty}^{\infty}  \langle e^{-i H_B t} B_{\alpha} e^{i H_B t} B_{\beta} \rangle\, dt \ , \\
S_{\alpha \beta}(\omega) & = \int_{\infty}^{\infty} d \omega' \gamma_{\alpha \beta}(\omega') \mathcal{P} \left( \frac{1}{\omega - \omega'} \right)\, d\omega' \ , 
\end{align} 
\ees
where $\mathcal{P}$ denotes the Cauchy principal value.  Under the assumption of Ohmic independent baths, we have:
\beq
\gamma_{\alpha \beta}(\omega) = \delta_{\alpha \beta} \frac{ 2 \pi g^2 \eta \omega}{1 - e^{- \upbeta \omega}} e^{-|\omega|/\omega_c} \ , 
\label{eq:gamma-ab}
\eeq
where $\upbeta$ is the inverse temperature, $\eta$ is a parameter (with units of time squared) characterizing the bath, and $\omega_c$ is an ultraviolet cut-off, which we set to $8\pi \, \mathrm{GHz}$ to satisfy the assumptions made in deriving the ME \cite{ABLZ:12-SI}.  Note that the only remaining free dimensionless parameter is 
\beq
\kappa \equiv g^2 \eta / \hbar^2
\label{eq:kappa}
\eeq 
(we have reintroduced the factor of $\hbar$ here), which controls the effective system-bath coupling.
We choose to work with the ME in Eq.~\eqref{eqt:NRWA} instead of its counterpart in (completely positive) Lindblad form because it is numerically more efficient to calculate the evolution.  Although it does not guarantee positivity of the density matrix, we always make sure to work in a parameter regime where we do not observe any violations of positivity.  In the ME simulations presented in the text, we truncate the spectrum to the lowest 56 instantaneous energy eigenstates to keep computational costs within reason.  We have checked that increasing this number for the smallest $\alpha$ regime does not substantially change our conclusions.
%

\begin{figure*}[t]
\subfigure[]{\includegraphics[width=0.98\columnwidth]{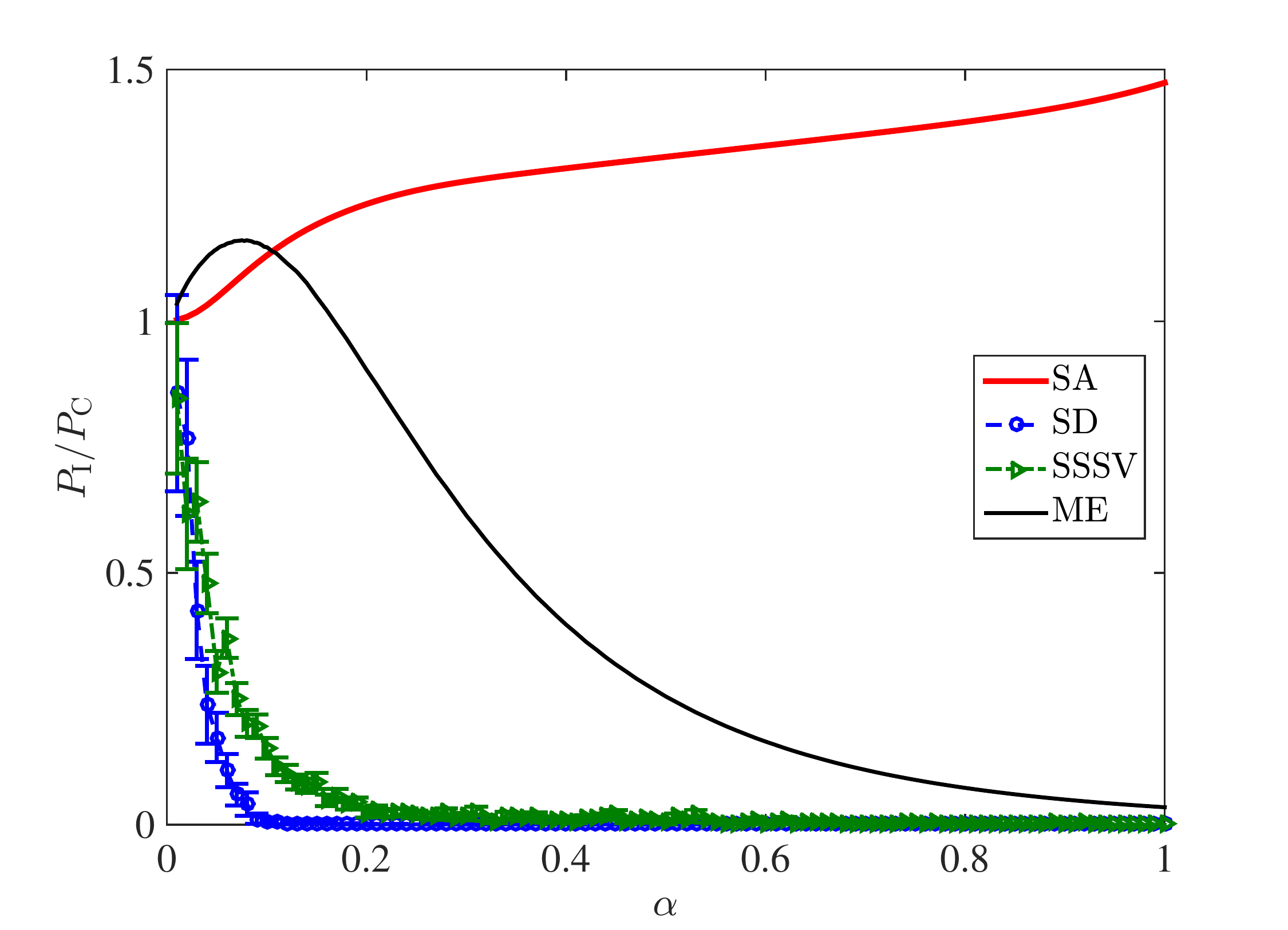} \label{fig:RATIOCOLL8EXTENDED}}
\subfigure[]{\includegraphics[width=0.98\columnwidth]{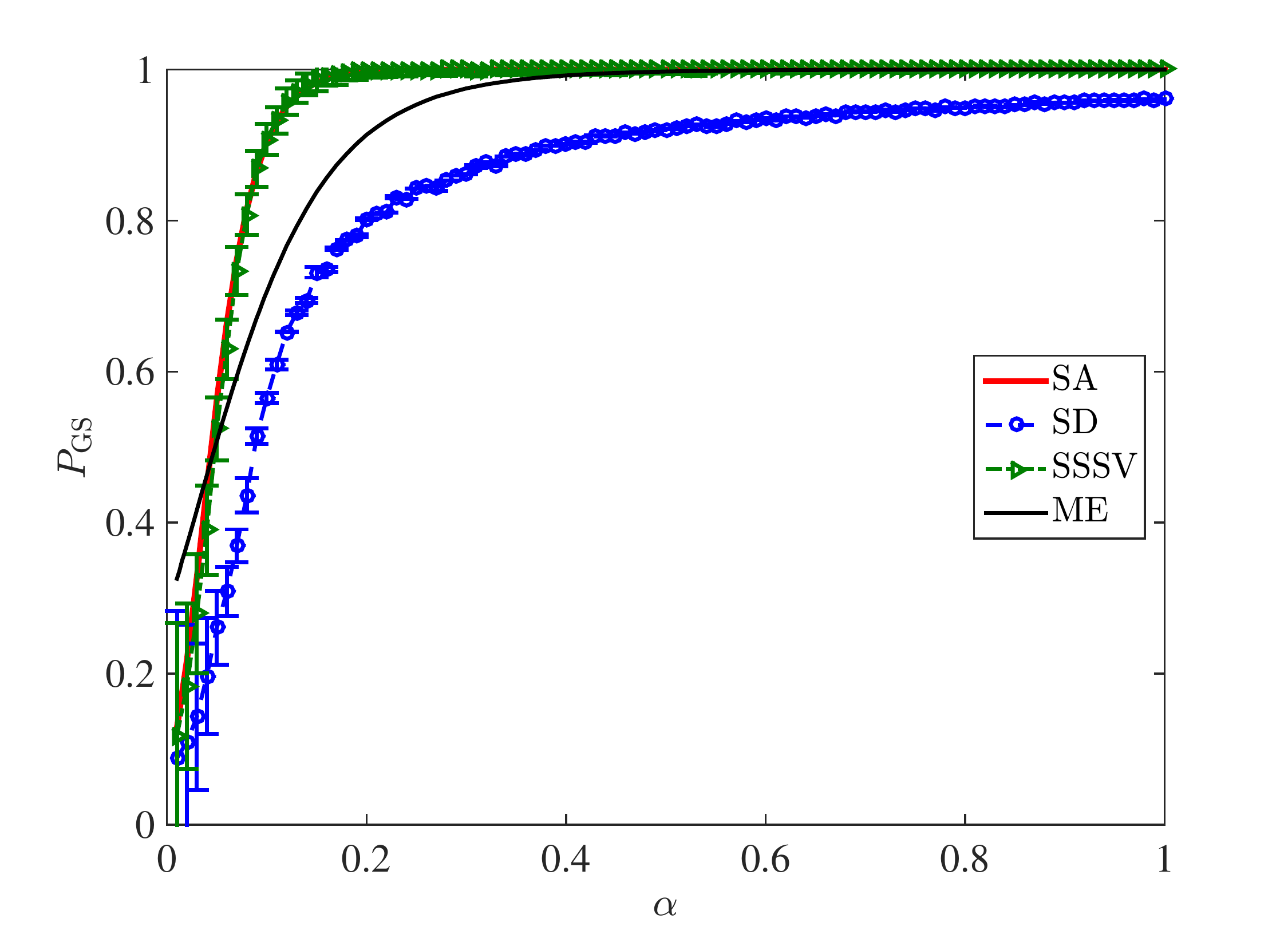}    \label{fig:PvsAlpha}}
\caption{Numerical results distinguishing the quantum ME and classical SA, SD, and SSSV models. (a) Results for the ratio of the isolated state population to the average population in the cluster-states ($P_\textrm{I}/P_\textrm{C}$), and (b) the  ground state probability ($P_\textrm{GS}$), as a function of the energy scale factor $\alpha$, at a fixed annealing time of $t_f=20 \,\mu s$. The error bars represent the 95\% confidence interval. Two striking features are the ``ground state population inversion" between the isolated state and the cluster (the ratio of their populations crosses unity), and the manifestly non-monotonic behavior of the population ratio, which displays a maximum.  At the specific value of the system-bath coupling used in our simulations ($\kappa = 1.27 \times 10^{-4}$), it is interesting that the ME underestimates the magnitude and position of the peak in $P_\textrm{I}/P_\textrm{C}$ but qualitatively matches the experimental results shown in Fig.~\ref{fig:DW2-Uncalibrated2}, capturing both the population inversion and the presence of a maximum even in the absence of noise.  In contrast to the ME results shown, the SA, SSSV, and SD results for the population ratio are not in qualitative agreement with the ME. Specifically, all three classical models miss the population inversion and maximum seen for the ME.  Simulation parameters can be found in Appendix~\ref{app:sims}.}
\label{fig:quantumtothermal}
\end{figure*}

\section{Comparing the models in the noiseless case}
\label{app:MoreNoiselessData}
In this section we present additional numerical findings for SA and SD in the absence of noise on the local fields and couplings. The SD model is explained below.  

Since we presented evidence in the main text that noise and cross-talk play an important role in the experimental DW2 results, the results presented in this section are limited to a comparison between the models, which behave quite differently in the ideal case. We may expect the ME results to match a future quantum annealer with better noise characteristics and no cross-talk. Therefore we present our findings by contrasting each of the classical models in turn with the ME simulations.

\subsection{SA} 
\label{app:SA2}

The main result showing the dependence of $P_\textrm{I} / P_\textrm{C}$ as a function of the energy scale $\alpha$ for $N=8$ qubits is summarized in Fig.~\ref{fig:quantumtothermal}.  We note first that the total ground state probability $P_{\textrm{GS}} = P_{\textrm{I}} + 16 P_{\textrm{C}}$ decreases monotonically as $\alpha$ is decreased. This reflects an increase in thermal excitations, whereby the ground state population is lost to excited states, and confirms that $\alpha$ acts as an effective inverse temperature knob. 

However, in contrast to SA, the ME result for $P_\textrm{I} / P_\textrm{C}$ is non-monotonic in $\alpha$; see Fig.~\ref{fig:RATIOCOLL8EXTENDED}. Initially, as $\alpha$ is decreased from its largest value of $1$, the ratio $P_\textrm{I} / P_\textrm{C}$ increases and eventually becomes larger than 1; i.e., the population of the isolated state becomes enhanced rather than suppressed.  For sufficiently small $\alpha$, the ME $P_\textrm{I} / P_\textrm{C}$ ratio turns around and decreases towards $1$. The SA results also converge to $1$ as $\alpha \rightarrow 0$ but do not display a maximum.%
%
%

A close examination of Fig.~\ref{fig:RATIOCOLL8EXTENDED} shows that even in the ``relatively classical" small $\alpha$ region ($\alpha \lesssim 0.1$) the curvature of $P_{\mathrm{I}} / P_{\mathrm{C}}$ for the ME results [$d^2(P_{\mathrm{I}} / P_{\mathrm{C}})/d\alpha^2 <0$] is inconsistent with the curvature of the SA result [$d^2(P_{\mathrm{I}} / P_{\mathrm{C}})/d\alpha^2 >0$], as seen in Fig.~\ref{fig:PvsAlpha}. We can show that the positive curvature of SA is a general result as long as the initial population is uniform.
To see this, we expand the SA Markov-chain transition matrix [Eq.~\eqref{eq:Markov-matrix}] in powers of $\alpha$:
\beq
\mathrm{T}(\alpha) = \mathrm{T}_0 + \alpha \mathrm{T}_1 + \dots \, ,
\eeq
where the SA state vector $\vec{p}(K)$ at the $K$th time-step is given by $\vec{p}(K)^T = \vec{p}(0)^T \mathrm{T}(\alpha)^K$, where the $T$ superscript denotes the transpose.

At $\alpha = 0$, all transitions are equally likely, so $(\mathrm{T}_0)_{ij} = 1 / N$.  The first order term satisfies:
\bes
\begin{align}
\left(\mathrm{T}_1 \right)_{i \to j} &= \min \left( 0 , - \beta \left( E_{j} - E_i \right) \right) \ , \ i \neq j \ , \\
 \left(\mathrm{T}_1 \right)_{i \to i } &= - \sum_{j}  \min \left( 0 , - \beta \left( E_{j} - E_i \right) \right) \ .
 \end{align}
\ees
The first order term has the property that $\sum_j \left(\mathrm{T}_1 \right)_{ij} = 0$. Therefore, applying the transition matrix $K$ times, we have to first order in $\alpha$:
\beq
\vec{p}(K)^T = \vec{p}(0)^T (\mathrm{T}_0)^K  + \alpha \vec{p}(0)^T\sum_{i=0}^{K-1} (\mathrm{T}_0)^{K-1-i} \mathrm{T}_1  (\mathrm{T}_0)^i  + \dots 
\eeq
Using the fact that we start from the uniform state $p_i(0) = 1/N$, we have $ \vec{p}(0)^T \mathrm{T}_1 = 0$, but also that
\beq
\left(\mathrm{T}_1 \mathrm{T}_0 \right)_{ij}= \frac{1}{N} \sum_{k} \left(\mathrm{T}_1\right)_{ik } =0 \ .
\eeq 
Therefore,
\beq
\vec{p}(K) = \vec{p}(0) + O(\alpha^2) \ .
\eeq
This in turn implies that for SA, 
\beq
\frac{P_{\mathrm{I}}}{P_{\mathrm{C}}} = 1+ \alpha^2 f + O(\alpha^3) \ .
\eeq
Since we showed in Appendix~\ref{app:B1} that this quantity is greater than or equal to 1 for SA, this implies that $f \geq 0$, and hence the curvature $d^2(P_{\mathrm{I}} / P_{\mathrm{C}})/d\alpha^2$ at $\alpha=0$ is positive.

However, we emphasize that this argument for the positivity of the initial curvature of ${P_{\mathrm{I}}}/{P_{\mathrm{C}}}$ requires that the initial state be uniform.  If a non-uniform initial state is chosen, there will be a non-zero linear term in $\alpha$ for $P_{\mathrm{I}}/P_{\mathrm{C}}$, which prevents us from concluding anything about the curvature.

\subsection{Results for the SD model}
\label{sec:SD-no}

Using the DW2 operating temperature and annealing schedules for $A(t)$ and $B(t)$, we find that the SD model does not match the ME data. This can be seen in Fig.~\ref{fig:PvsAlpha}, where the SD population ratio (the dashed blue line) fails to reproduce the qualitative features of the ME result, in particular the ground state population inversion peak. Another illustration of the same failure of the SD model is given in Fig.~\ref{fig:O3BoxPlot}(a), which shows the distribution of $M^z$ for the core and outer qubits for different values of $\alpha$. We expect the core spins to align in the $\ket{1111}$ state for sufficiently small $\alpha$ (i.e., to each have $M^z = -1$), when the isolated state becomes enhanced. However, as can be seen from Fig.~\ref{fig:O3BoxPlot}(a) the median of the core spins is in fact never close to $M^z = -1$ for small $\alpha$ values, so that the enhancement of the isolated state is missed by the SD model.  Furthermore, the model shows a preference for a particular cluster-state, the one with all of the outer spins in the $\ket{1111}$ state ($M^z=-1$).  In the inset of Fig.~\ref{fig:O3BoxPlot}(a), this can be seen in that the median of the data occurs always below $M^z = 0$. The explanation is provided in Appendix~\ref{app:O3-open}.

\subsection{Results for the SSSV model}

\begin{figure*}[t] %
\begin{center}
\subfigure[\ SD model]{\includegraphics[width=0.48\textwidth]{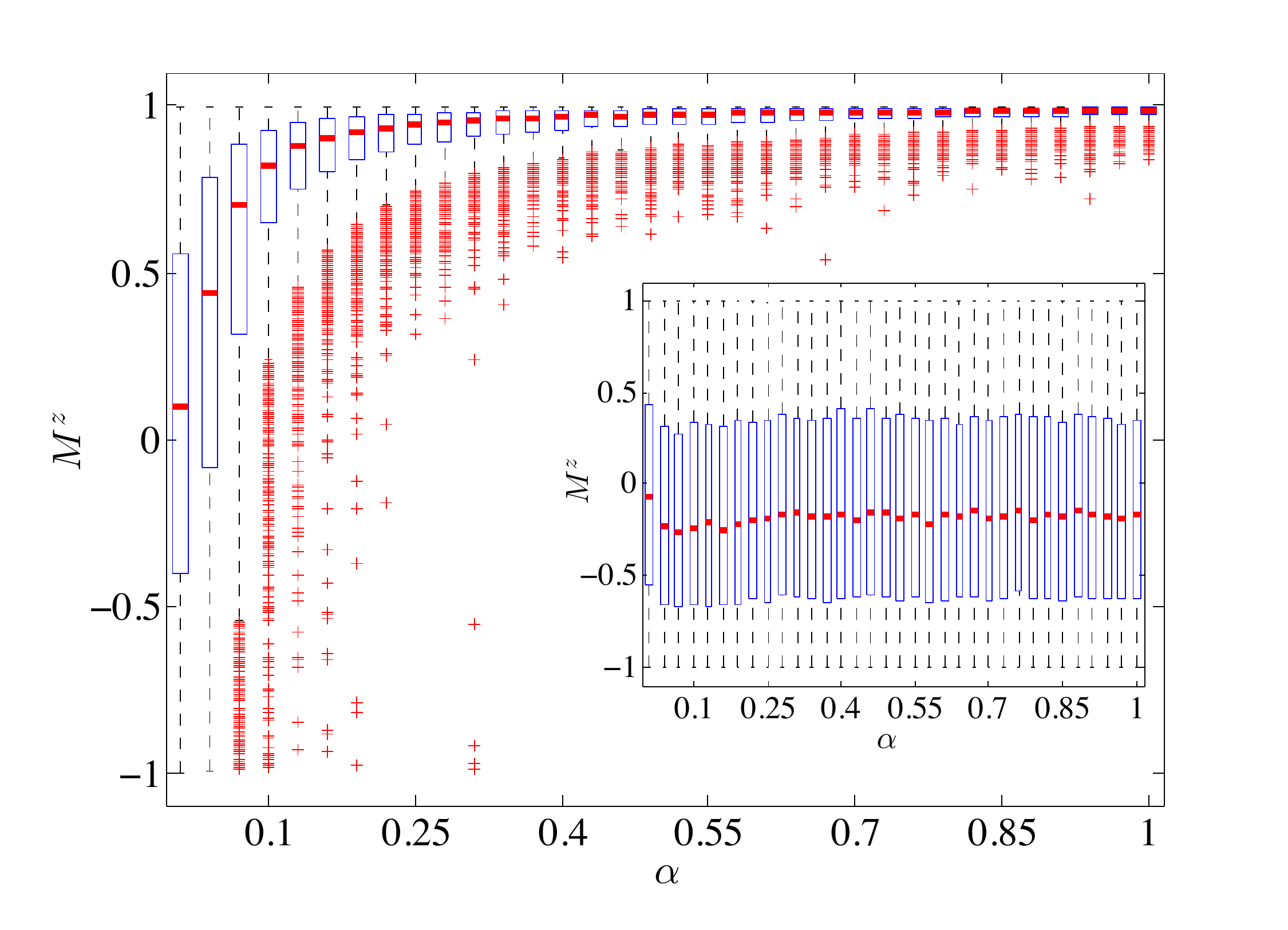}\label{fig:SD-box}}
\subfigure[\ SSSV model]{\includegraphics[width=0.48\textwidth]{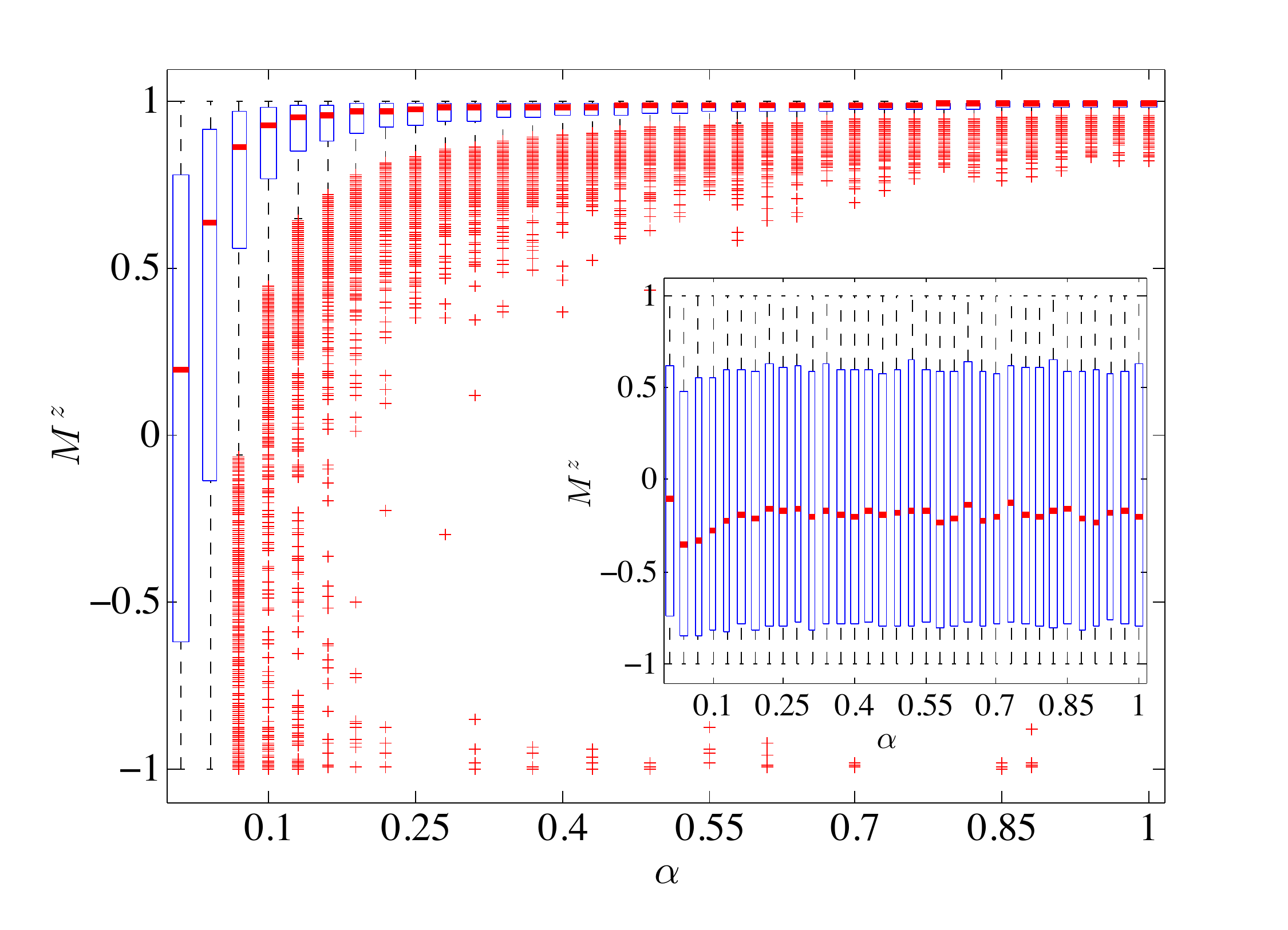}\label{fig:SSSV-box}}
\caption{Statistical box plot of the average $z$ component for all core qubits (main plot) and all outer qubits (inset) at $t= t_f = 20 \mu s$. 
(a) The SD model. The data are taken for $1000$ runs with Langevin parameters $k_B T / \hbar = 2.226 \, \mathrm{GHz}$ (i.e., $17$mK, to match the operating temperature of the DW2) and $\zeta= 10^{-3}$.   In Appendix~ \ref{app:O3-open} we show that the results do not depend strongly on the choice of $\zeta$. (b) The SSSV model. The data are taken for $1000$ runs with parameters $k_B T / \hbar = 1.382\, \mathrm{GHz}$ (i.e., $T=10.56$mK, as in Ref.~\cite{SSSV}) and $5\times10^5$ sweeps.}
\label{fig:O3BoxPlot}
\end{center}
\end{figure*}

To test whether this model matches the results of our quantum signature Hamiltonian we use similar parameters as given in Ref.~\cite{SSSV}, apart from the annealing schedule, for which we used that of the DW2. Reference~\cite{SSSV} found the best agreement with the DW1 data from Ref.~\cite{q108} for a temperature of $10.6$mK, lower than the $17$mK operating temperature of the DW1, and for a total of $1.5 \times 10^5$ Monte Carlo update steps per spin (sweeps). We found negligible differences when we used the operating temperature  of the DW2 ($17$mK) for the SSSV model, or when we varied the number of sweeps. As can be seen in Fig.~\ref{fig:quantumtothermal}, the SSSV model does not reproduce the ground state population inversion and maximum seen in the experimental data.  In fact the SSSV results are quantitatively similar to the SD model, even in showing a preference for a specific cluster-state, as shown in Fig.~\ref{fig:O3BoxPlot}(b).  Furthermore, the SSSV model does not reproduce the ground state population inversion even after the number of qubits is increased to $40$ [see Fig.~\ref{fig:N=40}], which is particularly significant as it shows that the essential quantum features that result in the disagreement are retained beyond the initial ``small" $N=8$ problem size.  

\begin{figure}[t] 
\includegraphics[width=1.1\columnwidth]{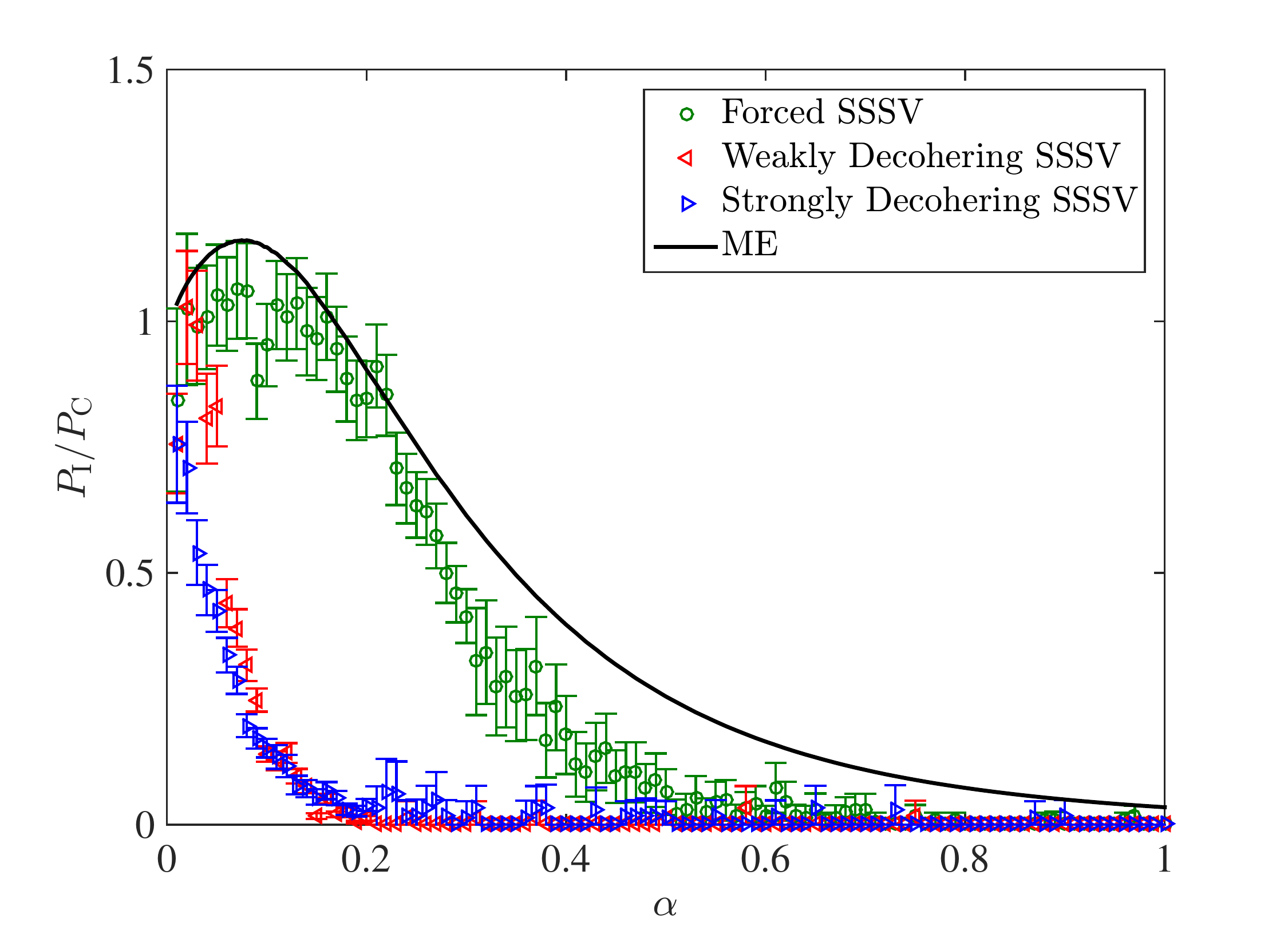} 
\caption{``Forced" and strongly or weakly ``decohering" SSSV models. Shown are the  results for the ratio of the isolated state population to the average population in the cluster-states ($P_\textrm{I}/P_\textrm{C}$) as a function of the energy scale factor $\alpha$, for $N=8$ and at a fixed annealing time of $t_f=20 \,\mu s$.  The error bars represent the 95\% confidence interval.  For reference the plot also includes the curves for the ME from Fig.~\ref{fig:RATIOCOLL8EXTENDED}.  Additional parameters for the modified SSSV models: $g^2 \eta = 10^{-6}$ for the strongly decohered model and $g^2 \eta = 2.5 \times 10^{-7}$ for the weakly decohered and forced models. }
\label{fig:mod-SSSV}
\end{figure}

\section{ME \textit{vs} Modified SSSV models with ``decoherence" from O(2) rotors to Ising spins}
\label{sec:dec-SSSV}

In this section we consider variants of the SSSV model where the O(2) rotors are first mapped to qubits and then allowed to decohere. The rationale is that the SSSV ``qubits" may be too coherent, and we wish to account for single-qubit decoherence effects.

\subsection{Strongly decohering SSSV model}
\label{sec:strong-dec-SSSV}
Because of the large deviation of SSSV from the ME at small $\alpha$ observed in Fig.~\ref{fig:RATIOCOLL8EXTENDED}, we propose to modify the model to fix this.  In order to raise the $P_{\mathrm{I}}/P_{\mathrm{C}}$ value, we note that SA, which uses effectively incoherent qubits, has $P_{\mathrm{I}}/P_{\mathrm{C}} \geq 1$.  Therefore, we might consider the scenario where the qubits become more incoherent as $\alpha$ becomes smaller, until in the limit of vanishing $\alpha$ they fully decohere and become Ising spins in the computational basis, we might be able to reproduce similar behavior.  To model this we replace the $x$-component of the magnetization vector of each spin by $M^x_i = e^{-t/\tau_\alpha} \sin\theta_i$ and leave the $z$-component unchanged, i.e., $M^z_i = \cos\theta_i$. This is equivalent to a model of single-qubit dephasing in the computational basis, via the mapping to the density matrix $\rho_i = \frac{1}{2}I+\vec{M}_i\cdot\vec{\sigma}$, where $\vec{M}_i = (M^x_i,0,M^z_i)$ and $\vec{\sigma} = (\sigma^x_i,\sigma^y_i,\sigma^x_i)$. This can be visualized as a gradual squashing of the Bloch sphere (restricted to the $x-z$ plane) into an ellipsoid (ellipse) with major axis in the $z$-direction and a shrinking minor ($x$-)axis. It is also equivalent to replacing the transverse field amplitude $A(t)$ in Eq.~\eqref{eqt:SA-SD} by $A(t)e^{-t/\tau_\alpha}$ while leaving the magnetization unchanged, i.e., decreasing the time-scale over which the transverse field plays a role.

Next we ensure that $\tau_\alpha$ is monotonically increasing with $\alpha$. In this manner, for $t\ll t_f$ the range is almost that of the fully ``coherent" SSSV, while for $t\lesssim t_f$ the range is restricted to that of the ``incoherent" SA. The ``decoherence" time $\tau_\alpha$ dictates how quickly this transition from one extreme to the other occurs, and to incorporate its $\alpha$-dependence we set $\tau_\alpha = 1 / 2 \gamma_\alpha(0)$, where  $\gamma_\alpha(0) =2\pi g^2 \eta(\alpha\upbeta)^{-1}$ is the dephasing rate used in our ME calculations [the general expression for $\gamma(\omega)$ is given in Eq.~\eqref{eq:gamma-ab}], with a rescaled inverse temperature, i.e., $\alpha \upbeta$ instead of $\upbeta$, to capture the idea that $\alpha$ acts to rescale the energy, or equivalently the inverse temperature. Thus in this model $\tau_\alpha =\alpha \frac{\upbeta}{4\pi g^2 \eta}$. Note that we only replaced $\upbeta$ with $\alpha \upbeta$ here and not anywhere else in the simulations, so the physical temperature is still given by $\upbeta^{-1}$.

Figure~\ref{fig:mod-SSSV} presents the results of this ``strongly decohering SSSV" model. The results are similar to the original SSSV model. Thus, ``decoherence" of the coherent O(2) spins fails to improve the agreement with the ME results.

\subsection{Weakly decohering SSSV model}
We can consider a weaker version of this dephasing model, which attempts to mimic dephasing in the energy eigenbasis of the ME model.  When the transverse field Hamiltonian dominates over the Ising Hamiltonian, the dephasing occurs in the $z$-component of the magnetization, and when the Ising Hamiltonian dominates over the transverse field Hamiltonians, the dephasing occurs in the $x$-component of the magnetization.  Explicitly, this translates to replacing the magnetization components of the spin by
\beq
M_i^x = \sin \theta_i \ , \ M_i^z = e^{-t/\tau_\alpha} \cos \theta_i \ , 
\eeq
if $A(t) \geq \alpha B(t)$ and by 
\beq
M_i^x = e^{-(t-t_c)/\tau_\alpha} \sin \theta_i \ , \ M_i^z = e^{-t_c/\tau_\alpha} \cos \theta_i \ , 
\eeq
if $ A(t) < \alpha B(t)$, where $t_c$ is the transition time satisfying $A(t_c) = \alpha B(t_c)$.  As can be seen in Fig.~\ref{fig:mod-SSSV}, this model also fails to capture the ME results.

\subsection{A modified SSSV model with a forced transition from O(2) rotors to Ising spins}
\label{sec:forced-SSSV}

To try to get better agreement of a classical model with the ME we finally consider a somewhat contrived model which simply forces a transition to SA with Ising spins. To implement this, instead of uniformly drawing $\theta_i\in [0,\pi]$ as in the SSSV model, we draw $\theta_i\in [0,\frac{\pi}{2} e^{-t/\tau_\alpha}]\cup [\pi - \frac{\pi}{2} e^{-t/\tau_\alpha},\pi]$, where $\tau_\alpha$ is selected just as in the decohered SSSV model (Appendix~\ref{sec:dec-SSSV}). This can be visualized as a restriction of the range of angles to gradually shrinking top and bottom parts of the Bloch sphere (again restricted to the $x-z$ plane). We call this a ``forced SSSV" model since it does not originate from a natural model of decoherence. 

Figure~\ref{fig:mod-SSSV} also presents the results of this forced SSSV model. In contrast to the original SSSV model result [Fig.~\ref{fig:quantumtothermal}(b)], the population ratio now rises to $1$ for $\alpha > 0$.  In this regard the forced SSSV model qualitatively captures the tendency toward ground state population inversion. However, it does not exhibit a pronounced ground state population inversion, and this appears to be a robust feature that is shared by other forced SSSV models we have tried (with different ``forcing" rules).
Furthermore, it exhibits a noticeable drop in $P_\textrm{I}/P_\textrm{C}$ at $\alpha\approx 0.1$, and the fraction of ground state population is almost one in the ground state population inversion regime, in contrast to the ME.  In this sense even the forced SSSV model does not agree with the ME data, and further evidence to this effect is presented in the next subsection. 

\subsection{Distance from the Gibbs state}
\label{sec:equil}

\begin{figure}[t] 
\includegraphics[width=0.95\columnwidth]{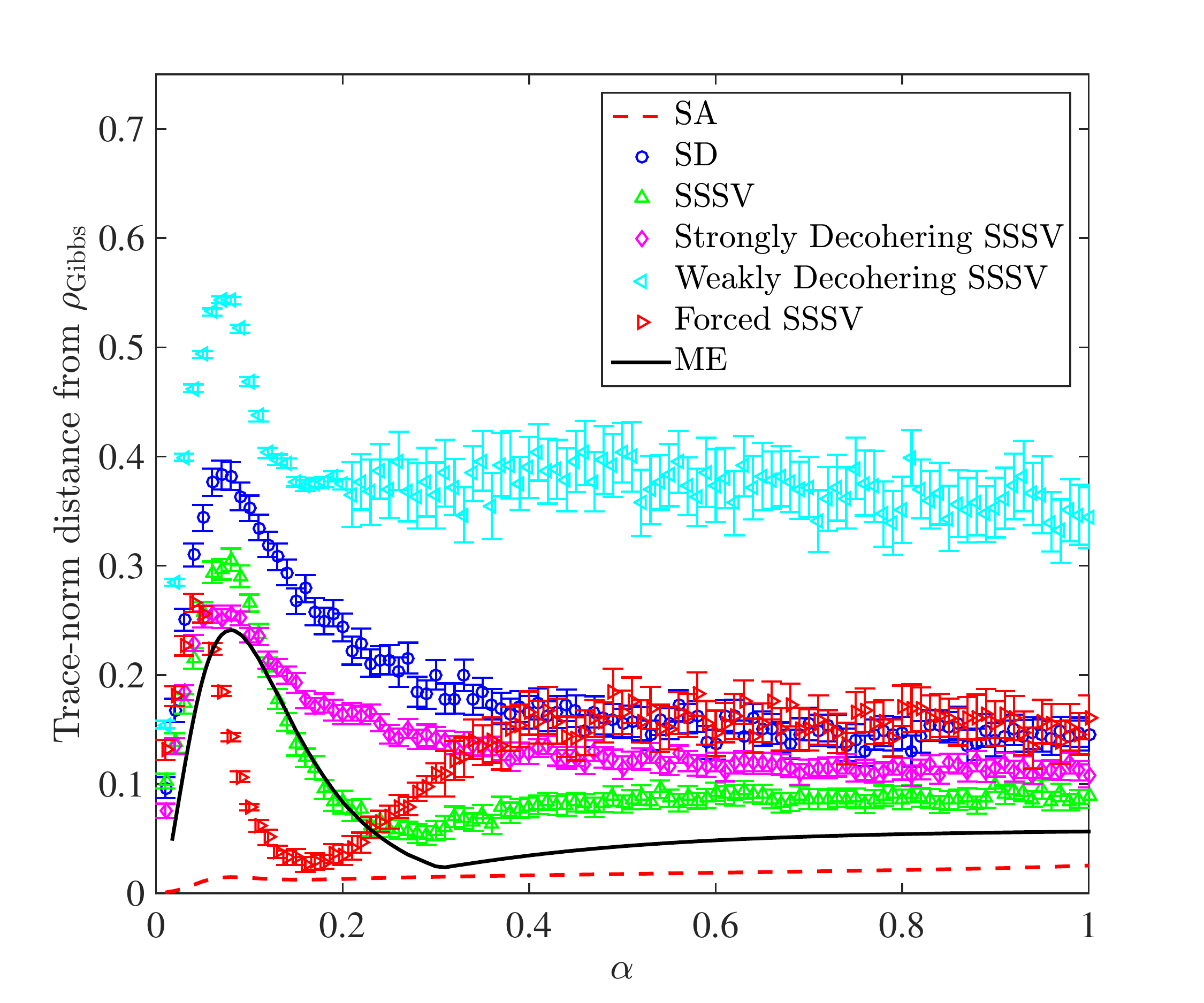} 
\caption{Trace-norm distance of the ME, SA, SD, SSSV, weakly and strongly decohered SSSV, and forced SSSV states from the $T = 17$mK Gibbs state at $t_f=20 \,\mu s$ and $N=8$. The error bars represent the 95\% confidence interval.  Three regions are clearly distinguishable for the ME: (1) $1 \geq \alpha \gtrsim 0.3$, where $\mathcal{D}$ is decreasing as $\alpha$ decreases; (2) $0.3 \gtrsim \alpha \gtrsim 0.1$, where $\mathcal{D}$ is increasing as $\alpha$ decreases; (3) $0.1 \gtrsim \alpha \geq 0$, where $\mathcal{D}$ is again decreasing as $\alpha$ decreases. Both SA and SD lack the minimum at $\alpha\approx 0.3$. }
\label{fig:TraceDistance}
\end{figure}

How well does the system thermally equilibrate? In this section we consider how distinguishable the final density matrix $\rho(t_f)$ is from the thermal Gibbs state at $t_f$, using the standard trace-norm distance measure \cite{nielsen2000quantum}
\beq
\mathcal{D} \left(\rho(t_f),\rho_{\mathrm{Gibbs}} \right) = \frac{1}{2} \| \rho(t_f) - \rho_{\mathrm{Gibbs}} \|_1 \ ,
\eeq
where $\rho_{\mathrm{Gibbs}} = e^{- \upbeta H(t_f)}/\mathcal{Z}$ with $\mathcal{Z} = \textrm{Tr}e^{- \upbeta H(t_f)}$ the partition function, and $\|A\|_1 \equiv \Tr\sqrt{A^\dagger A}$ (the sum of the singular values of the operator $A$). Note the fact that in the Gibbs state all ground states are equiprobable, so that $P_\textrm{I}/P_\textrm{C}=1$. This simple observation helps to explain many of the experimental results. 

The trace-norm distance result is shown in Fig.~\ref{fig:TraceDistance} for the ME, and the six classical models.  Although most of the models exhibit a peak in the trace-norm distance like the ME, none of the six classical models exhibits a minimum like the ME does at the corresponding value of $\alpha$, thus confirming once more that there is a strong mismatch between these classical models and the ME. This is particularly noticeable for the ``forced" SSSV model, which as discussed above exhibited the best agreement with the \emph{ground state} features among the classical models (Fig.~\ref{fig:mod-SSSV}), but poorly matches the excited state spectrum at low $\alpha$, as can be inferred from Fig.~\ref{fig:TraceDistance}. Indeed, this model is designed to transition to SA at low $\alpha$, and it does so at $\alpha\approx 0.1$. It then deviates from SA at even lower $\alpha$ values, presumably since there is no transverse field at all in SA, but the transverse field remains active in the ``forced" SSSV model at any $\alpha>0$.  Furthermore, we observe that the ``weakly decohering SSSV" model has a higher trace-norm distance than all other models even at high $\alpha$, and this is due to its strong preference for a particular cluster state, which is a failure mode of the SD and SSSV models that was discussed earlier (see  Fig.~\ref{fig:O3BoxPlot}).

Let us now focus on the ME results and explain the three regions seen in Fig.~\ref{fig:TraceDistance}.

\emph{Large $\alpha$, region (1)}.  As $\alpha$ decreases from $1$ to $\approx 0.3$, 
since $\alpha$ is relatively large, thermal excitations are not strong enough to populate energy eigenstates beyond the lowest $17$ that eventually become the degenerate ground state.  Therefore, the system is effectively always confined to the subspace that becomes the final ground state, as can also be seen from the $P_{\textrm{GS}}$ data in Fig.~\ref{fig:RATIOCOLL8EXTENDED}. However, recalling that the isolated state has overlap with excited states higher in energy than the cluster-states for $t<t_f$ (see Appendix~\ref{app:pert-theory}), thermal excitations populate the isolated state. Thus as $\alpha$ decreases,  $P_{\mathrm{I}} / P_{\mathrm{C}}$ approaches $1$, which is also the ratio satisfied by the Gibbs state, and hence $\mc{D}$ decreases as observed. At the same time, Fig.~\ref{fig:PvsAlpha} shows that at $\alpha = 0.3$ both the SD and SSSV models have $P_\textrm{I}/P_\textrm{C} \approx 0$, i.e., these models fail to populate the isolated state.  This therefore suggests that the quantum spectrum makes it easier for the system to thermally hop from one eigenstate to another.  

\emph{Intermediate $\alpha$, region (2)}.
Fig.~\ref{fig:RATIOCOLL8EXTENDED} shows that $P_{\mathrm{GS}}$ begins to decrease from $1$ at $\alpha \approx 0.3$, meaning that thermal excitations are now strong enough to populate energy eigenstates beyond the lowest $17$.  A loss in ground state population to excited states results, and the growth of $P_{\mathrm{I}} / P_{\mathrm{C}}$ beyond $1$ seen in Fig.~\ref{fig:RATIOCOLL8EXTENDED} results in the increase of $\mc{D}$ observed in Fig.~\ref{fig:TraceDistance}.   At $\alpha \approx 0.1$, the maximum distance from the Gibbs state is reached.  Beyond this value of $\alpha$, the energy scale of the Ising Hamiltonian is always below the temperature energy scale, as shown in Fig.~\ref{fig:AnnealingSchedule}.

\emph{Small $\alpha$, region (3)}. As $\alpha \to 0$ there is only a transverse field left, which is gradually turned off. Thus the system approaches the maximally mixed state (which is the associated Gibbs state). In light of this, for $0.1 \gtrsim \alpha \geq 0$ the energy gaps are sufficiently small that there is a large loss of population from the ground state; thermal excitations become increasingly more effective at equilibrating the system, thus pushing it towards the Gibbs state.

\section{An alternative model for breaking the symmetry of the cluster states} \label{app:hvsJ}
\subsection{The effect of an $h$ vs $J$ offset}
\label{sec:control-errs}
%

%
\begin{figure}[t]
\includegraphics[width=3.2in]{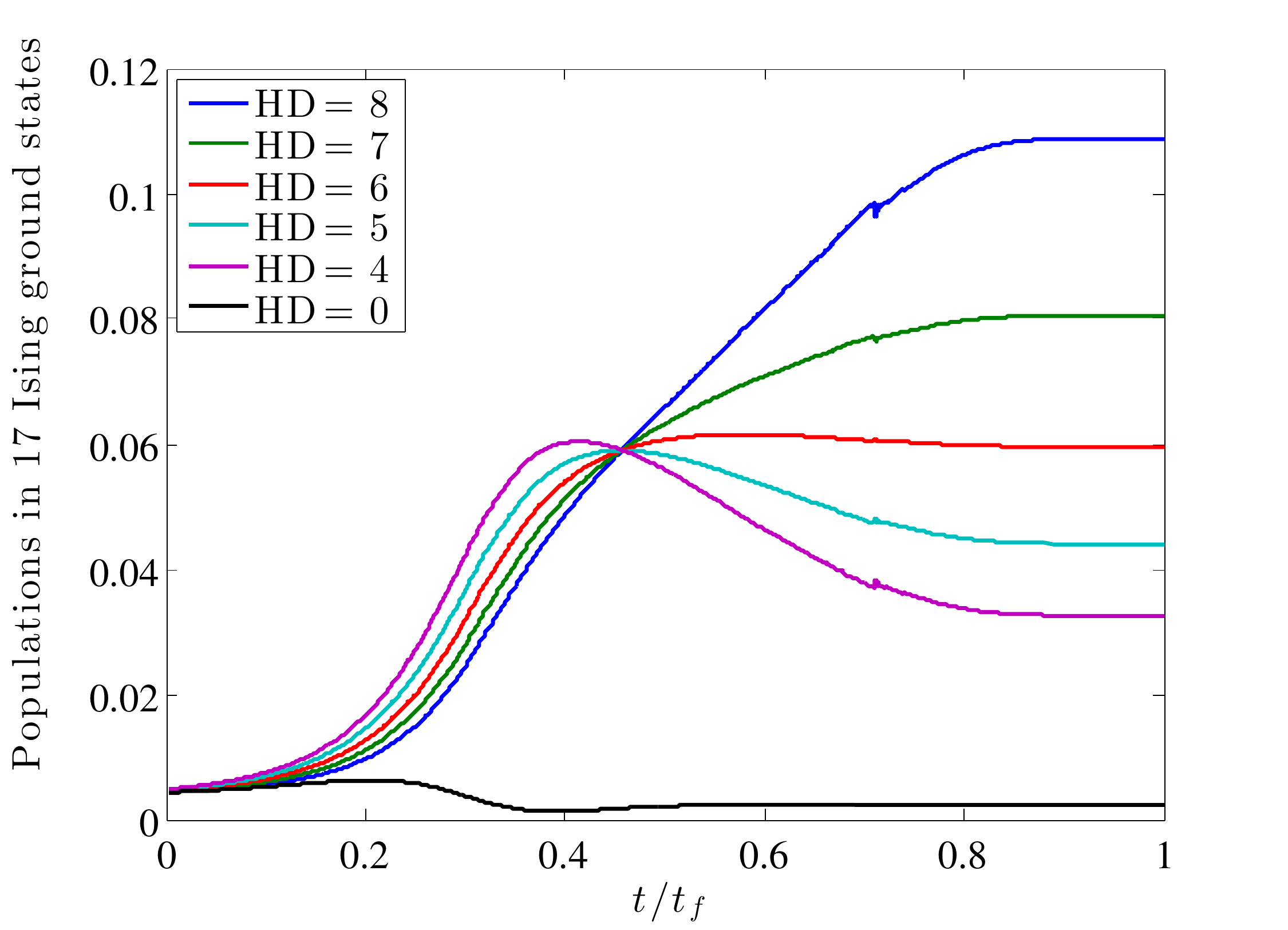}
\caption{Master equation results for the populations of the $17$ Ising ground states, with $\alpha=1$, $|h| = 0.981 |J|$, $t_f = 20\, \mu s$, and $\kappa = 1.27 \times 10^{-4}$. The cluster-states split by Hamming distance from the isolated state (bottom curve), in agreement with the experimental results shown in Fig.~\ref{fig:NotOpt-Opt}(a).}
\label{fig:ControlErrorEvolution}
\end{figure}
\begin{figure*}[t]
\begin{center}
\subfigure[]{\includegraphics[width=.48\textwidth]{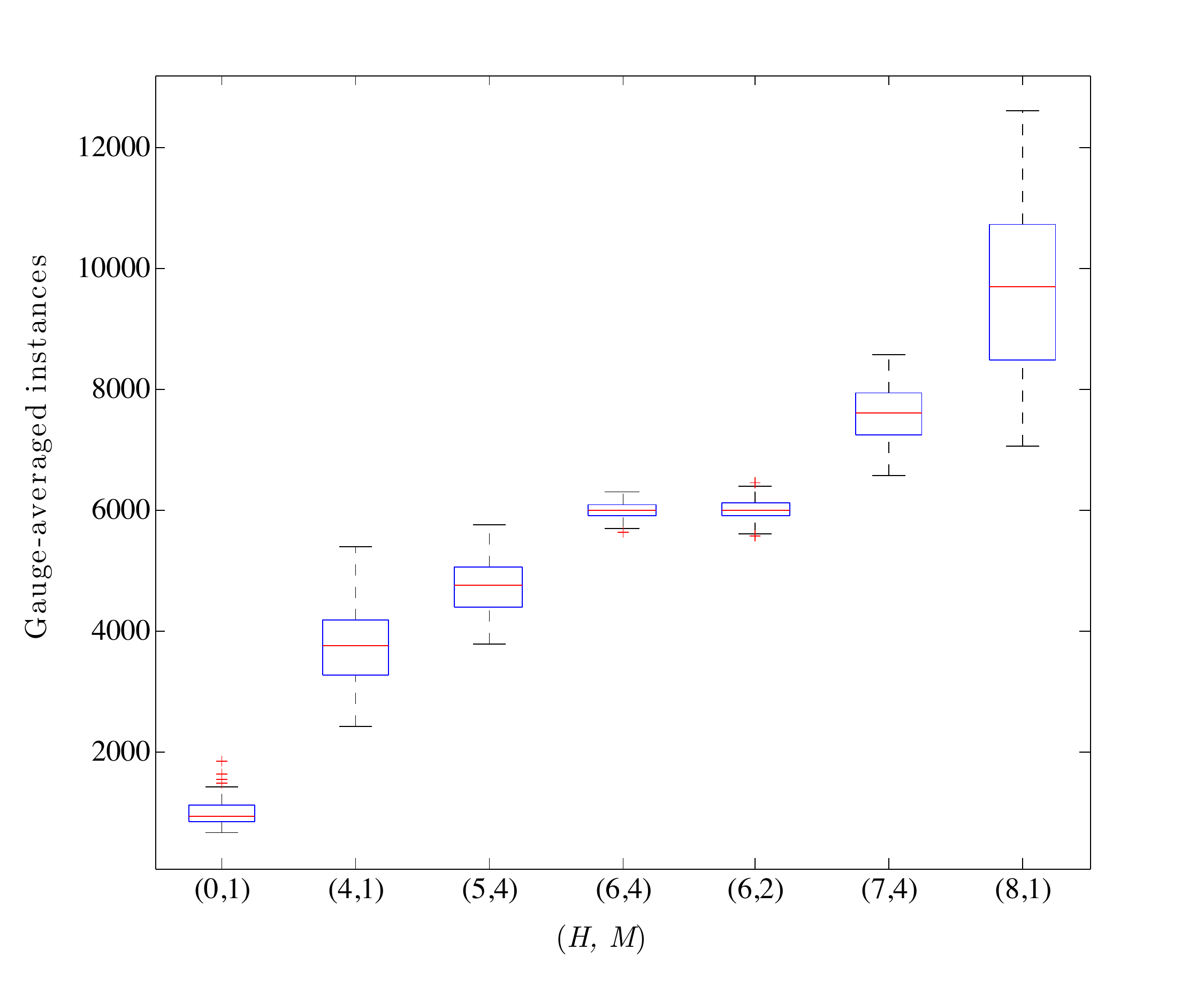}\label{fig:WalterNotOpt}}
\subfigure[]{\includegraphics[width=.48\textwidth]{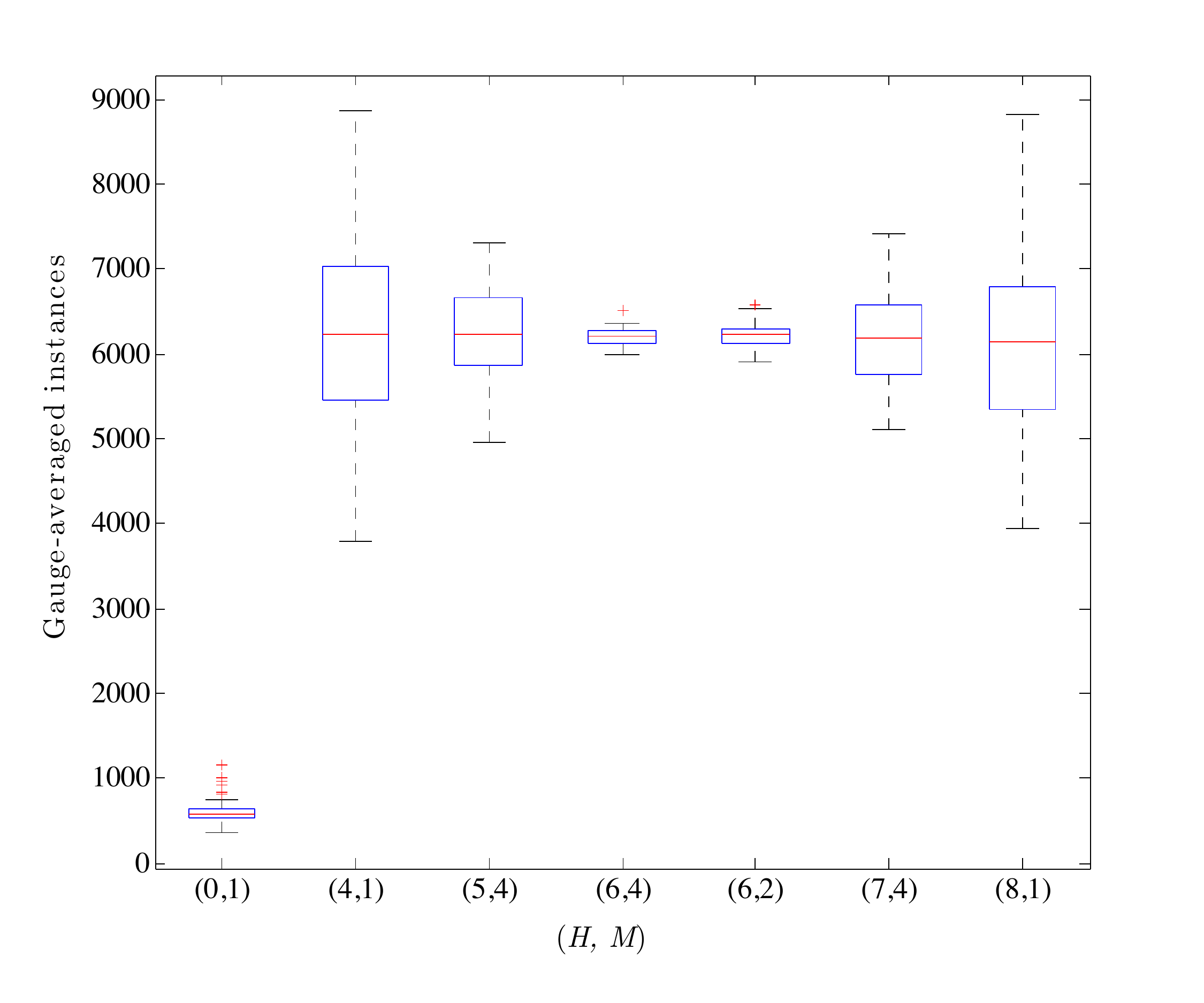}\label{fig:AnuragtOpt}}
\subfigure[]{ \includegraphics[width=.48\textwidth]{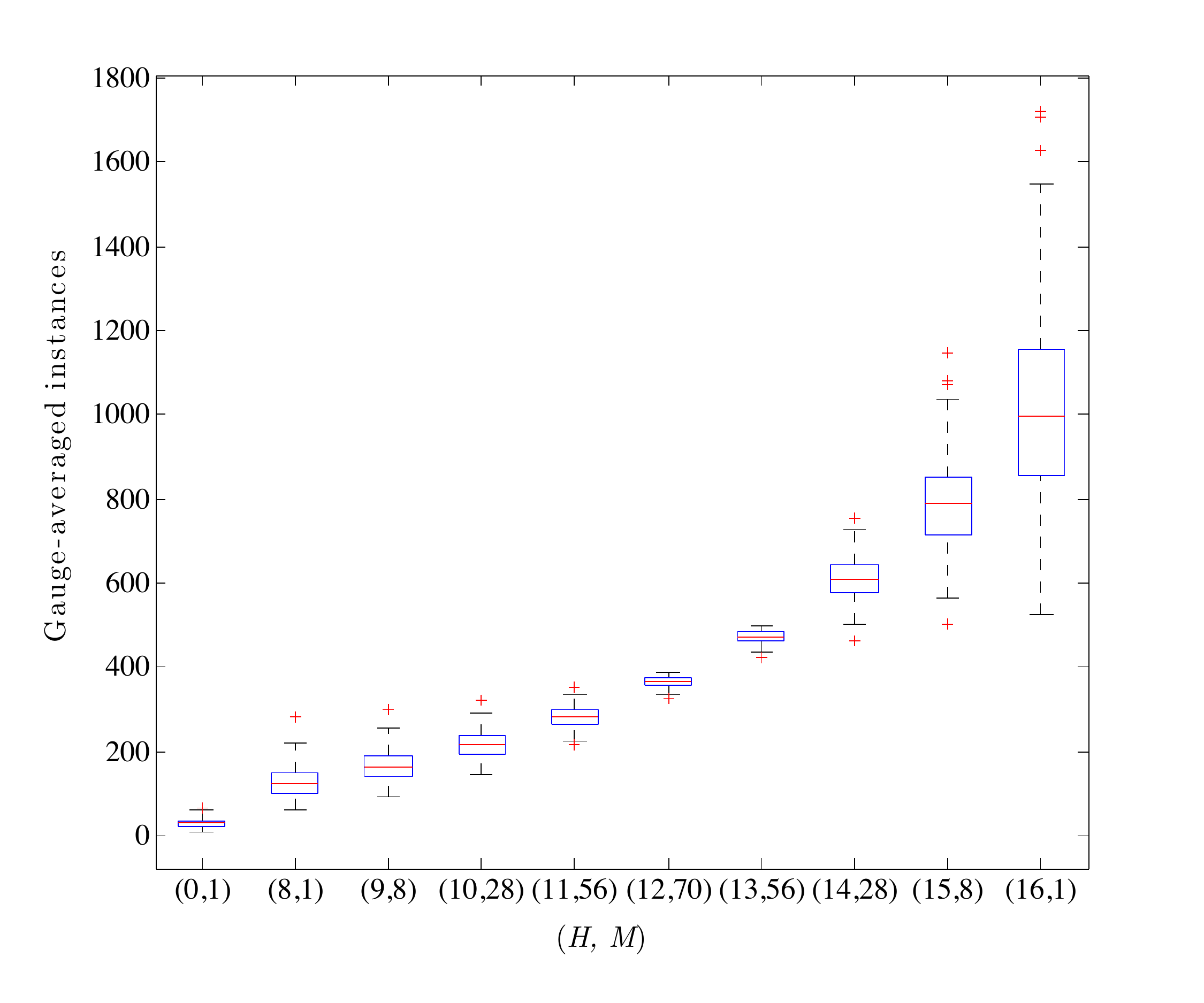}\label{fig:N16-not-opt}} 
 \subfigure[]{\includegraphics[width=.48\textwidth]{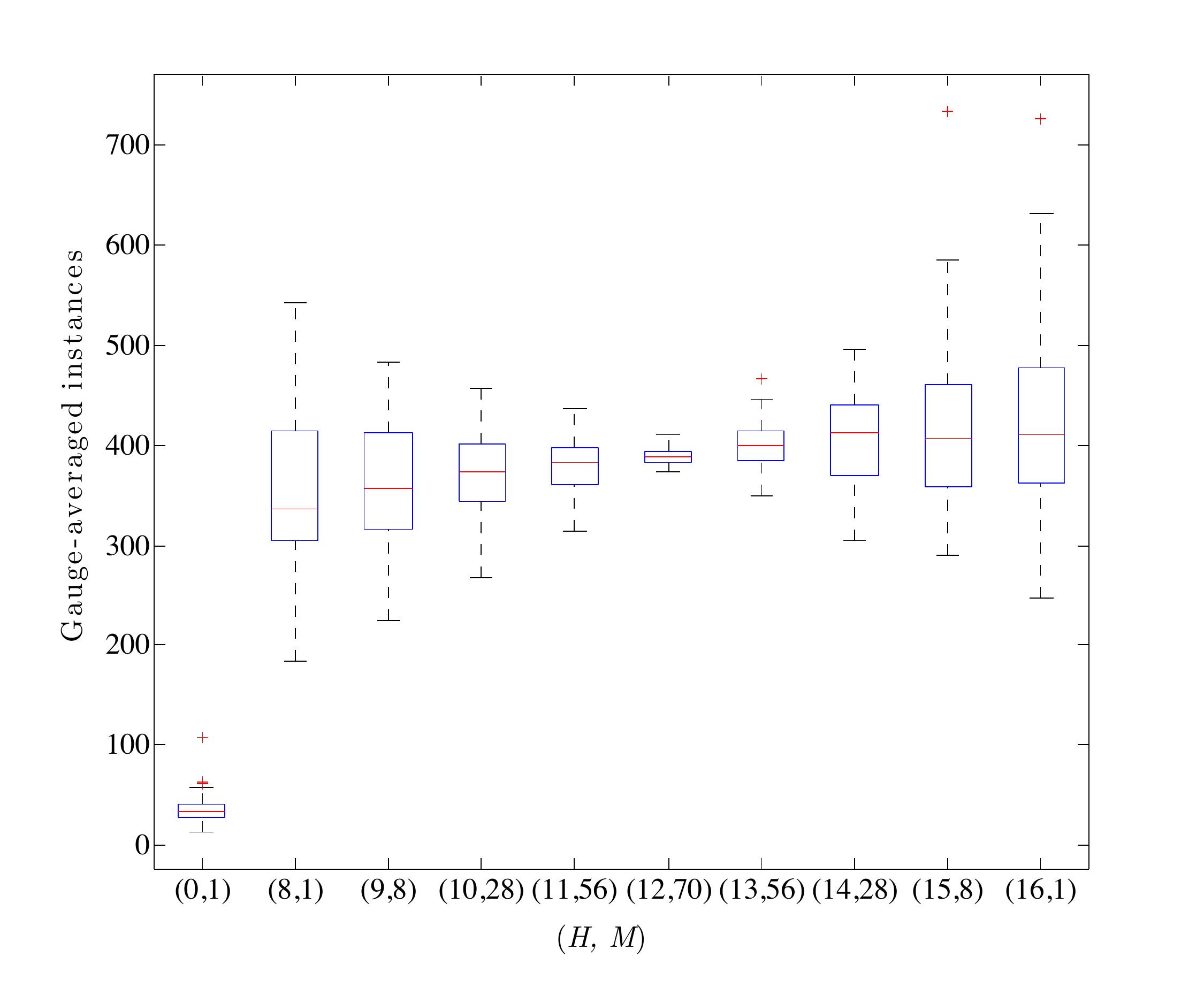}\label{fig:N16-opt}}
\end{center}
\caption{Statistical box plot of the gauge-averaged ground states population (a,c) before and (b), (d) after optimization of $J_{ij}$ as per Table~\ref{table:optimizedJ}, for (a),(b) $N=8$ and (c),(d) $N=16$, $\alpha = 1$ and $t_f=20\, \mu s$.
Only the $N=8,H=6$ case splits into two rotationally inequivalent sets. 
Note the clear step structure in the cluster-states ($H>0$) in (a),(c), while in (b),(d) the population of the cluster-states is fairly equalized (less so in the $N=16$ case since Table~\ref{table:optimizedJ} is optimized for $N=8$).
(a) Data taken with the random parallel embeddings strategy. (b) Data taken using the in-cell embeddings strategy, with the optimized values of the couplings given in Table~\ref{table:optimizedJ}. The same optimization removes the step structure from data taken with the random parallel embeddings strategy (not shown).}
\label{fig:NotOpt-Opt}
\end{figure*}

Under a closed system evolution all the cluster states end up with an identical population. The same is true if we compute the populations using the ME for an independent bath model. In the main text we discussed how cross-talk breaks the symmetry between the cluster states. In this section we discuss another mechanism, of making $|J|$ and $|h|$ unequal, that also breaks the symmetry between the cluster states. This is shown in Fig.~\ref{fig:ControlErrorEvolution} using the ME, where the population of the cluster-state splits by Hamming distance from the isolated state.  Thus this can be viewed as an alternative explanation for the cluster state distribution observed on the DW2, although as we will show, the fit to the experiment is not as good as the cross-talk model discussed in the main text.

To understand the origin of this phenomenon, consider the following perturbation theory argument for the $N=8$ case. Assume that all the local fields are perturbed by $\delta h >0$ so that for the outer spins $h_i = -1 + \delta h$ ($1\leq i \leq 4$) and for the core spins $h_i = 1 - \delta h$ ($5\leq i \leq 8$).  Therefore $|h_i| < |J_{ij}|=1$ and the perturbation to $H_{\mathrm{I}}$
[Eq.~\eqref{eq:problem}] can be written as:
\beq
V = -\delta h \left( \sum_{i=1}^4 \sigma_i^z - \sum_{i=5}^8 \sigma_i^z  \right)
\eeq
All the cluster states have their core spins in the $\ket{0}$ state, so $V$ increases all their energies by $4 \delta h$.  The perturbation acting on the outer spins, however, breaks the degeneracy by Hamming weight.  The contribution from this term is given by $-(n_{0} - n_{1} )\delta h$ where $n_{0}$ or $n_{1}$  is the number of outer spins in the $\ket{0}$ or $\ket{1}$ state, respectively.  Therefore the energy of the 
$\ket{0000\,0000}$ state is unchanged (it becomes the unique ground state), while the energy of the $\ket{1111\,0000}$ state increases by $8\delta h$, so it becomes the least populated among the cluster states. Consequently the final population of the cluster states becomes ordered by Hamming distance from the isolated state $\ket{1111\,1111}$. 

Interestingly, the experimental data for the final populations of the cluster-states displays a pronounced ``step" structure, clearly visible in Fig.~\ref{fig:WalterNotOpt}. The observed steps correspond to an organization of the cluster-states in terms of their Hamming distance from the isolated state, and agrees with the ordering observed in Fig.~\ref{fig:ControlErrorEvolution} and the perturbation theory argument. Thus, the step structure can be explained if, in spite of the fact that for all gauges we set $|h_i|/|J_{ij}| = 1$, in reality there is a systematic error causing $|h_i|<|J_{ij}|$. Such an error would arise if the ratio of $B(t)|h_i|$ and $B(t)|J_{ij}|$ is not kept fixed throughout the annealing, where $B(t)$ is the annealing schedule shown in Fig.~\ref{fig:AnnealingSchedule}. Moreover, such an error would not be unexpected, as the local fields (an inductance) and couplers (a mutual inductance) are controlled by physically distinct devices \cite{Johnson:2010ys}. 

%
\begin{table}[t] 
\begin{center}
\begin{tabular}{|c|c|c|c|} \hline 
$|h_i|$ & $|J_{ij}|$ & \% Change & Absolute Change \\
\hline
 1 & 0.9810 & -1.90 & -0.0190 \\
    \hline
 6/7 & 0.8440 & -1.53 & -0.0131 \\
    \hline
 5/7 & 0.7040 & -1.44 & -0.0103 \\
    \hline
 4/7 & 0.5655 & -1.04 & -0.0059 \\
    \hline
 3/7 & 0.4265 & -0.48 & -0.0021 \\
    \hline
 2/7 & 0.2850 & -0.25 & -0.0007 \\
    \hline
 1/7 & 0.1420 & -0.60 & -0.0009 \\
   \hline
 \end{tabular}
 \end{center}
 \caption{Optimized $|J_{ij}|$ values for a given $|h_i|$ value, yielding the flat population structure shown in Fig.~\ref{fig:AnuragtOpt}. The systematic corrections are of the order of $1\%$, smaller than the random control errors of $5\%$ at $\alpha=1$.} 
 \label{table:optimizedJ}
 \end{table}

\begin{figure}[t]
\begin{center}
\includegraphics[width=0.48\textwidth]{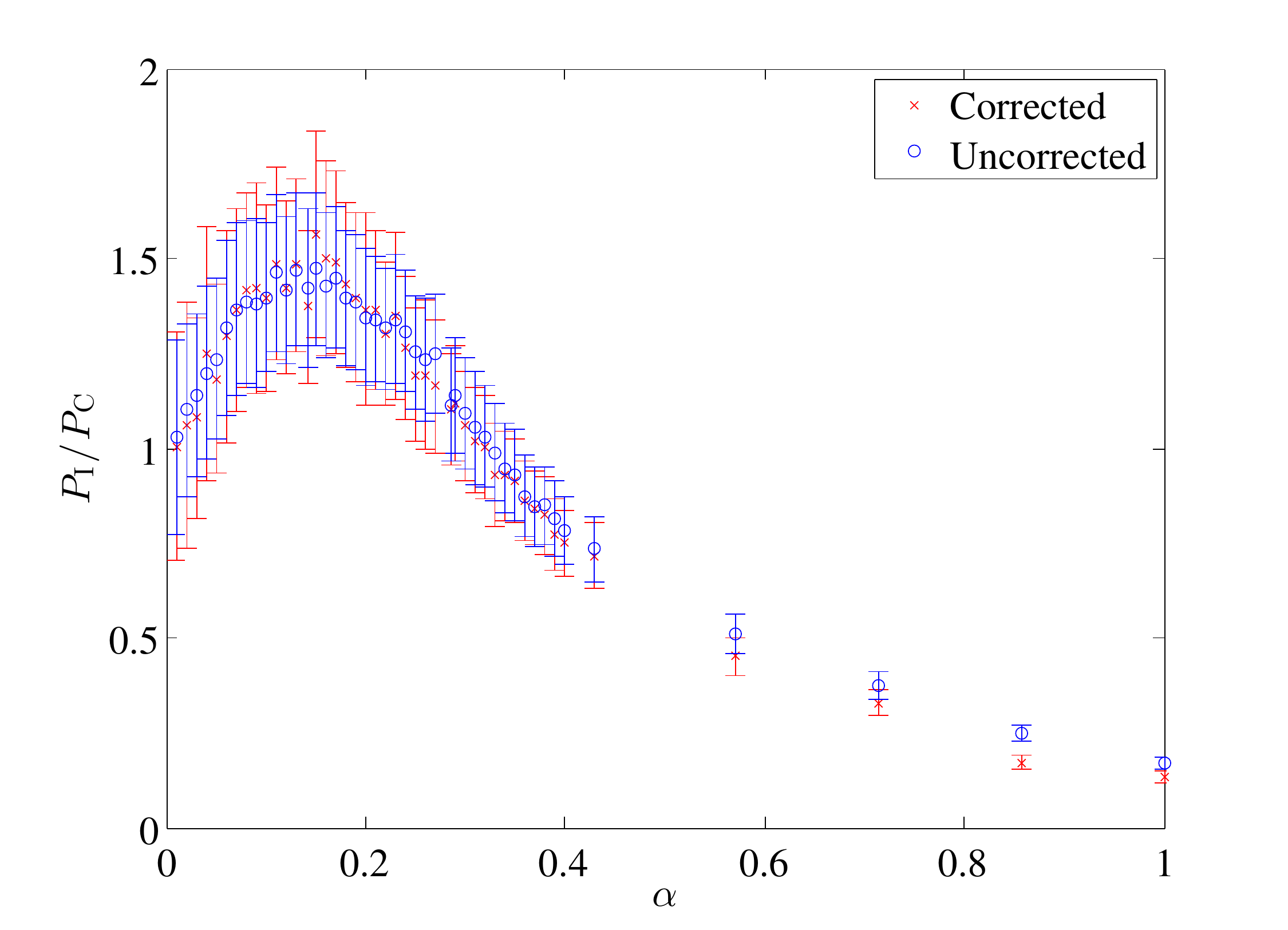}
\end{center}
\caption{Ratio of isolated state population to average cluster-state population as a function of the energy scale factor $\alpha$, for $t_f = 20 \mu s$ and $N=8$.  Shown are the ratios calculated with both uncorrected and corrected values of $J$ (as per Table~\ref{table:optimizedJ}), the latter tuned to flatten the steps seen in the population of the cluster states.  Error bars represent the standard error of the mean value of the ratio estimated using bootstrapping.}
\label{fig:CorrectedvsUncorrected}
\end{figure}

A natural question is whether we can mitigate this type of error. To do so we introduce a simple optimization technique. Specifically, we can compensate for $|h_i| < |J_{ij}|$ and fine-tune the values of $h$ and $J$ to nearly eliminate the step structure for each value of the energy scale factor $\alpha$. We show this in Fig.~\ref{fig:AnuragtOpt}. The corresponding optimized values are given in Table~\ref{table:optimizedJ}, where the compensation reverses the inequality to $|h_i| > |J_{ij}|$. 
We have further checked that the same fine-tuning technique suppresses the step structure seen for larger $N$, as shown in Fig.~\ref{fig:N16-not-opt}.  The step structure in the $N=16$ case is even more pronounced than in the $N=8$ case. By adjusting the value of $J$ while keeping $|h|=1$ we can reduce the step structure, as shown in Fig.~\ref{fig:N16-opt}. (We note that this gives rise to the interesting possibility of using this ``step-flattening" technique to more precisely calibrate the device.)
This control error has little effect on the suppression or enhancement of the isolated state, as can be seen in Fig.~\ref{fig:CorrectedvsUncorrected}. 

\subsection{Using the distribution of cluster states to rule out the noisy SSSV model}
\label{sec:noisy-SSSV}
As we saw in Figs.~\ref{fig:NotOpt-Opt}, without the Table~\ref{table:optimizedJ} correction, the DW2 results at $\alpha = 1$ exhibit a non-uniform distribution over the cluster states. We now demonstrate that the noisy SSSV model is incapable of correctly capturing this aspect of the experimental results. We do so for various scenarios differing in how we treat the control error.

\begin{figure*}[t]
   \subfigure[\ SSSV, offset $h = 0.97 J$]{ \includegraphics[width=0.32\textwidth]{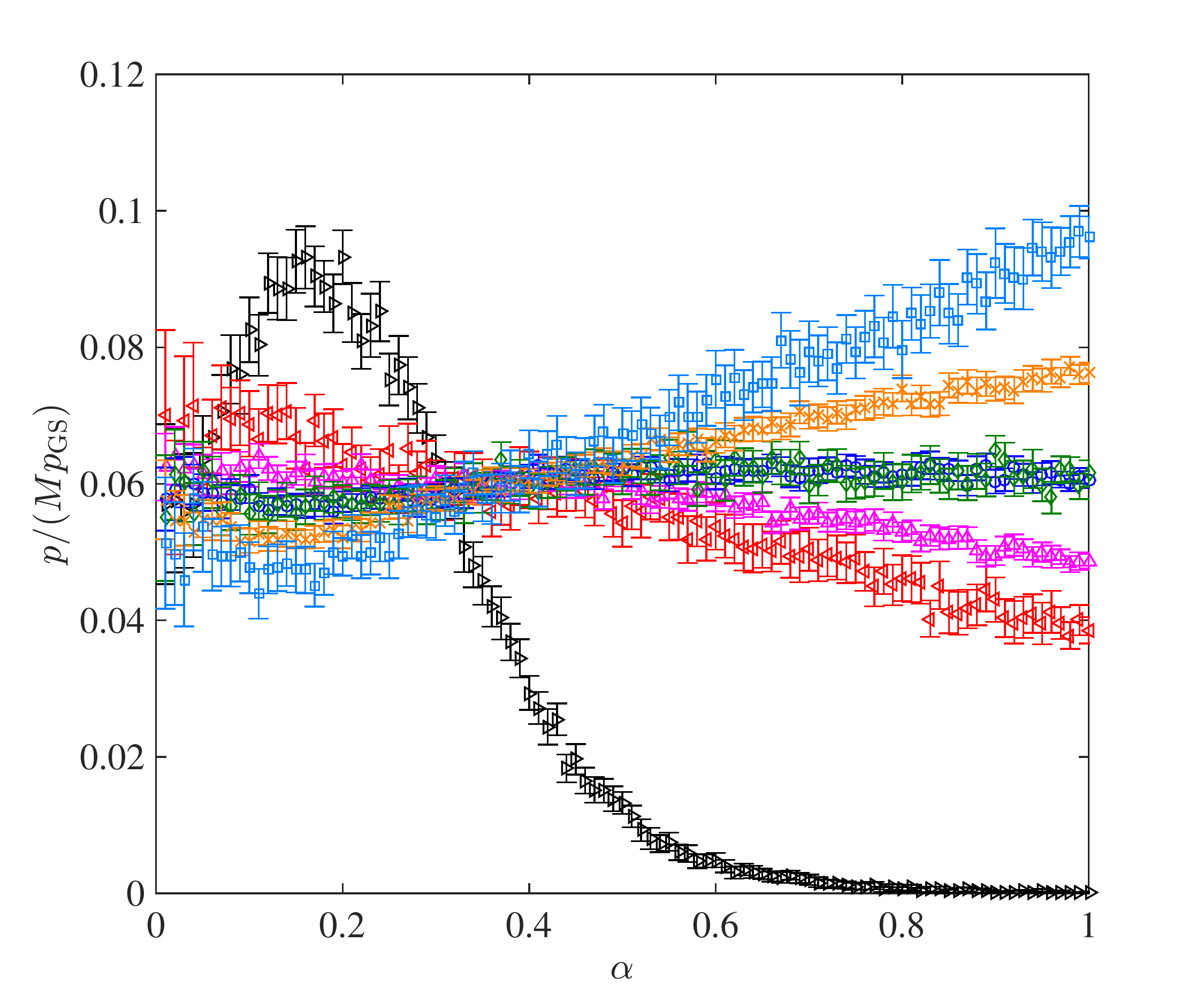}\label{fig:SSSV-calibrated1}} 
   \subfigure[\ ME, offset $h = 0.985 J$]{ \includegraphics[width=0.32\textwidth]{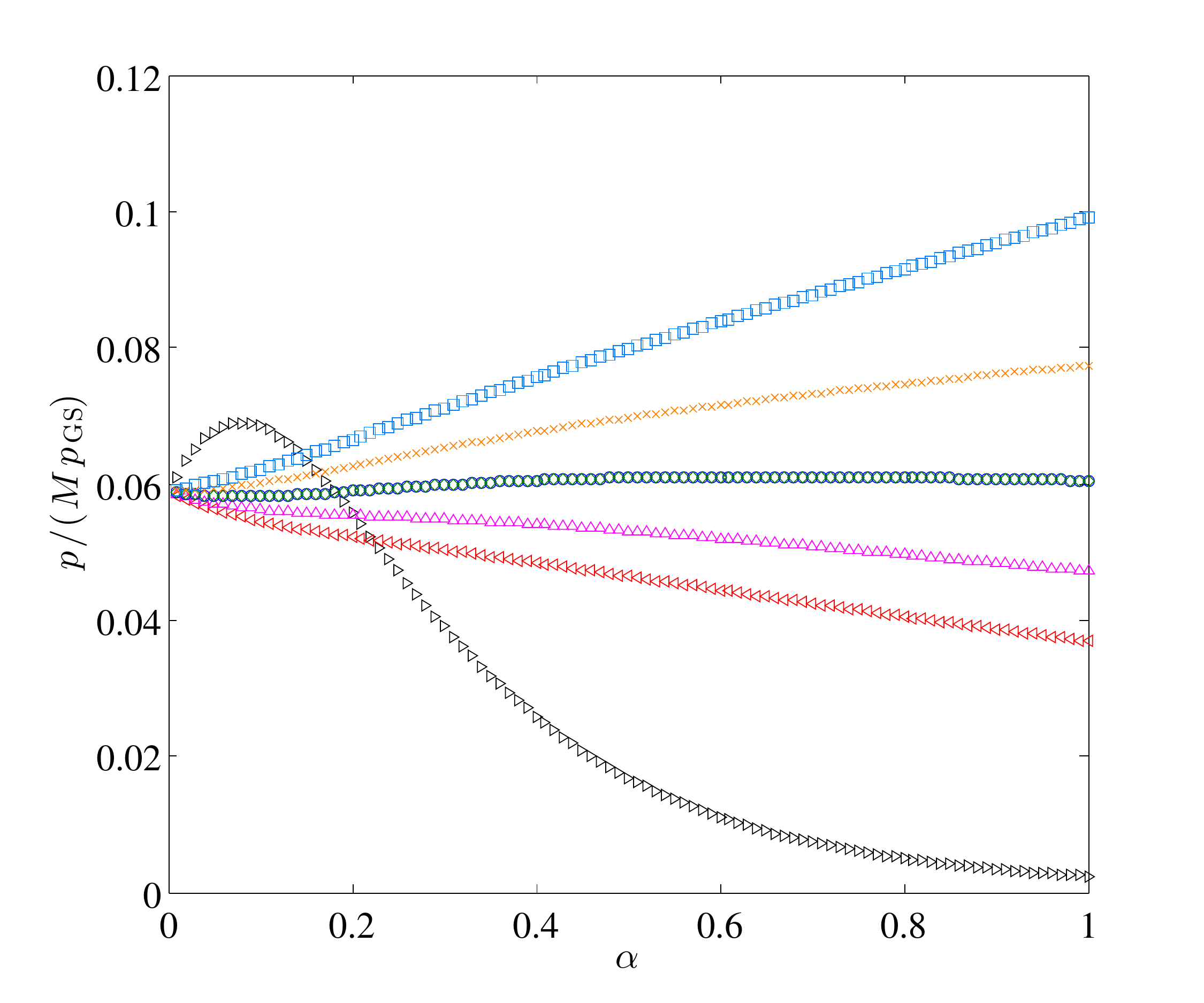} \label{fig:ME-calibrated1}} 
    \subfigure[\ Perturbation theory, offset $h = 0.975 J $]{ \includegraphics[width=0.32\textwidth]{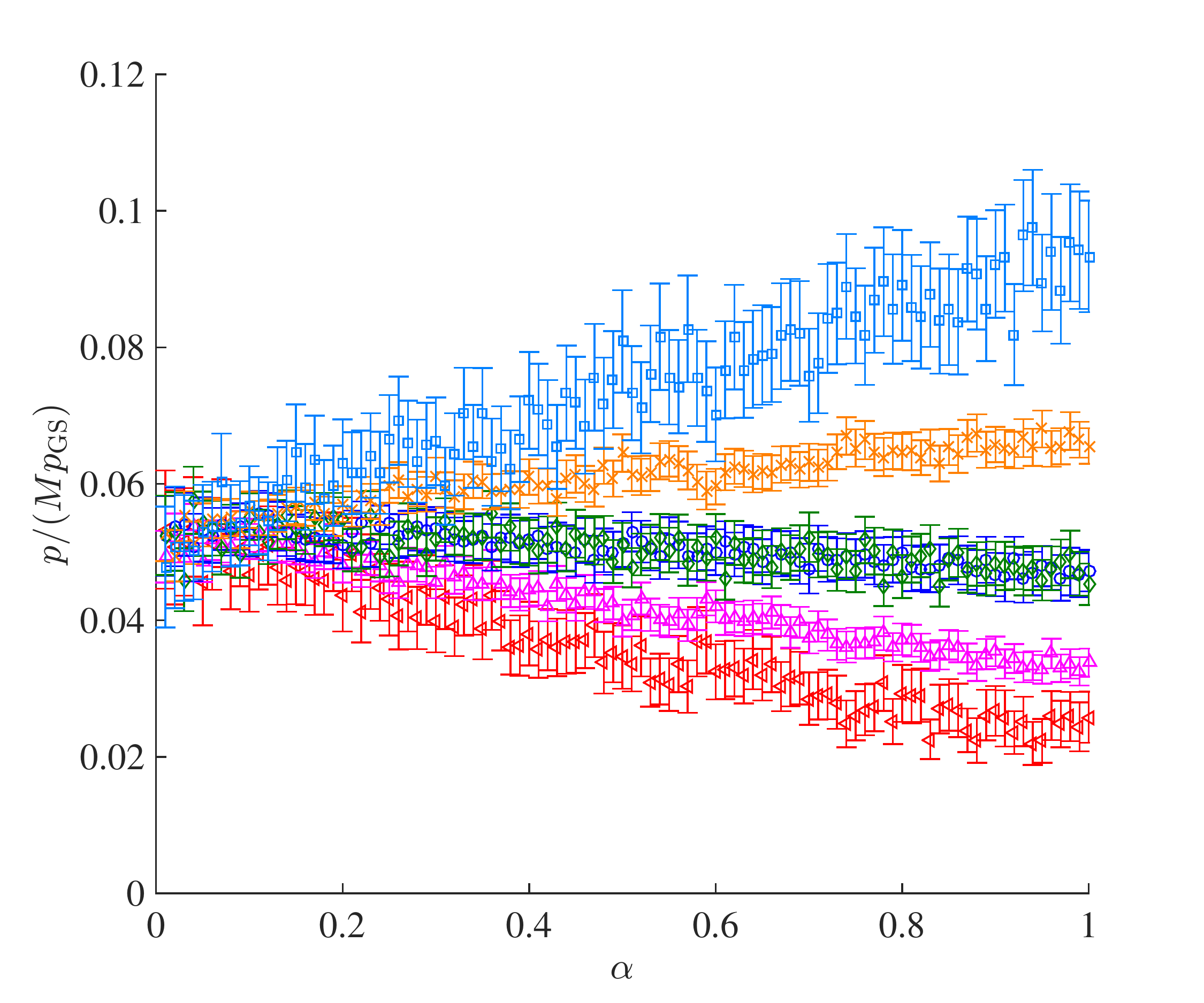} \label{fig:PT-calibrated1}} 
\caption{(a) Noisy SSSV; (b) ME, (c) perturbation theory [Eq.~\eqref{eq:pertH}] with a population ordering correction (offset of $h$ vs $J$) at $\alpha = 1$. The error bars represent the 95\% confidence interval.  The SSSV model now has the right ordering of the cluster states but clearly disagrees with the DW2 result [Fig.~\ref{fig:DW2-Uncalibrated2}] for $\alpha \lesssim 0.3$, near where the isolated state has its maximum. The ME result is in qualitative agreement with the DW2 result except that the cluster state populations do not equalize for small $\alpha$, which is a consequence of not including the $\alpha$-dependence of the offset.}
   \label{fig:Calibration1}
\end{figure*}  
\begin{figure*}[t]
   \subfigure[\ DW2, offset $J = 0.981 h$]{ \includegraphics[width=0.32\textwidth]{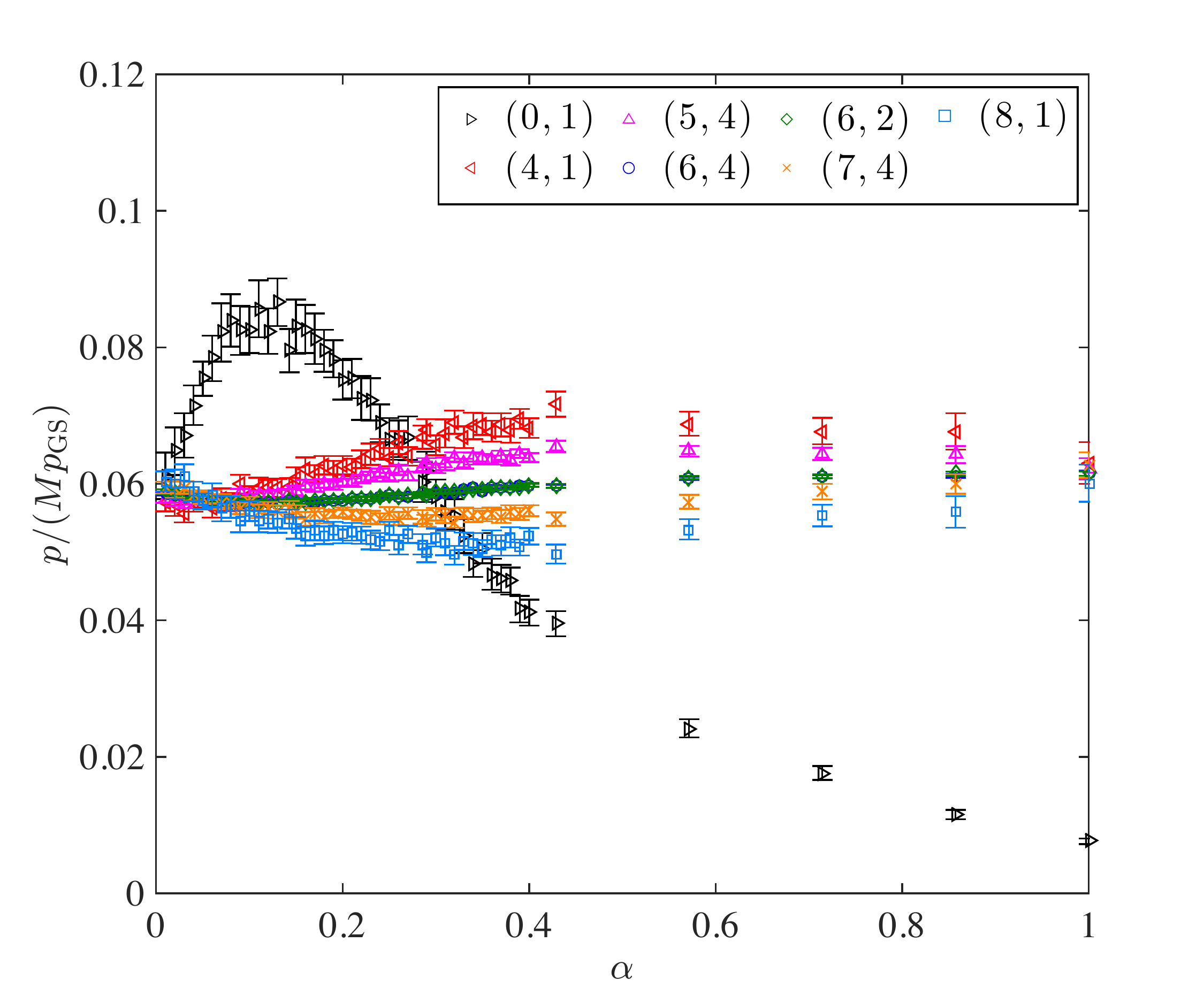}\label{fig:DW2-calibrated2}} 
   \subfigure[\ SSSV, offset $h = 0.988J$]{ \includegraphics[width=0.32\textwidth]{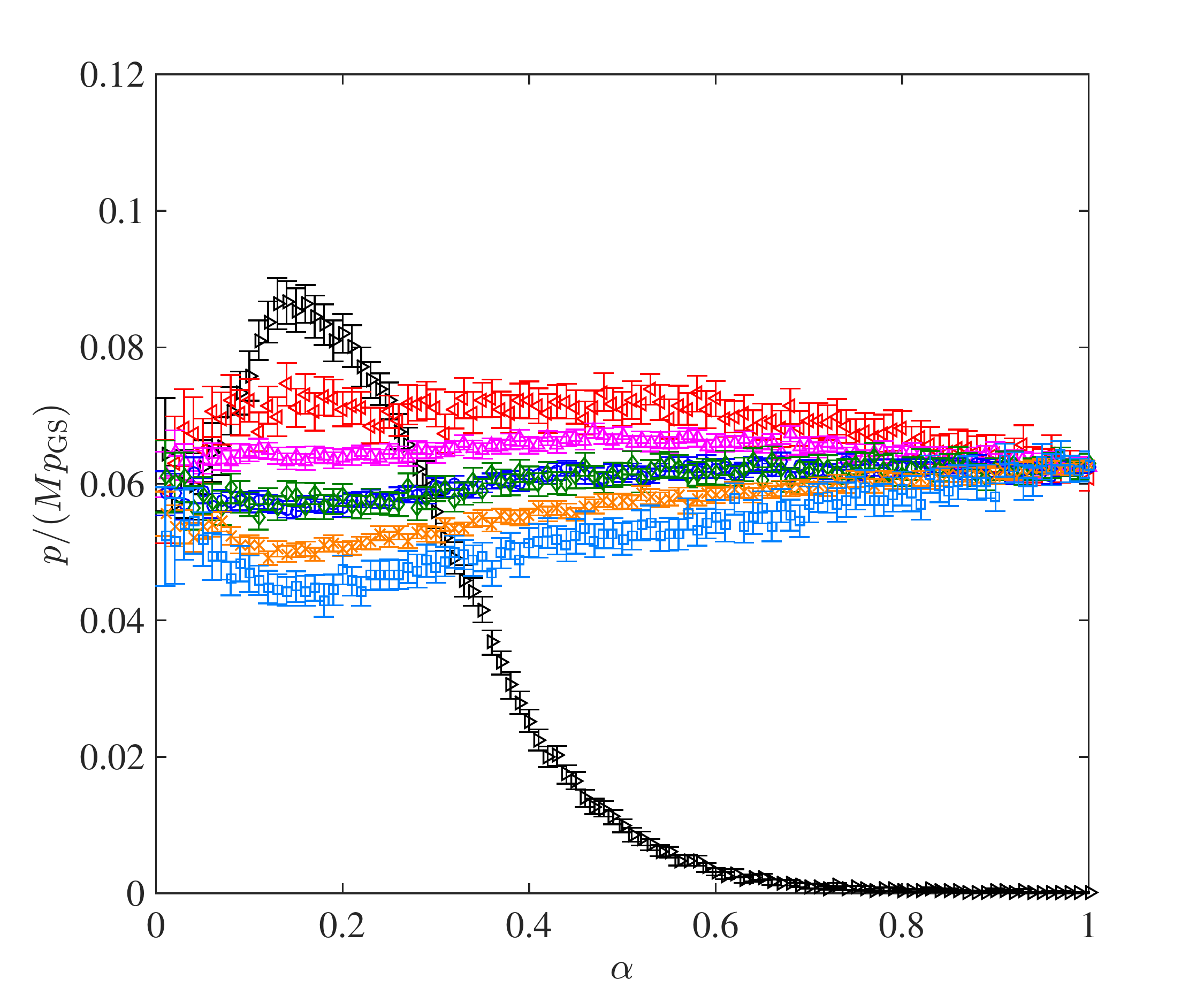} \label{fig:SSSV-calibrated2}} 
   \subfigure[\ SSSV, offset $h = 0.94J$]{ \includegraphics[width=.32\textwidth]{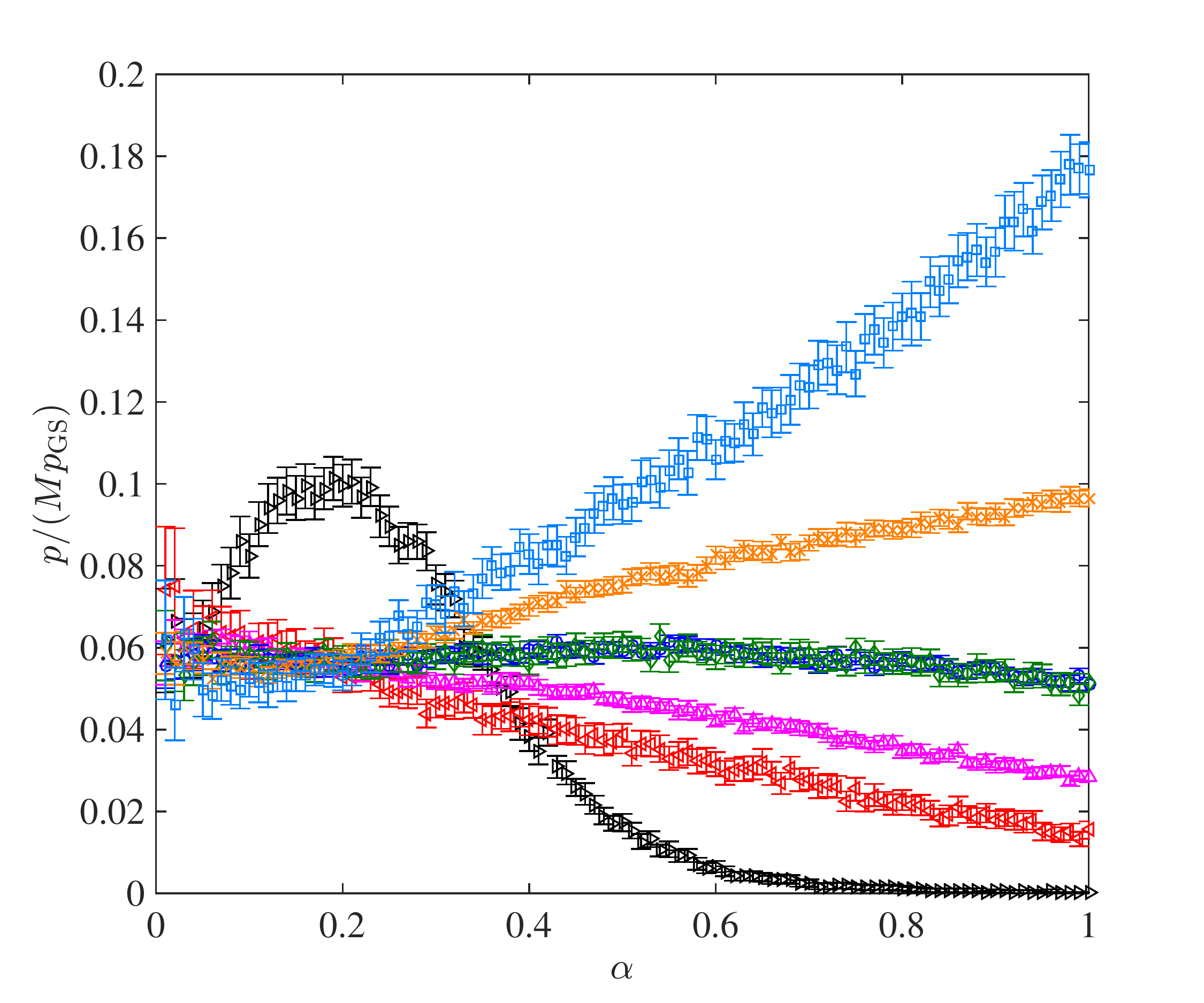} \label{fig:SSSV-calibrated3}} 
\caption{(a) DW2 and (b) noisy SSSV model with a population equalizing correction at $\alpha = 1$. (c) Noisy SSSV model with offset chosen to equalize the cluster state populations at $\alpha=0.2$.  The error bars represent the 95\% confidence interval.}
\label{fig:Calibration3}
\end{figure*} 
\begin{figure*}[ht]
     \subfigure[\ Perturbation theory, offset $h = 0.975 J $]{\includegraphics[width=0.34\textwidth]{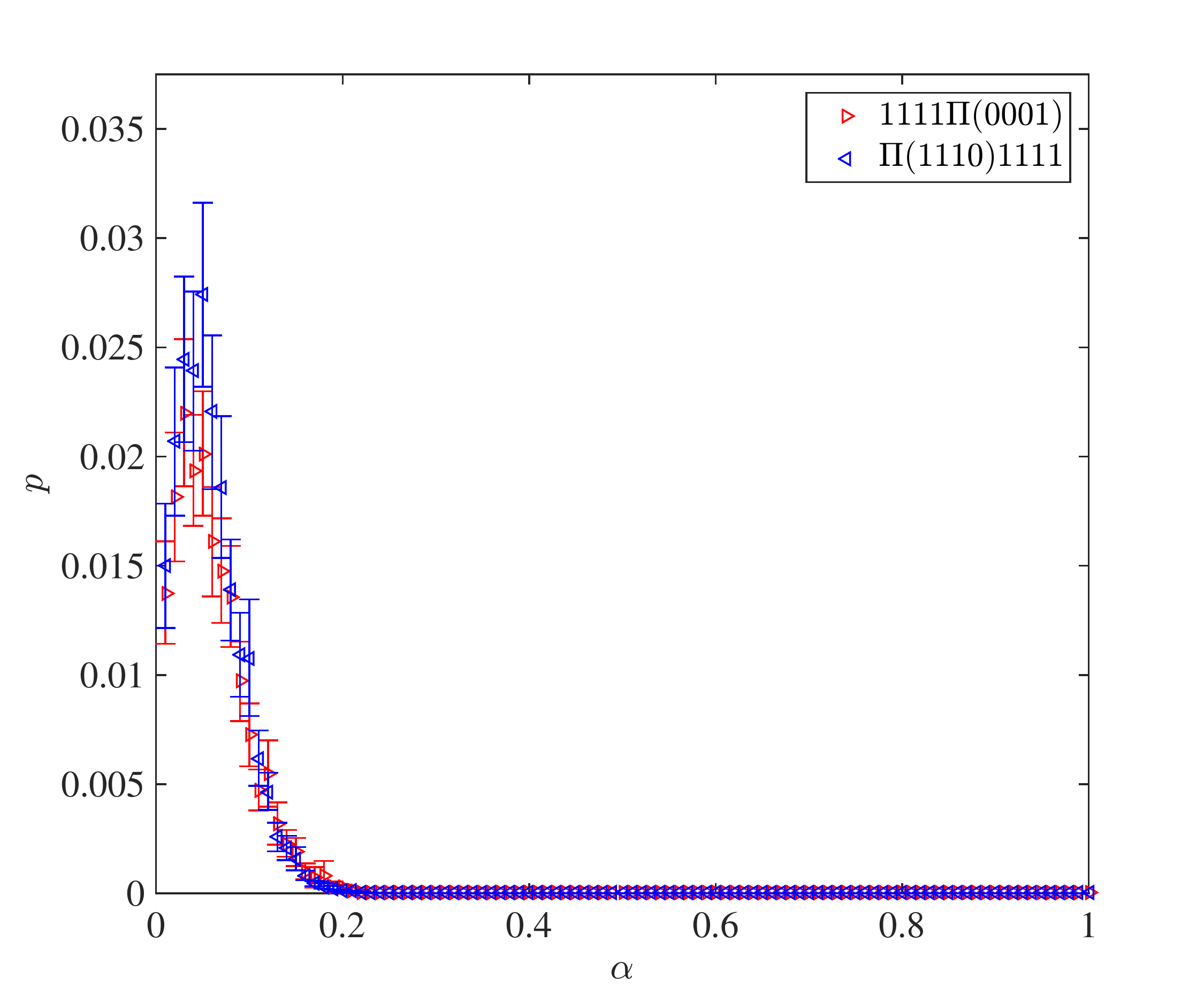}\label{fig:pert-ES}}
   \subfigure[\ SSSV, offset $h = 0.97J$]{ \includegraphics[width=0.34\textwidth]{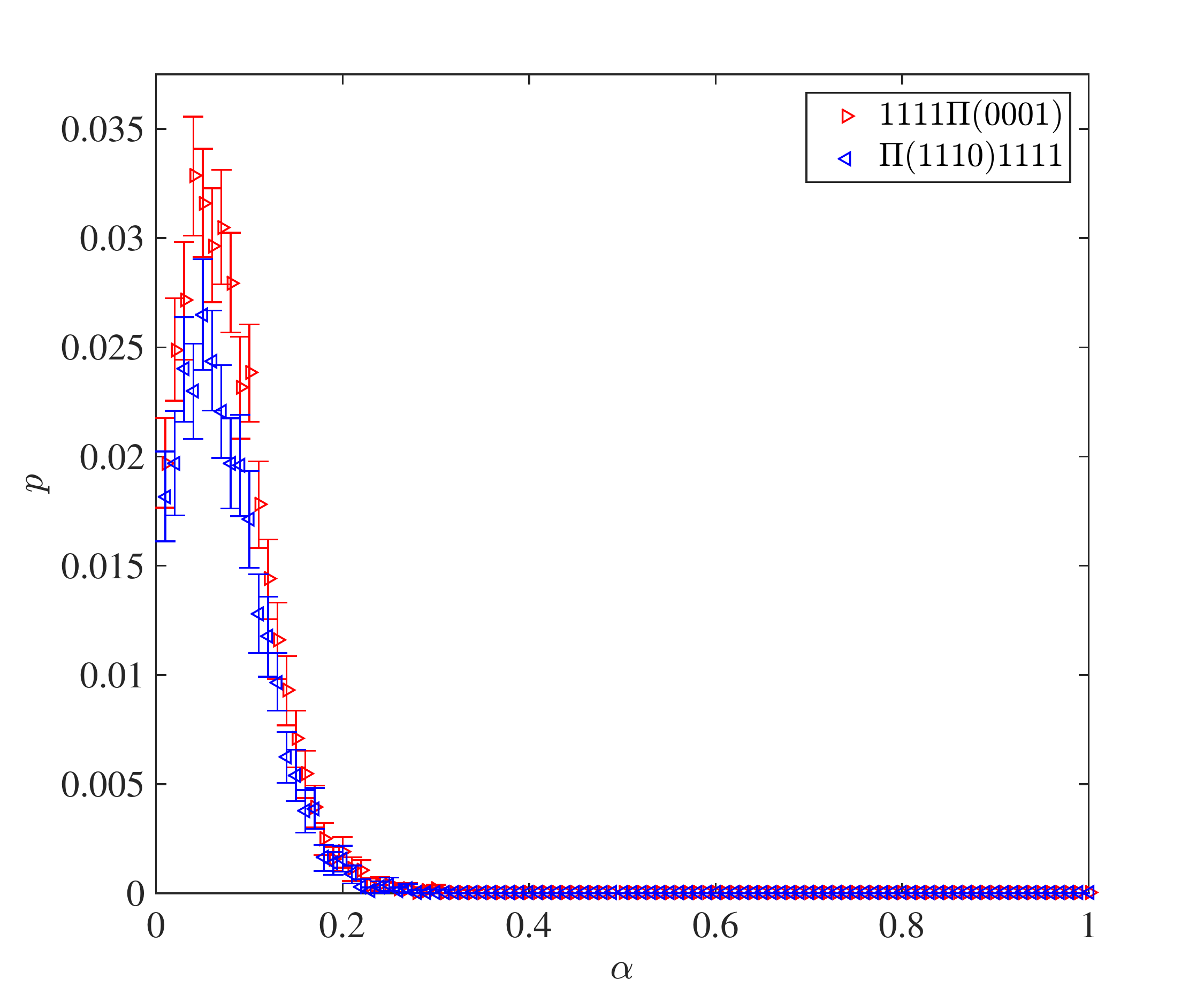} \label{fig:SSSV-ES-calibrated1}} 
   \subfigure[\ SSSV, offset $h = 0.988J$]{ \includegraphics[width=0.34\textwidth]{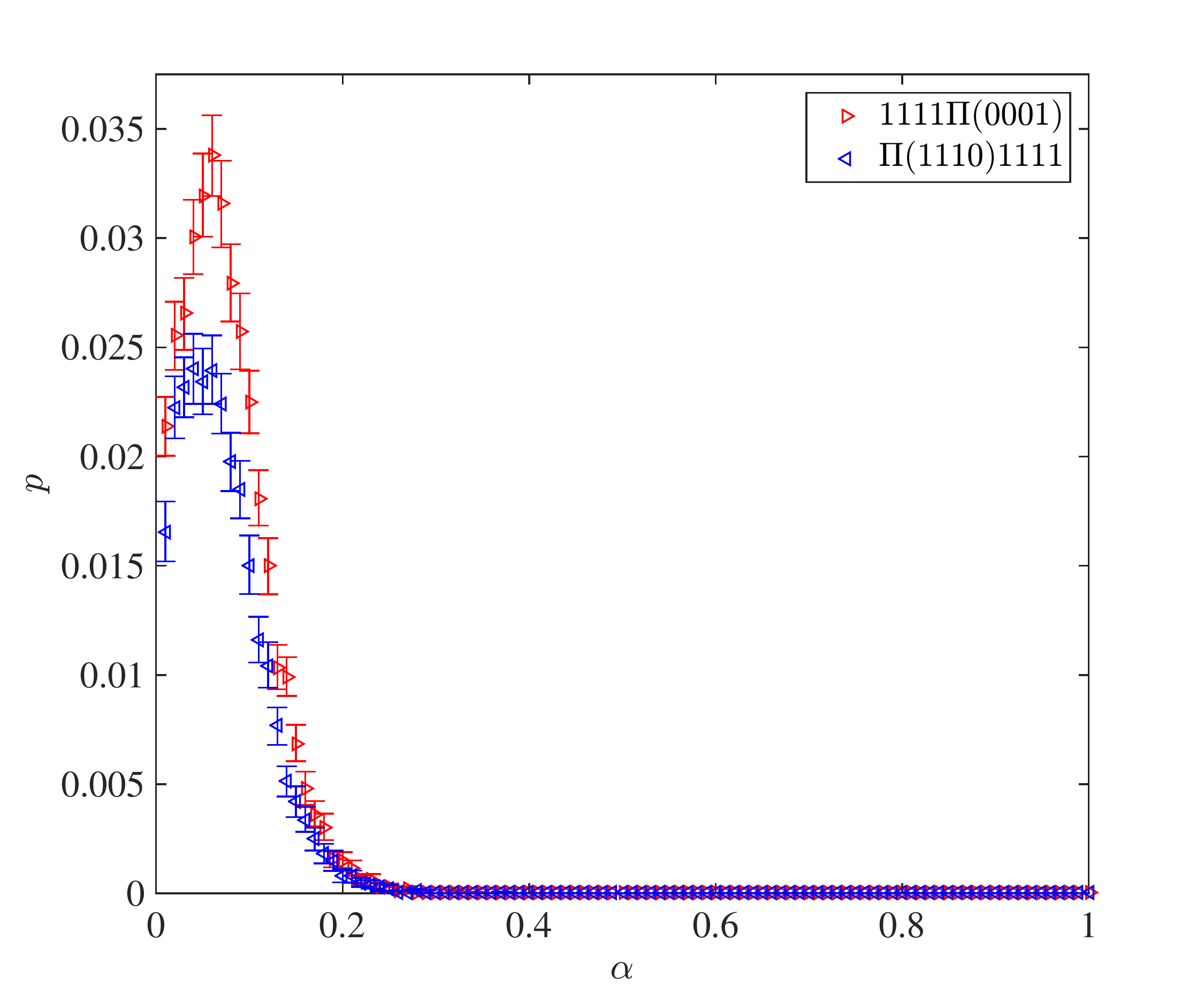} \label{fig:SSSV-ES-calibrated988}} 
     \subfigure[\ SSSV, offset $h = 0.94J$]{ \includegraphics[width=0.34\textwidth]{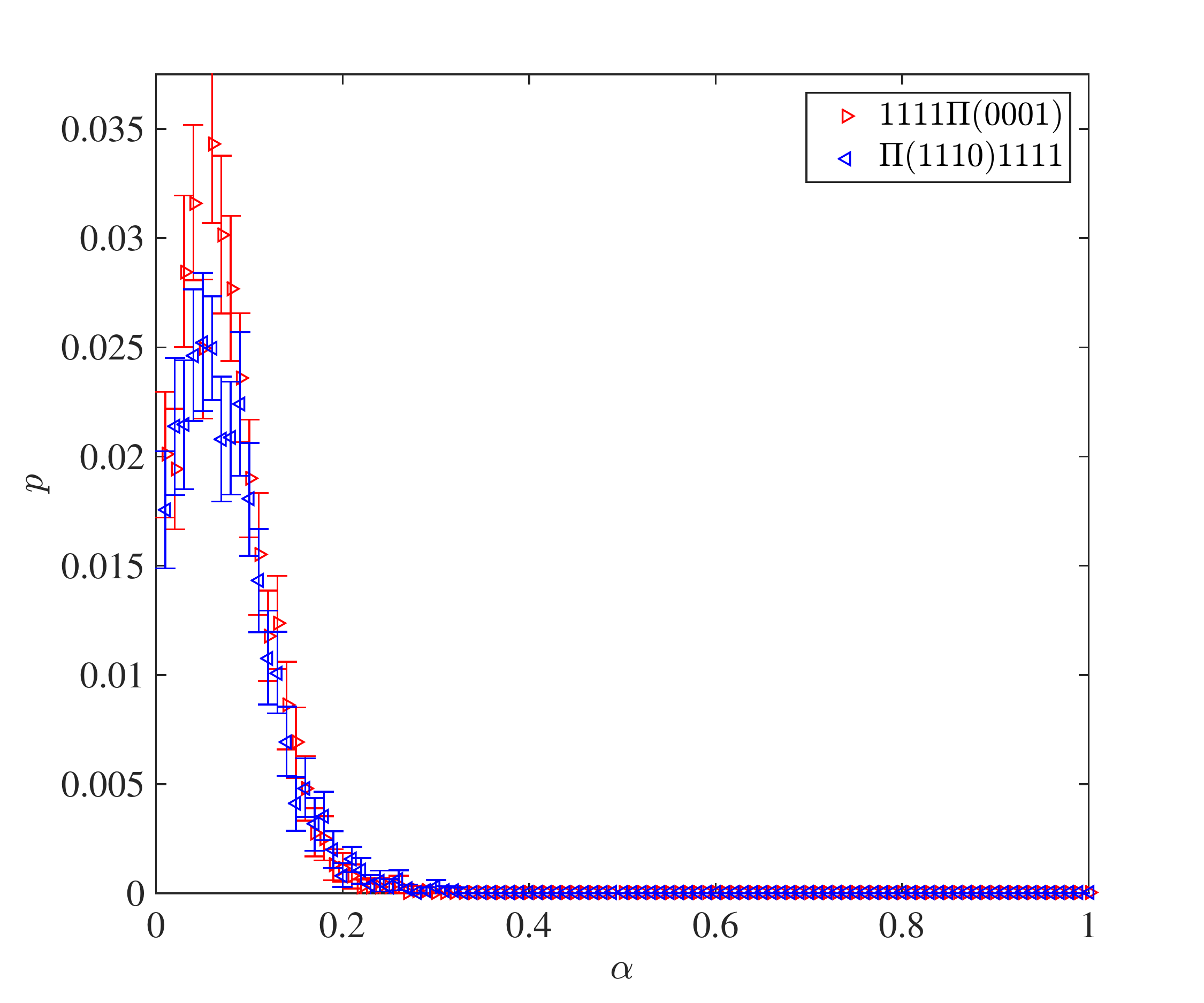} \label{fig:SSSV-ES-calibrated94}} 
\caption{Subset of the first excited state populations for (a) perturbation theory [as in Eq.~\eqref{eq:pertH}], (b), (c) and (d) SSSV with offsets matching Figs.~\ref{fig:SSSV-calibrated1}, \ref{fig:SSSV-calibrated2} and \ref{fig:SSSV-calibrated3} respectively. Panel (f) shows the case with qubit cross-talk discussed in Sec.~\ref{sec:cross-talk}. The $\Pi$ symbol denotes all permutations. Whereas the perturbation theory result for a QA Hamiltonian [Eq.~\eqref{eq:pertH}] reproduces the correct ordering, none of the three SSSV cases shown does. These three SSSV cases were chosen to optimize the fit for the cluster state populations.  The error bars represent the 95\% confidence interval.}
   \label{fig:Calibration4}
\end{figure*}

\subsubsection{Population ordering correction at $\alpha = 1$}

The noisy SSSV and ME results after calibrating $h$ and $J$ so as to match the DW2 ordering of the distribution of cluster states at $\alpha=1$ are shown in Fig.~\ref{fig:SSSV-calibrated1} and \ref{fig:ME-calibrated1}, respectively.  Like the DW2, the ME cluster state populations converge as $\alpha$ goes to zero, whereas the SSSV populations converge up to $\alpha \approx 0.3$ and then diverge again. Therefore, with this calibration, the noisy SSSV model fails to capture the DW2 cluster state populations at low $\alpha$.

We can understand the equalization of the cluster state populations in the ME as follows: as $\alpha$ is made smaller, the spacing of the quantized, discrete energy levels (when the Ising Hamiltonian dominates) shrinks with $\alpha$.  Thermal excitations between the levels will be less suppressed, allowing for a redistribution of the population.  We check this intuition with a generalization of our  perturbation theory argument that was used explain the suppression of the isolated state in the closed system setting (see Appendix~\ref{app:pert-theory}).  We diagonalize the Hamiltonian:
\beq
H= \alpha (H_{\mathrm{I}} + 0.01 H_X) + \eta
\label{eq:pertH}
\eeq
where $H_{\mathrm{I}}$ is the detuned Ising Hamiltonian, $\eta$ is Gaussian noise (independent of $\alpha$) introduced on the couplings and local fields, and $H_X$ is the transverse field, whose small magnitude models the end of the annealing evolution.  We then populate the lowest $17$ energy eigenstates of this Hamiltonian by a Boltzmann distribution, i.e., $p_n = e^{- \upbeta E_n} / Z$, where $Z = \sum_{n=1}^{256} e^{- \upbeta E_n} $.  We pick $\upbeta/\hbar = 10.7$ns and choose a calibration of $h$ and $J$ in order to best match the DW2 results at $\alpha = 1$.  The cluster state populations are then extracted from their overlap with these Boltzmann populated $17$ energy levels.  As shown in Fig.~\ref{fig:PT-calibrated1}, this perturbation theory argument reproduces the behavior of the ME for the cluster states very well: it shows the cluster state populations converging to an equal population as $\alpha$ goes to zero.  This at least suggests that the intuition presented above is consistent.  However, this method does not reproduce all the data.  The isolated state shows a very large population (it is off the scale of the graph), which does not match the ME or the DW2 results.  This is not entirely surprising since the Boltzmann distribution of course does not take into account the annealing evolution.

The reason that the ME does not exhibit a uniform population on the cluster states for small $\alpha$ [as seen in the DW2 results of Fig.~\ref{fig:DW2-Uncalibrated2}] was addressed in Sec.~\ref{sec:cross-talk}, where we discussed a cross-talk mechanism that generates an $\alpha$-dependence of $h$ and $J$. With this dependence the ME reproduces this feature of the DW2 data as well. 

\subsubsection{Population equalizing correction at $\alpha = 1$ or $\alpha = 0.2$}
As an example of a different calibration procedure, we can calibrate the DW2 and the noisy SSSV model to have equal populations at $\alpha = 1$.  As shown in Fig.~\ref{fig:DW2-calibrated2} and \ref{fig:SSSV-calibrated2}, we observe that initially as $\alpha$ is decreased, both SSSV and DW2 behave in a similar manner whereby the cluster state populations diverge, but whereas DW2 converges again for small $\alpha$ and is almost uniform at $\alpha \approx 0.15$, the noisy SSSV model populations do not start to reconverge until much closer to $\alpha = 0$.  Therefore, once again, we find a qualitative difference between the noisy SSSV model and the DW2.  We note that for this calibration (i.e., having the cluster states equal at $\alpha=1$), the ME would not need to be offset and would be as shown in Fig.~\ref{fig:ME-Uncalibrated}. 

Since the no-offset DW2 results [Fig.~\ref{fig:DW2-Uncalibrated2}] show the cluster state populations equalizing at $\alpha\approx 0.2$ we can alternatively attempt to  calibrate the noisy SSSV model to match the no-offset DW2 results at this value of $\alpha$, i.e., we can choose an offset for SSSV such that it has an almost equal population at $\alpha = 0.2$.  This is shown in Fig.~\ref{fig:SSSV-calibrated3}.  The cluster states continue to diverge as $\alpha$ decreases, while they diverge in the opposite order as $\alpha$ grows.  If we continue this procedure, i.e., make the populations equal for smaller and smaller $\alpha$, this requires a larger offset which will make the staircase structure at $\alpha = 1$ even further pronounced, further increasing the mismatch with the DW2 in this regime. 

\subsubsection{Excited states ordering}
As in the main text, we now go beyond the ground subspace and consider an $8$-dimensional subspace of the subspace of first excited states. We arrange these according to permutations of the core or outer qubits, i.e., we group the states as $\ket{1111\, \Pi(0001)}$ and $\ket{\Pi(1110)\, 1111}$, where $\Pi$ denotes a permutation. As shown in Fig.~\ref{fig:Calibration4}, the DW2 prefers the set $\ket{\Pi(1110)\, 1111}$, and the perturbation theory analysis based on the noisy quantum signature Hamiltonian [Eq.~\eqref{eq:pertH}] agrees. However, for all values of the offset considered in the previous two subsections, the noisy SSSV model prefers the set $ \ket{1111\, \Pi(0001)}$, as seen in Fig.~\ref{fig:SSSV-ES-calibrated1}-\ref{fig:SSSV-ES-calibrated94}. 

\section{Effect of varying the annealing time or the total number of spins} 
\label{app:moreThermal}

\begin{figure}[t]
\includegraphics[width=0.98\columnwidth]{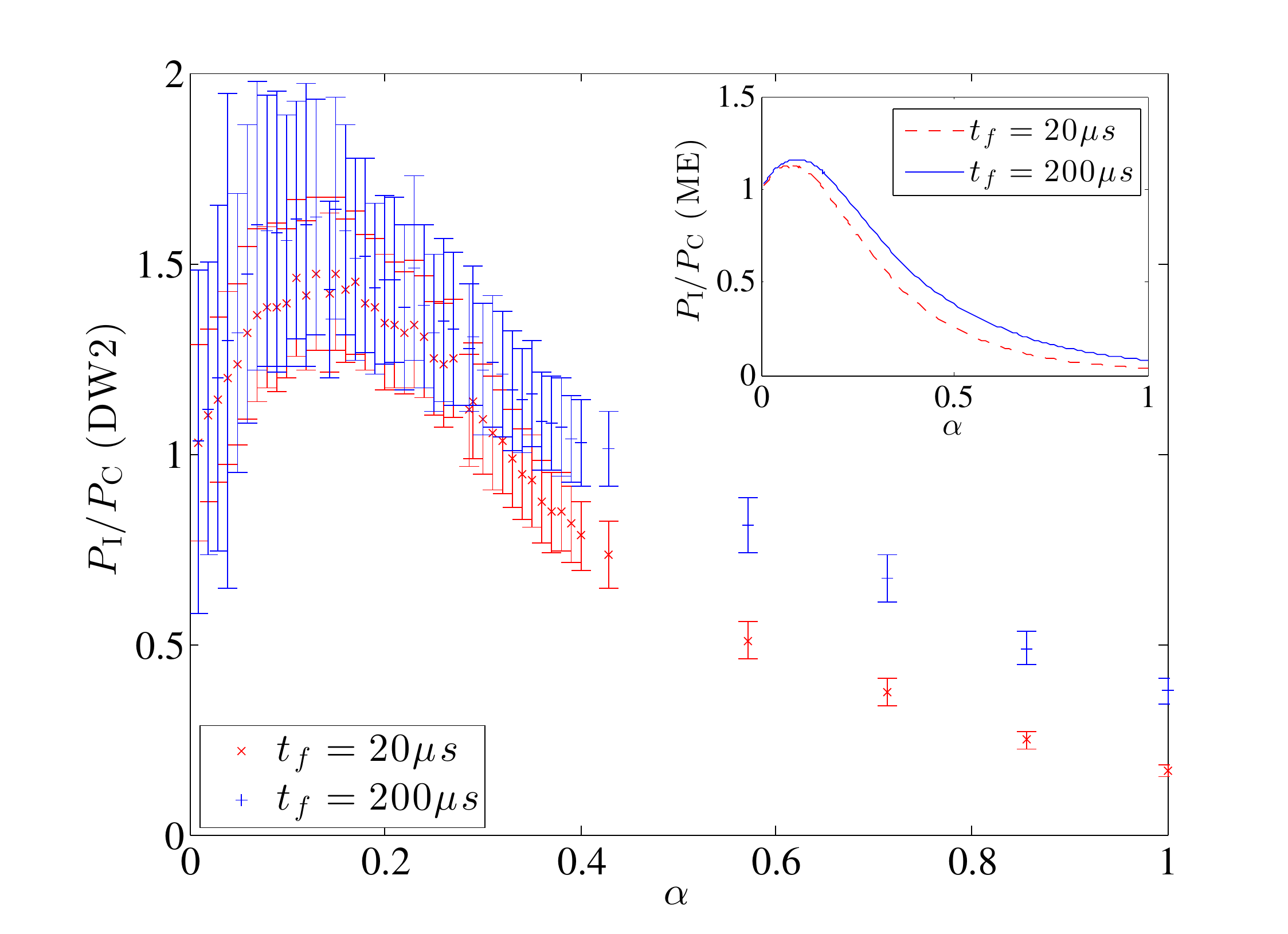}
\caption{Ratio of the isolated state population to the average population in the cluster ($P_{\mathrm{I}}/P_{\mathrm{C}}$) as a function of the energy scale factor $\alpha$, for two different values of $t_f$, and $N=8$. The inset shows the ME results. Data were collected using the ``in-cell embeddings" strategy (see Appendix~\ref{app:data-collection} for details).  Error bars are one standard deviation above and below the mean.}
\label{fig:8-12-16-EIGS}
\end{figure}
\begin{figure}[ht]
\includegraphics[width=\columnwidth]{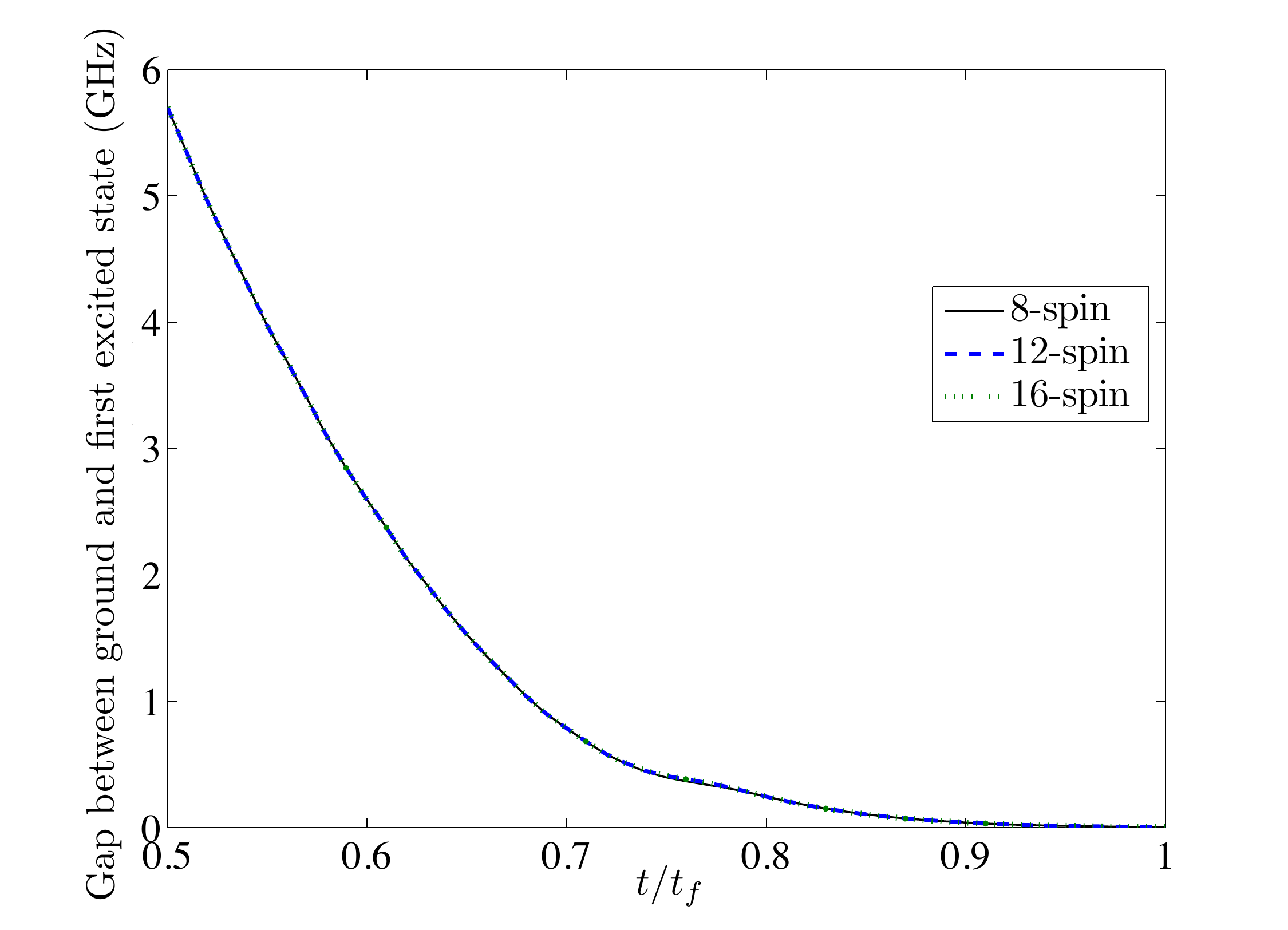}
\caption{Numerically calculated instantaneous energy gap between the ground and first excited state for the $8$, $12$ and $16$ spin Hamiltonians. The  gap vanishes since the first excited state becomes part of the $2^{N/2}+1$-fold degenerate ground state manifold at $t=t_f$.}
\label{fig:8-12-16-EIGS-SM}
\end{figure}
\begin{figure*}[t]
\subfigure[\, $N=12$]{\includegraphics[width=0.48\textwidth]{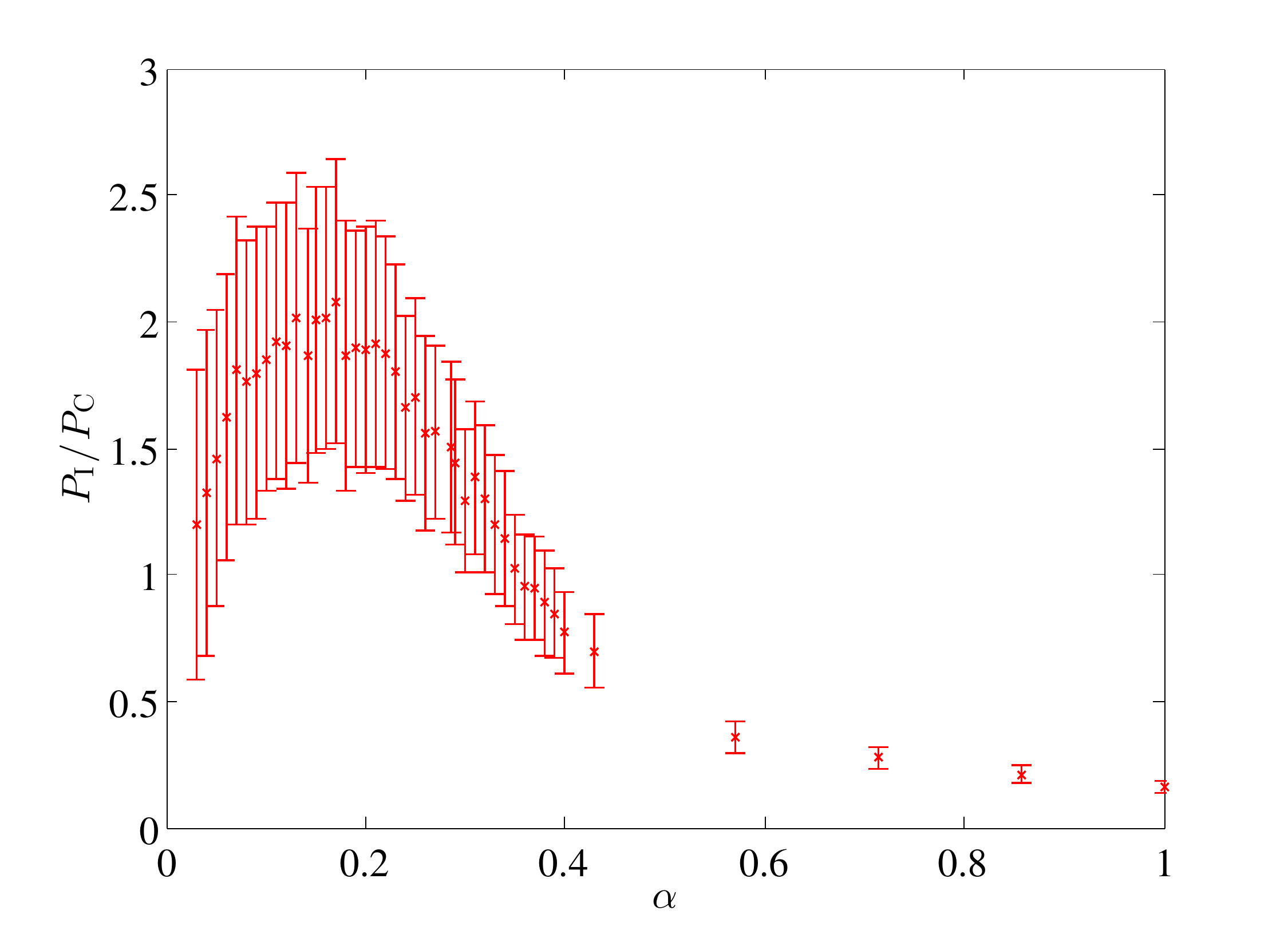}}
\subfigure[\, $N=16$]{\includegraphics[width=0.48\textwidth]{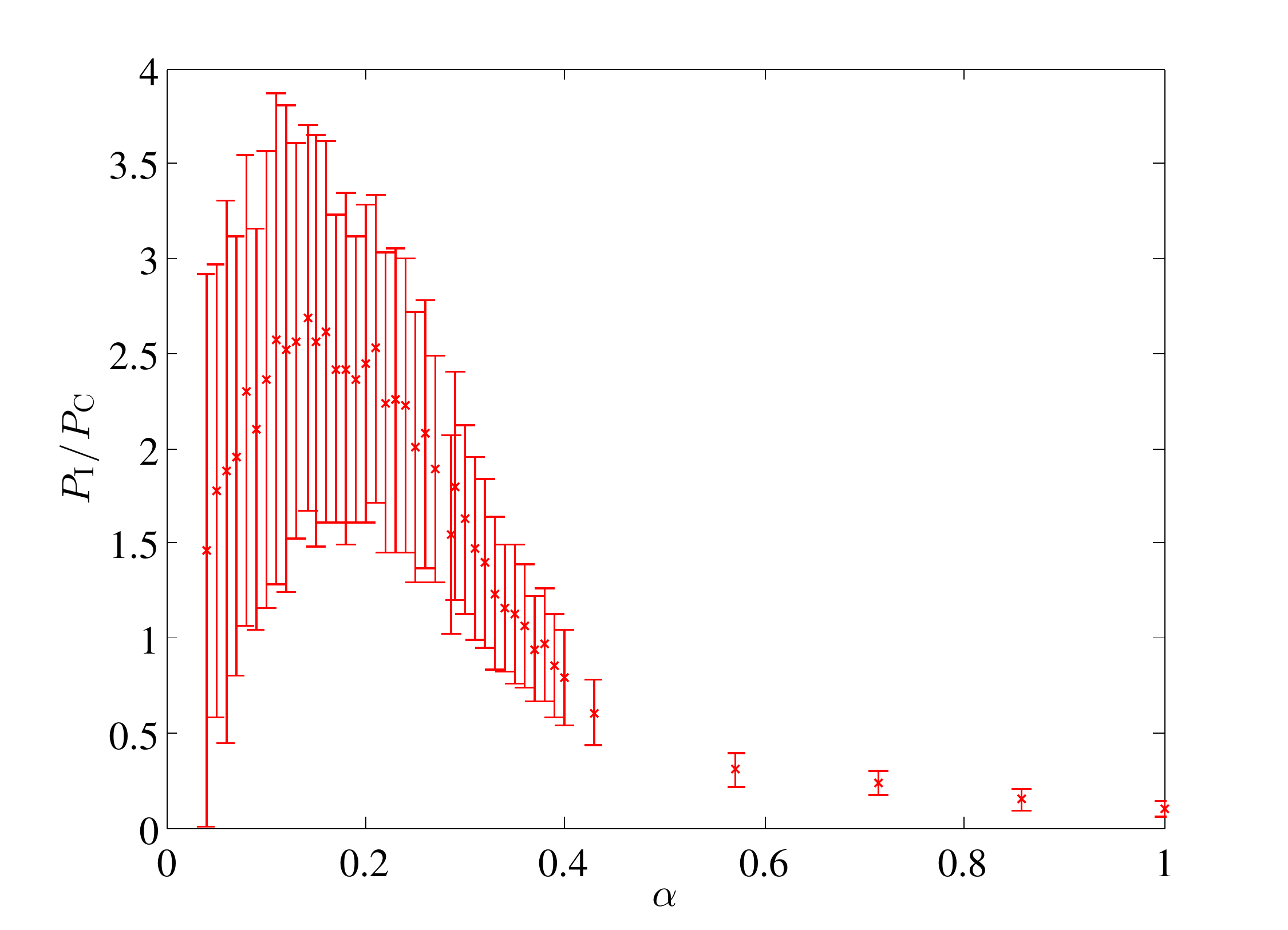}}
\subfigure[\, $N=20$]{\includegraphics[width=0.48\textwidth]{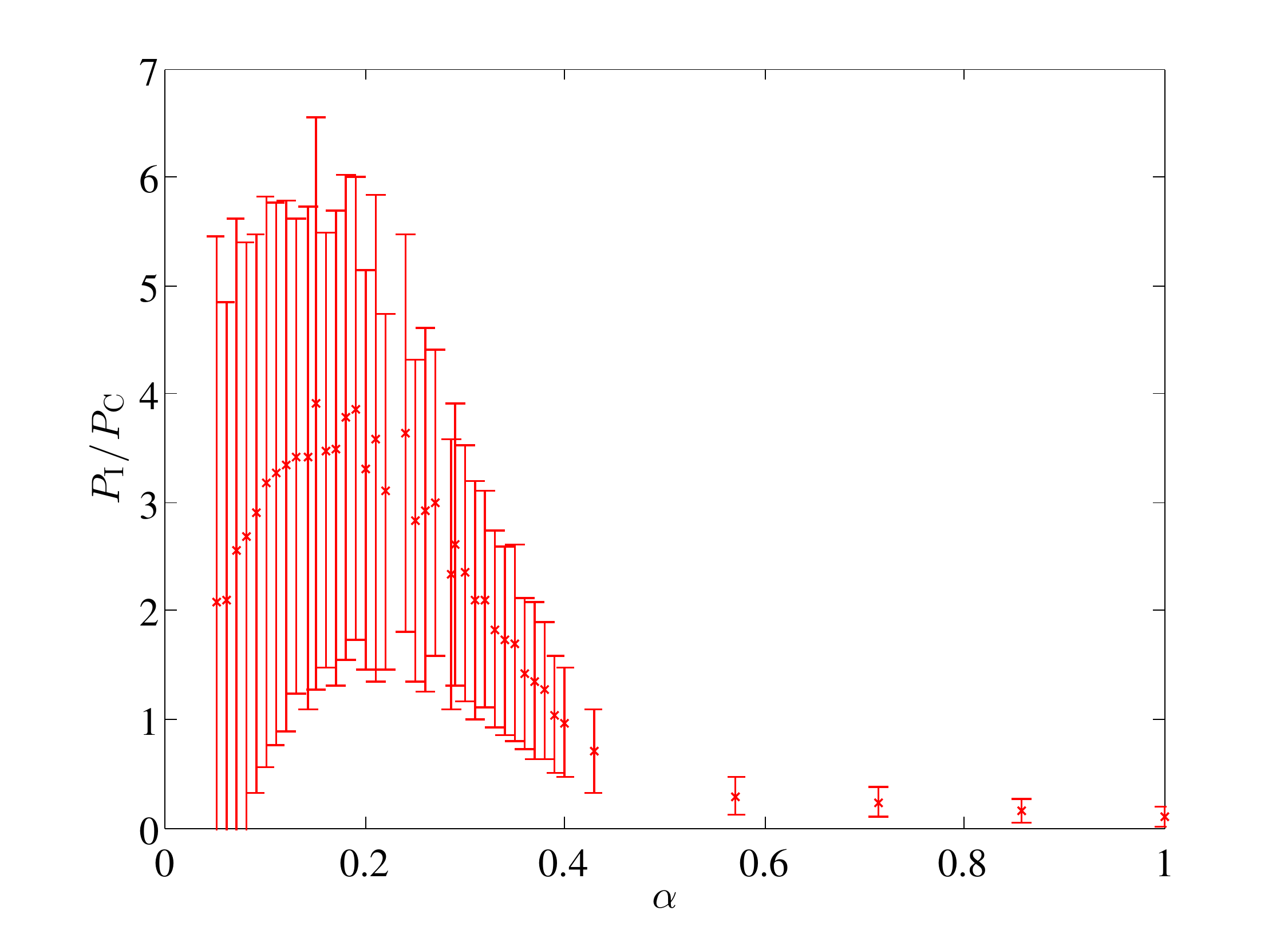}}
\subfigure[\, $N=40$]{\includegraphics[width=0.48\textwidth]{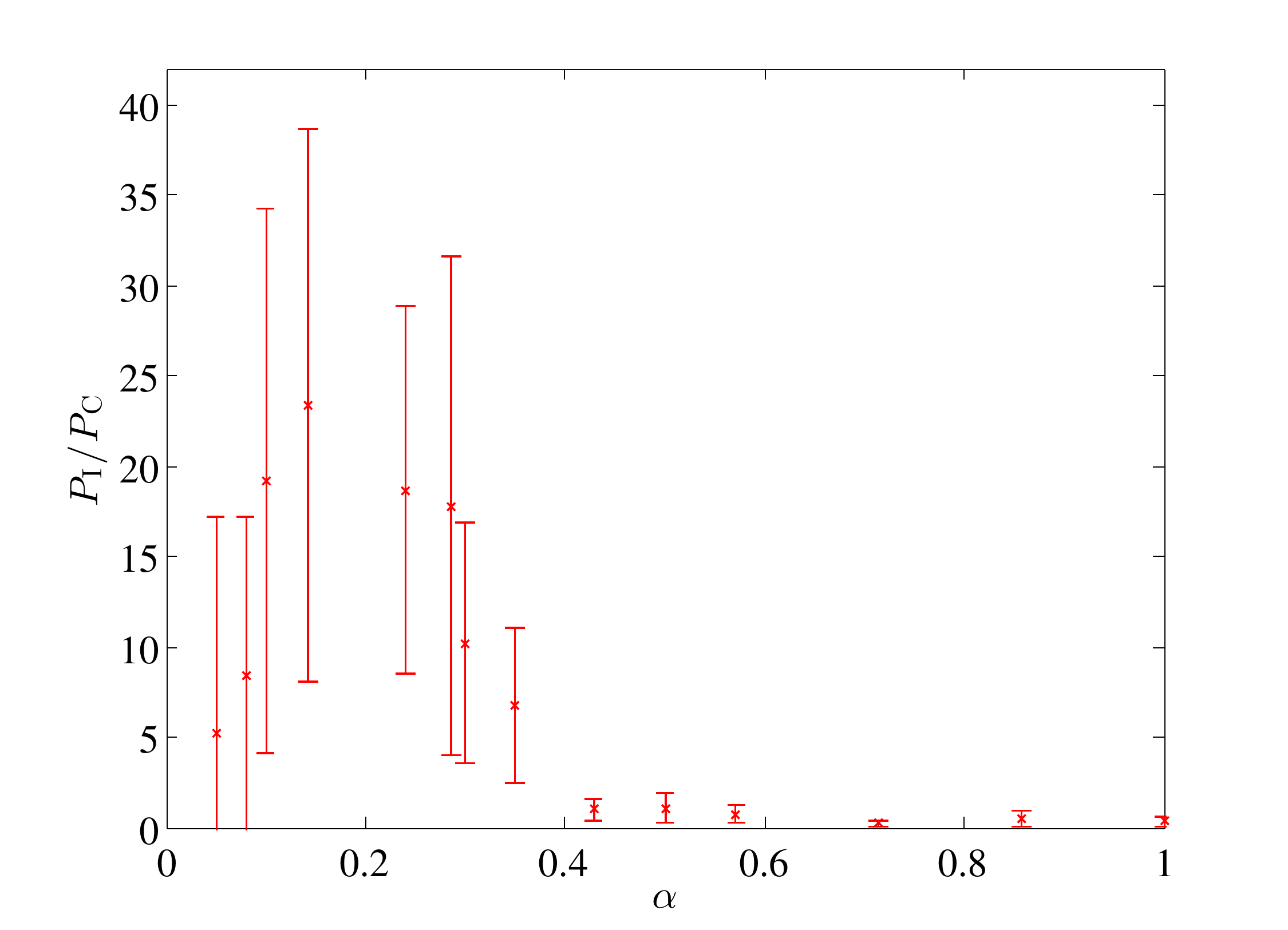} \label{fig:N=40}}
\caption{Ratio of the isolated state population to the average population in the cluster-states ($P_\textrm{I}/P_\textrm{C}$) as a function of the energy scale factor $\alpha$, for different values of $N$, at a fixed annealing time of $t_f=20 \,\mu s$. The non-monotonic dependence of the population ratio on $\alpha$ is observed for all values of $N$. The growth of the $P_\textrm{I}/P_\textrm{C}$ peak with increasing $N$ is consistent with the discussion presented in Appendix~\ref{sec:increase-N}. The increasingly large error bars are due to the smaller amount of data collected as $N$ grows. For $N=12,16,20$ data were collected using the ``random parallel embeddings'' strategy and for $N=40$ using the ``designed parallel embedding" strategy (see Appendix~\ref{app:data-collection} for details).  Error bars are one standard deviation above and below the mean.}
\label{fig:12-16-20-40Ratio}
\end{figure*}

In the main text we discussed the effect of varying the energy scale $\alpha$ of the final Hamiltonian as a means to control thermal excitations. In this section we consider two alternative approaches, namely varying the annealing time or the total number of spins and provide our experimental results.

\subsection{Increasing the total annealing time $t_f$}
\label{sec:increase-t_a}
As reported in Ref.~\cite{q-sig} (which only studied the $\alpha=1$ case), increasing the annealing time reduced the suppression of the isolated state, which is consistent with the effect of increased thermal excitations. To understand this, note that in general the requirement of high ground state fidelity generates a competition between adiabaticity (favoring long evolution times) and suppression of thermal effects (favoring short evolution times) \cite{PhysRevLett.95.250503}. Since the shortest annealing time of the DW2 ($20\mu$s) is already much longer than the inverse of the minimal gap ($\sim (25\, \textrm{GHz})^{-1}$ at $\alpha=1$; see Fig.~\ref{fig:GapVsTime}), increasing the annealing time does not suppress non-adiabatic transitions, but does increase the probability of thermal fluctuations. 
Figure~\ref{fig:8-12-16-EIGS} shows that, as expected, 
an increase in the annealing time is consistent with stronger thermalization, and indeed, over the range of $\alpha$ where we observe suppression of the isolated state ($P_\mathrm{I}/P_\mathrm{C} < 1$), this suppression is weaker for the larger total annealing time. The ME result is in qualitative agreement with the experimental data: The larger annealing time curve is the higher of the two, and the peak values of $P_{\mathrm{I}}/P_{\mathrm{C}}$ at the two different annealing times coincide, which also agrees with the experimental result, within the error bars. 

\subsection{Increasing the number of spins $N$}
\label{sec:increase-N}
There are two important effects to keep in mind when considering larger numbers of spins $N$ (even). First, increasing the number of spins does not change our previous argument that the instantaneous ground state has vanishing support on the isolated state towards the end of the evolution.  We showed this explicitly using first order perturbation theory in Appendix~\ref{app:pert-theory}.  This means that we should still expect that $P_\mathrm{I}/P_\mathrm{C} < 1$ for QA, unless thermal excitations dominate.  Second, the degeneracy of the instantaneous first excited state grows with $N$, while the energy gap to the ground state remains fixed with $N$. The latter is illustrated in Fig.~\ref{fig:8-12-16-EIGS-SM}.
Consequently there is an enhancement of the thermal excitation rate out of the instantaneous ground state into the first excited state, eventually feeding more population into the instantaneous excited states that have overlap with the isolated state. Thus we expect $P_\mathrm{I}/P_\mathrm{C}$ to grow with $N$ (as we indeed find experimentally; see Fig.~\ref{fig:12-16-20-40Ratio}).

We have studied the simplest extensions beyond $N=8$ (examples are shown in Fig.~\ref{fig:12-16-20spinIsing}) with $12$, $16$ and $40$ spins. We expect the same qualitative features observed for $N=8$ to persist, and this is confirmed in Fig.~\ref{fig:12-16-20-40Ratio}, which displays the same qualitative non-monotonic behavior as a function of $\alpha$. The main difference is that the enhancement of the isolated state (when $P_\textrm{I}/P_\textrm{C}>1$) becomes stronger as $N$ is increased. This is a manifestation of the growth, with $N$, in the number of excited states connected to the isolated state as compared to the number connected to the cluster-states. This implies that the excitation rate due to thermal fluctuations is proportionally larger for the isolated state than for the cluster states.

Going to even larger $N$ on the DW2 is prohibitive, since it requires the number of readouts to be $O(2^{N/2})$ in order to collect a statistically significant amount of data.  This is due to the growth of the number of cluster states as described in Eq.~\eqref{eq:degen}. 
\section{Derivation of the O(3) spin-dynamics model} 
\label{app:derivation}
\subsection{Closed system case}
\label{app:O3-closed}
Here we present the standard path integral derivation of the O(3) model, which is closely related to the O(2) SD model of Ref.~\cite{Smolin}.  
Let us introduce the tensor product state of coherent spin-$1/2$ states 
\beq
\ket{\Omega(t)} = \otimes_{i} \left( \cos (\theta_i(t)/2) \ket{0}_i + \sin(\theta_i(t)/2) e^{i \phi_i(t)} \ket{1}_i \right) .
\label{eq:Omegas}
\eeq
We consider the amplitude associated with beginning in $\ket{\Omega(0)} = \otimes_i \ket{\Omega_i(0)} $ and ending in $\ket{\Omega(t_f)}$,
\beq
\mathcal{A} = \bra{\Omega(t_f)}  T_+ e^{-\frac{i}{\hbar} \int_0^{t_f} H(t) dt} \ket{ \Omega(0)} ,
\eeq
where $T_+$ represents time-ordering. We write the integral in terms of a Riemann sum:
\beq
\int_0^{t_f} H(t) dt = \lim_{\nu \rightarrow \infty} \sum_{n=0}^{\nu-1} H(t_n) \Delta t ,
\eeq
where $\Delta t = t_f / \nu$ and $t_n = n \Delta t$, and then perform a Trotter slicing:
\beq
T_+ e^{-\frac{i}{\hbar} \int_0^{t_f} H(t) dt} = \prod_{n=0}^{\nu-1} e^{- \frac{i}{\hbar} H(t_n) \Delta t} + O(\Delta t^2)
\eeq
We now introduce an overcomplete set of spin-coherent states \eqref{eq:Omegas} between the Trotter slices
\beq
\ident = \int d \Omega  \ketbra{\Omega}{\Omega} ,
\eeq
where for general spin $S$
\beq
d \Omega = \prod_i \frac{2 S + 1}{4 \pi} \sin \theta_i d \phi_i d \theta_i ,
\eeq
so that we have:
\begin{align}
\mathcal{A} &= \int d \Omega_1 \cdots \int d \Omega_{\nu-1} \prod_{n=1}^{\nu} \bra{\Omega_{n}} e^{-\frac{i}{\hbar} H_{n-1} \Delta t} \ket{\Omega_{n-1}} \notag \\
&+ O(\Delta t^2)\, ,
\end{align}
where we have denoted $\Omega_\nu \equiv \Omega(t_f)$, $\Omega_0 \equiv \Omega(0)$, and $H_n \equiv H(t_{n})$.  To the same order of approximation we can write
\bes
\begin{align}
& \bra{\Omega_{n}} e^{-\frac{i}{\hbar} H_{n-1} \Delta t} \ket{\Omega_{n-1}}   \\
& \quad = \bra{\Omega_n} \left( \ident - \frac{i}{\hbar} H_{n-1} \Delta t \right)\ket{\Omega_{n-1}} + O(\Delta t^2)  \\
& \quad = \braket{\Omega_n |\Omega_{n-1}} \left( 1 - \frac{i \Delta t}{\hbar} \frac{ \bra{\Omega_n} H_{n-1} \ket{\Omega_{n-1}}}{\braket{\Omega_n|\Omega_{n-1}} } \right) + O(\Delta t^2) .
 \end{align}
\ees
Let us assume differentiability of the states $\Omega_{n}$ and the Hamiltonian $H_n$ so that we can write:
\bes
\begin{align}
\ket{\Omega_{n-1}} & = \ket{\Omega_n} - \Delta t \partial_t \ket{\Omega_n}  + O(\Delta t^2) \\
H_{n-1} &= H_n - \Delta t \partial_t H_n + O(\Delta t^2) .
\end{align}
\ees
Using this differentiability on the overlap, we have:
\bes
\begin{align}
\braket{\Omega_n |\Omega_{n-1}} & = \bra{\Omega_n} \left( \ket{\Omega_n} - \Delta t \partial_t \ket{\Omega_n} \right) + O(\Delta t^2) \\
& = 1 - \Delta t \bra{\Omega_n} \partial_t \ket{\Omega_n} + O(\Delta t^2) \\
& = \exp \left(  - \Delta t \bra{\Omega_n} \partial_t \ket{\Omega_n} \right)
\end{align}
\ees
Likewise, using this differentiability on the matrix element of the Hamiltonian, we have:
\bes
\begin{align}
& \Delta t \bra{\Omega_n} H_{n-1} \ket{\Omega_{n-1}}   \\
&= \Delta t \bra{\Omega_n} (H_n - \Delta t \partial_t H_n) (\ket{\Omega_n} - \Delta t \partial_t \ket{\Omega_n})  + O(\Delta t^2)  \\
& =\Delta t \bra{\Omega_n} H_n\ket{\Omega_n}  + O(\Delta t^2) 
\end{align}
\ees
Putting these results together, we have for the amplitude:
\bes
\begin{align}
\mathcal{A} & =  \int d \Omega_1 \cdots \int d \Omega_{\nu-1} \times \\
&e^{ \frac{i}{\hbar} \Delta t \sum_{n=1}^{\nu} \left( i \hbar \bra{\Omega_n} \partial_t \ket{\Omega_n} -  \bra{\Omega_n} H_{n} \ket{\Omega_{n}} \right)} + O(\Delta t^2) \\
& = \int \mathcal{D} \Omega \exp \left[ \frac{i}{\hbar} \int d t \left( i \hbar \bra{\Omega} \partial_t \ket{\Omega} -  \bra{\Omega} H(t) \ket{\Omega}  \right)\right] \\
&=  \int \mathcal{D} \Omega \ e^{\frac{i}{\hbar} S[\Omega]}
\end{align}
\ees
where we have taken the continuum limit such that $\Omega_n \to \Omega(t)$ and introduced the action
\beq
S[\Omega] = \int d t \mathcal{L} = \int d t \left( i \hbar \bra{\Omega} \partial_t \ket{\Omega} -  \bra{\Omega} H(t) \ket{\Omega}  \right)\, .
\eeq
For simplicity, let us now work in units of $\hbar = 1$.  Using Eq.~\eqref{eq:Omegas}
we can write the first term in the action as:
\beq
 i  \bra{\Omega} \partial_t \ket{\Omega} = - \frac{1}{2} \sum_{i} \left( 1- \cos \theta_i \right) \frac{d \phi_i}{dt}
\eeq
The Euler-Lagrange equations of motion 
\bes
\begin{align}
\frac{d}{dt}\left(\frac{\partial \mathcal{L}}{\partial \dot \phi_i}\right) - \frac{\partial \mathcal{L}}{\partial \phi_i} &= 0\, , \\
\frac{d}{dt}\left(\frac{\partial \mathcal{L}}{\partial \dot \theta_i}\right) - \frac{\partial \mathcal{L}}{\partial \theta_i} &= 0
\end{align}
\ees 
extremize the action and yield the semi-classical saddle point approximation:
\bes
\label{eq:saddle}
\begin{align}
\frac{1}{2} \sin \theta_i \frac{d}{dt} \theta_i - \frac{\partial}{\partial \phi_i} \bra{\Omega} H(t) \ket{\Omega} = 0\, , \\
- \frac{1}{2} \sin \theta_i \frac{d}{dt} \phi_i - \frac{\partial}{\partial \theta_i} \bra{\Omega} H(t) \ket{\Omega} = 0\, .
\end{align}
\ees
These are the equations of motion for the $O(3)$ model, where $\bra{\Omega} H(t) \ket{\Omega}$ plays the role of a time-dependent potential.  

\begin{figure*}[ht] %
   \centering
       \subfigure[\ Closed system]{ \includegraphics[width=0.98\columnwidth]{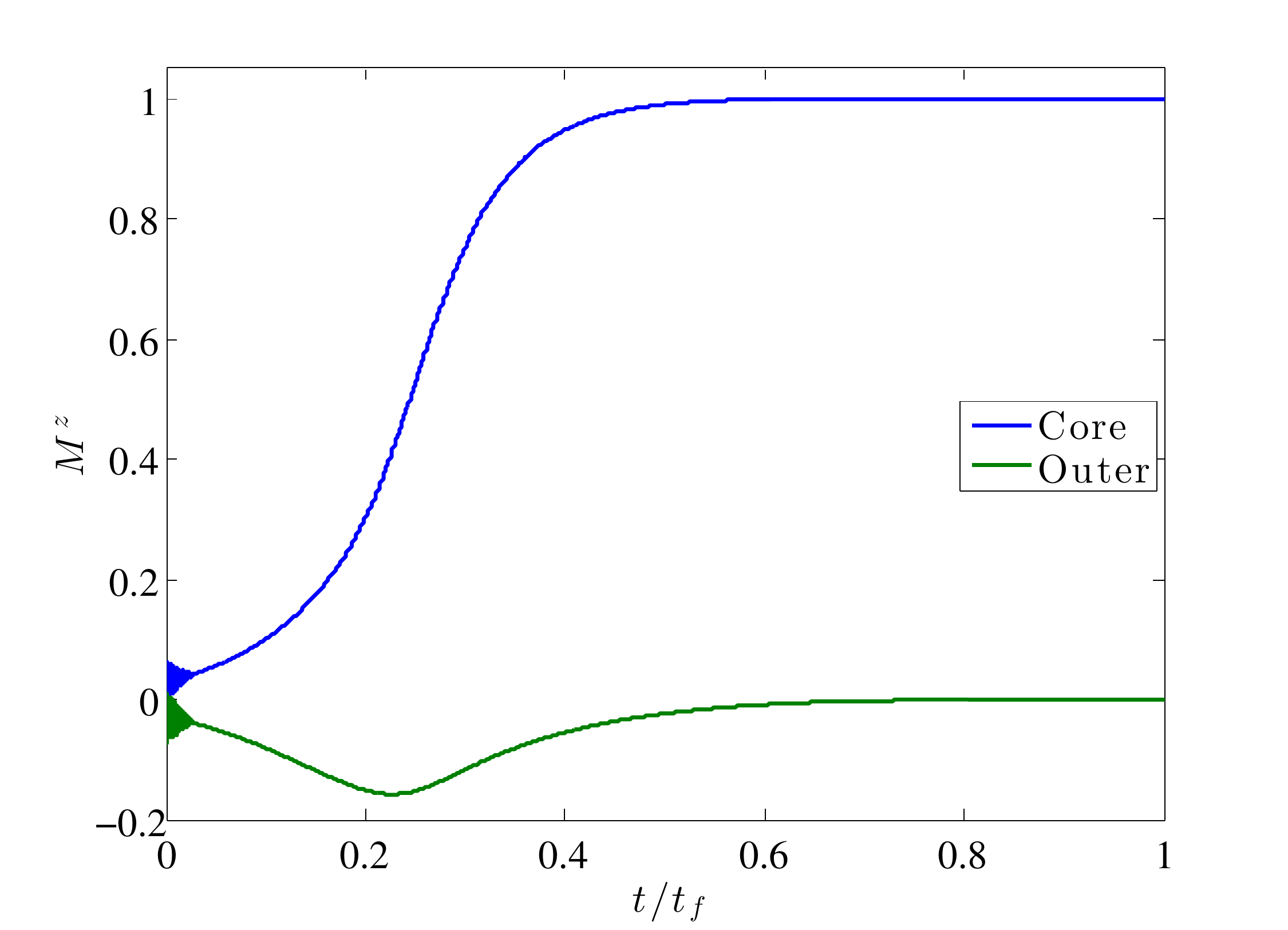} \label{fig:O3ClosedEvolution}}
        \subfigure[\ Open system]{ \includegraphics[width=0.98\columnwidth]{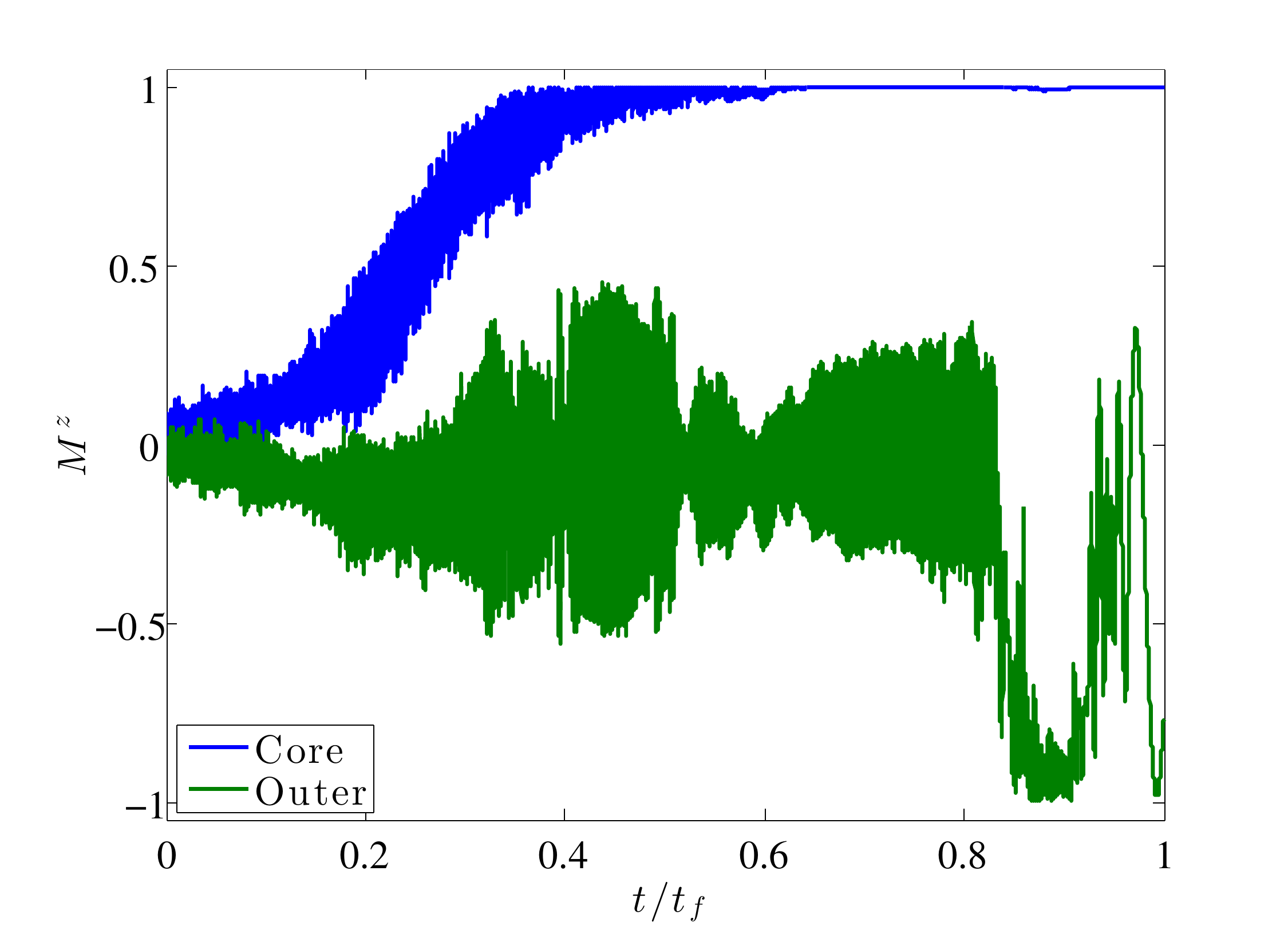} \label{fig:O3OpenEvolution}}
   \caption{\protect{Evolution of a core (blue) and outer (green) spin with $t_f = 20 \, \mu s$, subject to the O(3) SD model with $\alpha=1$. All spins start with $M^x = 1$, $M^y = M^z = 0$, i.e., point in the $x$ direction.  (a) Closed system case given by Eq.~\eqref{eqt:O3}. (b) Open system case given by Eq.~\eqref{eq:open-O3}. Rapid oscillations at the beginning of the evolution in (a) are because the initial conditions used are not exactly the ground state of the system [because of the finite $B(0)$].  In (b), Langevin parameters are $k_B T / \hbar = 2.226 \, \mathrm{GHz}$ and $\zeta = 10^{-6}$.}}
   \label{fig:O3Evolution}
\end{figure*}
\begin{figure*}[t] %
   \centering
       \subfigure[\ Core spins]{ \includegraphics[width=0.98\columnwidth]{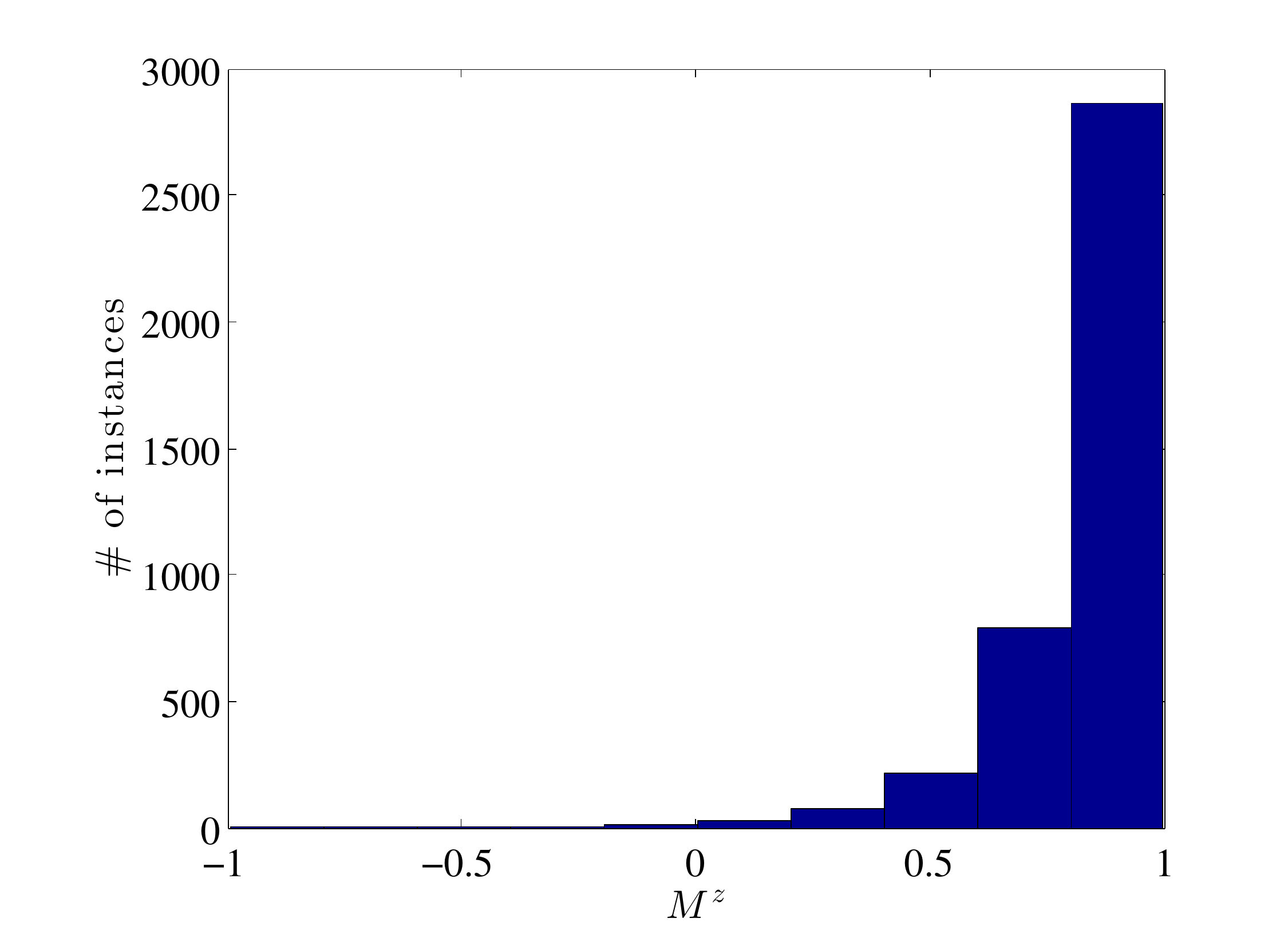} \label{fig:O3CoreAHist}}
        \subfigure[\ Outer spins]{ \includegraphics[width=0.98\columnwidth]{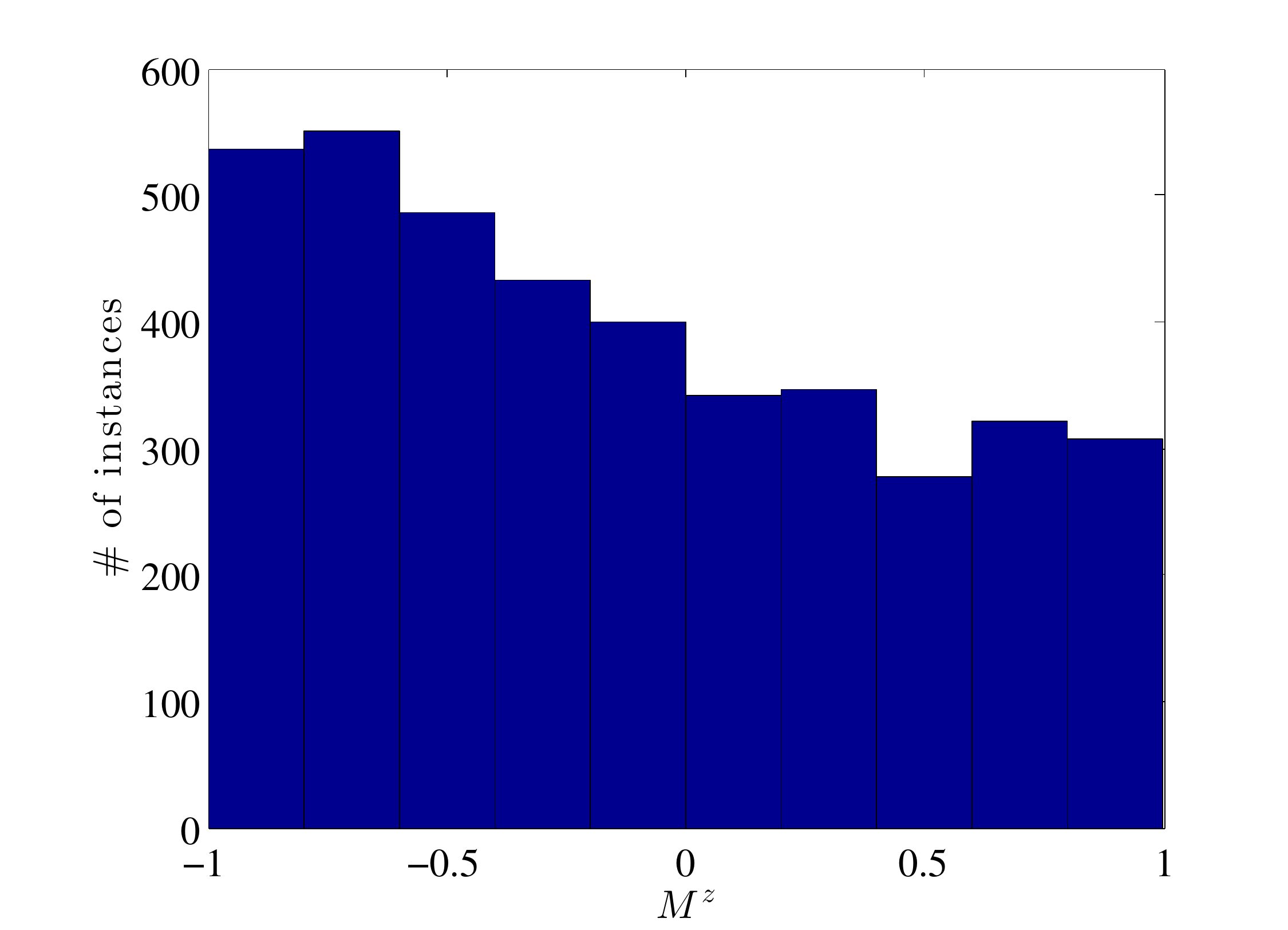} \label{fig:O3outerAHist}}
   \caption{\protect{Distribution of $M^z$ at the end of the evolution for all (a) core spins and (b) all outer spins 
for $\alpha = 1$.  Langevin parameters are $k_B T / \hbar = 2.226 \, \mathrm{GHz}$ and $\zeta = 10^{-3}$.  Data collected using $1000$ runs of Eq.~\eqref{eq:open-O3}.}}
   \label{fig:O3Histogram}
\end{figure*}

For a Hamiltonian of the form of Eq.~\eqref{eq:problem}, the equations of motion \eqref{eq:saddle} become, in terms of the magnetization $\vec{M} = \Tr \left( \vec{\sigma} \rho \right)$,
\bes
\label{eqt:O3}
\begin{align}
\frac{d}{dt} \dot{\vec{M}}_i &=   - \vec{H}_i \times \vec{M}_i \\
\vec{H}_i & \equiv  2 A(t) \hat{x}+ 2 \alpha B(t) \left( h_i  + \sum_{j \neq i}  J_{ij}  \vec{M}_j \cdot \hat{z}  \right) \hat{z} \, ,
\end{align}
\ees
where we have already included the $\alpha$ dependence. Using the DW2 annealing schedule in Fig.~\ref{fig:AnnealingSchedule} we plot the evolution of the spin system in Fig.~\ref{fig:O3Evolution}(a). This figure shows that the system evolves to a cluster state. Namely, the core spins have $M^z = 1$, i.e., are in the $\ket{0}$ state, and the outer spins have $M^z = 0$.  Since the outer spins have eigenvalues $\pm 1$ under $\sigma^z$ with equal probability, having the average equal zero is consistent with having an equal distribution among the cluster states. This suppression of the isolated state result is consistent with the QA evolution, and was used in Ref.~\cite{Smolin} to critique the conclusion of Ref.~\cite{q-sig} that the experimental evidence is consistent with quantum evolution. In the next subsection we discuss the effect of adding thermal noise and a dependence on the energy scale factor $\alpha$. 
%
%

\subsection{Open system case: Langevin equation}
\label{app:O3-open}
%
Now that we have our ``classical'' model, we introduce a thermal bath by extending the equations of motion to an appropriately generalized (Markovian) spin-Langevin equation \cite{Jayannavar:1990,PTPS.46.210},
\beq \label{eqt:LangevinO3}
 \frac{d}{dt} \vec{M}_i  = -\left( \vec{H}_i + \vec{\xi}(t) - \zeta  \frac{d}{dt} \vec{M}_i  \right) \times \vec{M}_i \ ,
 \eeq
with the Gaussian noise $\vec{\xi}$ satisfying $
\braket{\xi_i(t)} = 0$ and $\braket{\xi_i(t) \xi_i(t')} = 2 k_B T \zeta \delta(t- t')$.
One can simplify Eq.~\eqref{eqt:LangevinO3} by perturbatively inserting $ \frac{d}{dt} \vec{M}_i  =  -\vec{H}_i  \times \vec{M}_i$ into the ``friction'' term to get:
\beq
 \frac{d}{dt} \vec{M}_i  = -\left( \vec{H}_i + \vec{\xi}(t) + \zeta  \vec{H}_i \times \vec{M}_i  \right) \times \vec{M}_i \ ,
\label{eq:open-O3}
\eeq
which gives rise to a ``Landau-Lifshitz'' friction term \cite{LL:1935,PTPS.46.210} and is the evolution equation \eqref{eq:spin-Langevin}.  

An example of the resulting evolution for $\alpha=1$ is shown in Fig.~\ref{fig:O3Evolution}(b).  Note that the $M^z$ value of the outer spins does not converge to $0$, unlike the closed system case shown in Fig.~\ref{fig:O3Evolution}(a). This is not accidental: While the core spins prefer the $\ket{0}$ state [$M^z=1$, Fig.~\ref{fig:O3Histogram}(a)], the outer spins prefer the $\ket{1}$ state, i.e., the median occurs at $M^z < 0$, as is clearly visible in Fig.~\ref{fig:O3Histogram}(b). An explanation in terms of the effective Ising potential between a core-outer spin pair is given in the main text (Sec.~V) for the noisy SSSV model, but the same applies to the SD model.

The dependence on $\alpha$ is given in Fig.~\ref{fig:O3BoxPlot}.  We observe that as $\alpha$ is decreased, the median value of the core spins and outer spins does not significantly change.  However, we do observe a very slow decrease away from $M^z = 1$ for the core spins.  A larger effect is the appearance of more outliers as $\alpha$ decreases, which is consistent with the system being able to explore states away from the cluster states. However, we emphasize that the majority of the states observed are cluster states and not the isolated state (in contradiction with the ME results). 

\begin{figure*}[h] %
\centering
\subfigure[\ $\zeta = 10^{-1}$]{ \includegraphics[width=0.98\columnwidth]{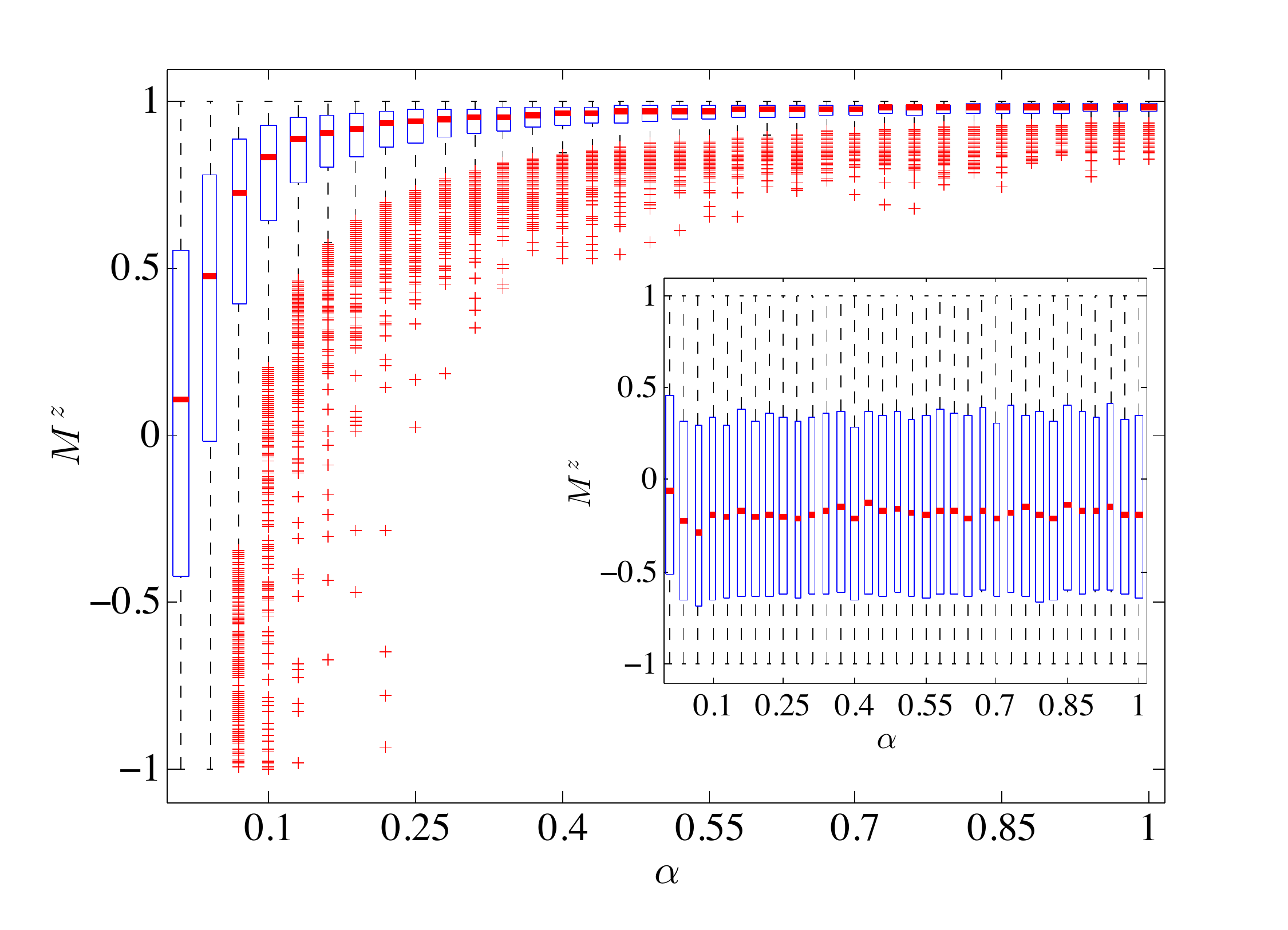} 
\label{fig:O3CoreA1}}	
\subfigure[\ $\zeta = 10^{-5}$]{ \includegraphics[width=0.98\columnwidth]{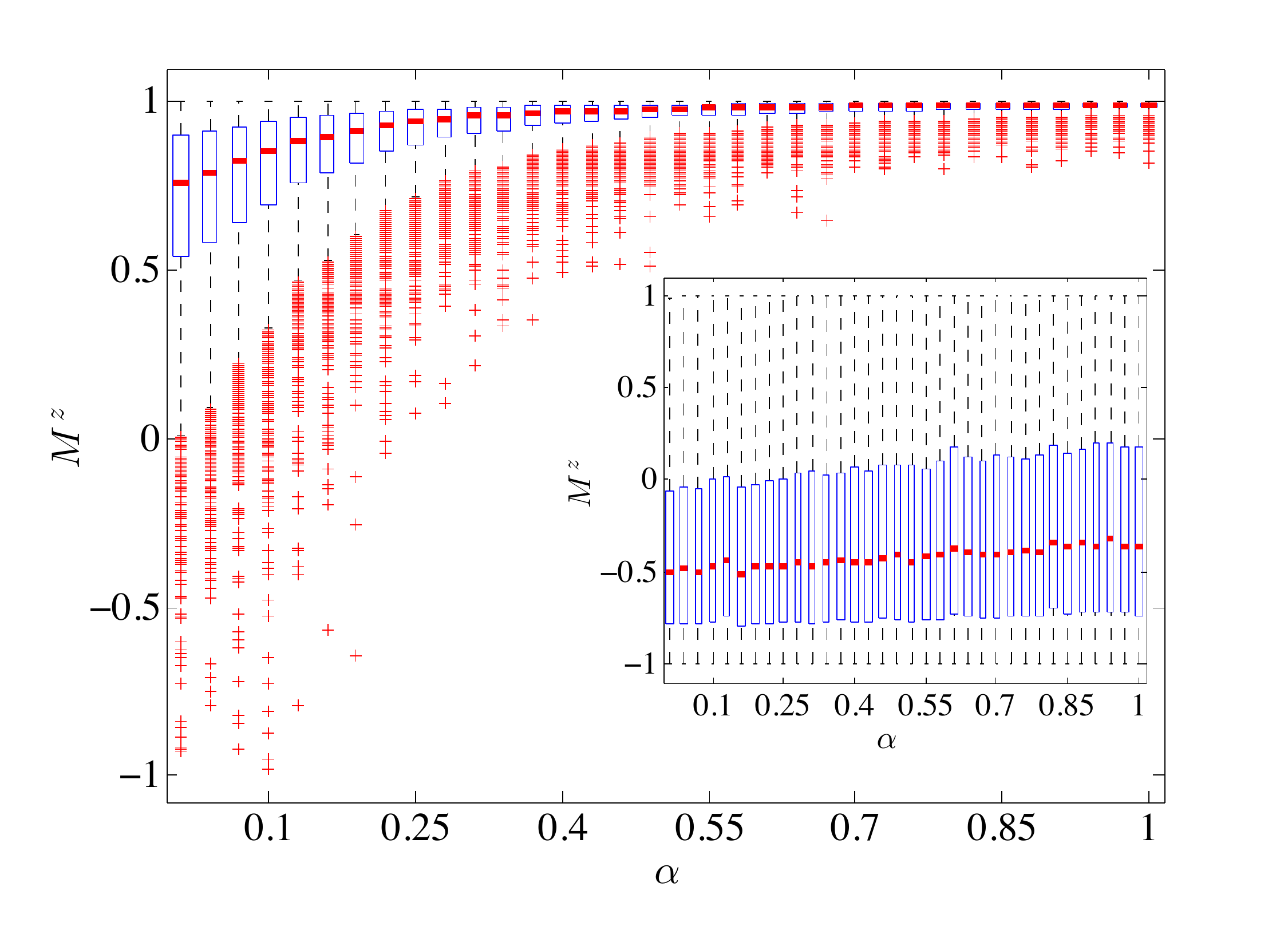} \label{fig:O3CoreA2}}
\caption{Statistical box plot of the $z$ component for all core qubits (main plot) and all outer qubits (inset) at $t= t_f = 20 \mu s$. 
The data is taken for $1000$ runs of Eq.~\eqref{eq:open-O3} with Langevin parameters $k_B T / \hbar = 2.226 \, \mathrm{GHz}$ (to match the operating temperature of the DW2) and (a) $\zeta=10^{-1}$ and (b) $\zeta=10^{-5}$.  The $\zeta=10^{-3}$ is shown in Fig.~\ref{fig:O3BoxPlot}(a). This illustrates that the results do not depend strongly on the choice of $\zeta$.} \label{fig:O3BoxPlotsup1}
\end{figure*}

We have checked the dependence of our results on the friction parameter $\zeta$.  In Fig.~\ref{fig:O3BoxPlotsup1}(a) we see that for sufficiently large $\zeta$ ($> 10^{-3}$), the median values are not affected significantly by changing $\zeta$.  For sufficiently small $\zeta$ [Fig.~\ref{fig:O3BoxPlotsup1}(b)] we observe that the median value of the core spins does not deviate very far from $1$, and the median of the outer spins appears to shift even further towards $M^z_{\textrm {o}} = -1$.

\begin{figure*}[h] %
\centering
\subfigure[\ Core spins]{ \includegraphics[width=0.47\textwidth]{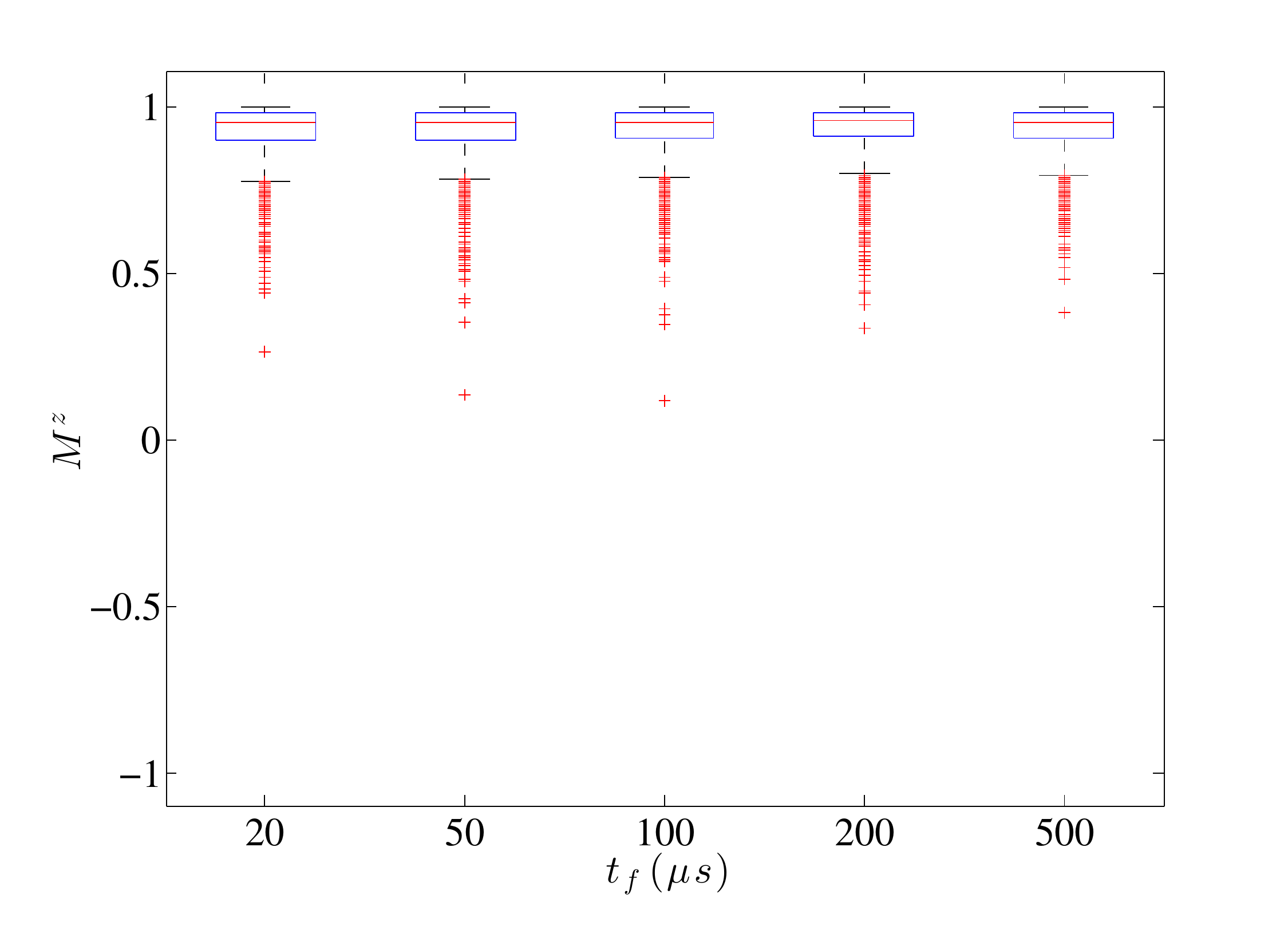} \label{fig:O3CoreAtf}}
\subfigure[\ Outer spins]{ \includegraphics[width=0.47\textwidth]{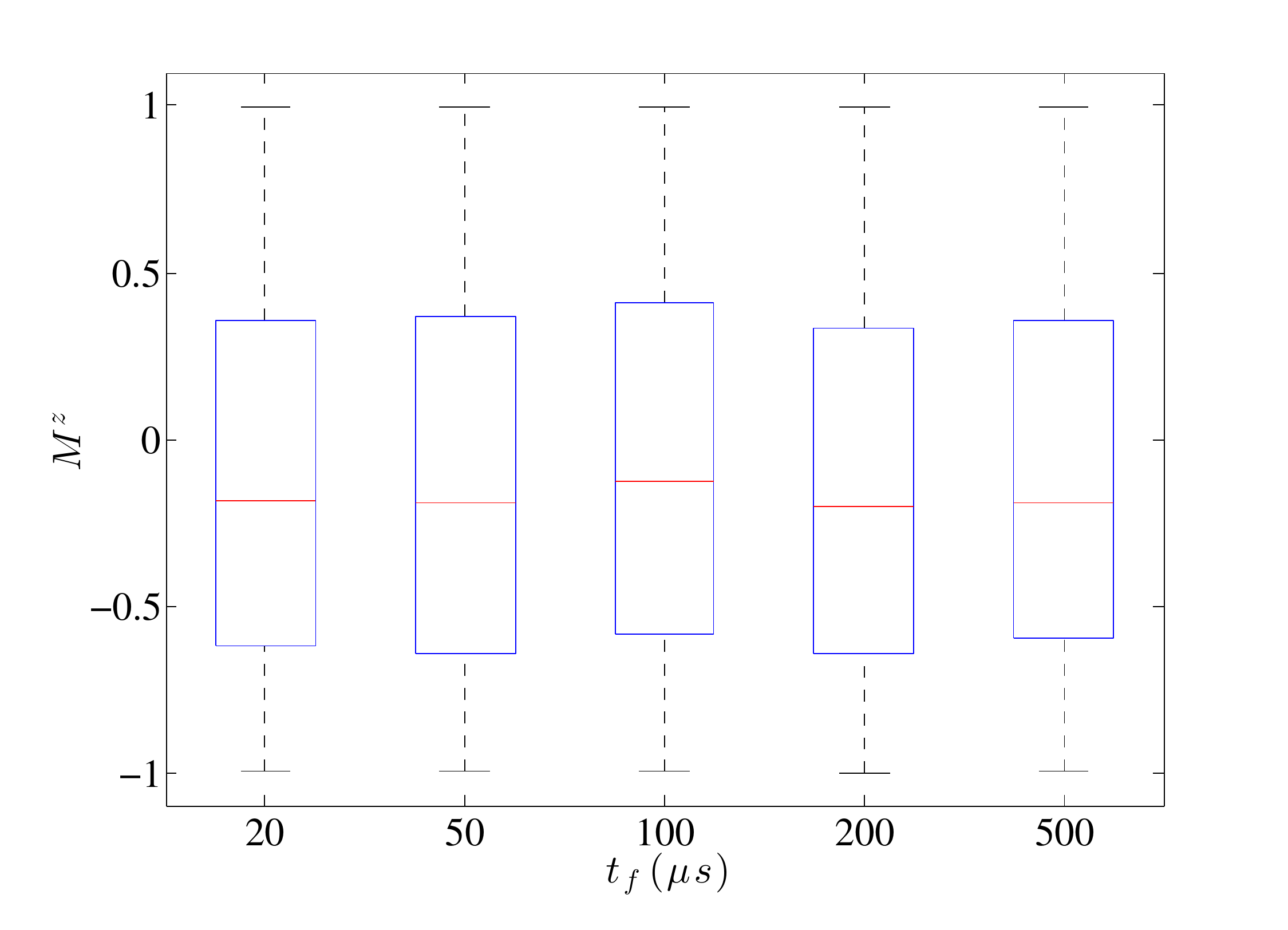} \label{fig:O3outerAtf}}
\caption{\protect{Distribution of $M^z$ at the end of the evolution for (a) all core spins and (b) all outer spins (i.e., the values of all core spins and all outer spins are included in each respective box plot) for $\alpha = 2/7$.  Langevin parameters are $k_B T / \hbar = 2.226 \, \mathrm{GHz}$ and  $\zeta = 10^{-3}$.  Data collected using $1000$ runs of Eq.~\eqref{eq:open-O3}.  Note that the $t_f$-axis scale is not  linear.}} \label{fig:O3BoxPlotsup2}
\end{figure*}

We have also checked the dependence on the annealing time. As shown in Fig.~\ref{fig:O3BoxPlotsup2},  there is no significant change in the median for either the core or the outer spins, suggesting that (over the range of annealing times studied) the system does not fully thermalize.

\end{document}